\documentclass{aa}

\usepackage{graphicx}
\usepackage{caption}
\usepackage{subcaption}
\usepackage{placeins}
\usepackage{float}

\usepackage{afterpage}
\usepackage{longtable}
\usepackage{stfloats}
\usepackage[varg]{txfonts}
\usepackage{lscape}
\usepackage{multicol}
\mathchardef\mhyphen="2D
\usepackage[colorlinks=true, allcolors=blue]{hyperref}

\usepackage{rotating}

\begin{document}
	\title{VLTI/PIONIER imaging of post-AGB binaries}

	\subtitle{An INSPIRING hunt for inner rim substructures in circumbinary discs\thanks{Based on observations collected at the European Southern Observatory under ESO programme 1103.D-0797.}}

	\author{ T.~De~Prins\inst{1,2,3,}
	\thanks{Corresponding author, \texttt{toon.deprins@kuleuven.be}}
	\and A.~Corporaal \inst{4} \and J.~Kluska \inst{1} \and D.~Kamath \inst{2,3} \and
	H.~Van~Winckel\inst{1} \and K.~Andrych \inst{2,3} \and J.~Alcolea \inst{5}
	\and N.~Anugu \inst{6} \and J.-P.~Berger \inst{7} \and V.~Bujarrabal \inst{8} \and
	I.~Gallardo~Cava \inst{5} \and S.~Kraus \inst{9} \and H.~Olofsson \inst{10}}

	\institute{Institute of Astronomy, KU Leuven, Celestijnenlaan 200D, 3001 Leuven, Belgium \and School of Mathematical and Physical Sciences, Macquarie University, Balaclava Road, Sydney, NSW 2109, Australia \and Astrophysics and Space Technologies Research Centre, Macquarie University, Balaclava Road, Sydney, NSW 2109, Australia \and European Southern Observatory, Alonso de Córdova 3107, Vitacura, Casilla 19001, Santiago, Chile \and Observatorio Astronómico Nacional (OAN-IGN), Alfonso XII 3, 28014 Madrid, Spain \and The CHARA Array of Georgia State University, Mount Wilson Observatory, Mount Wilson, CA 91023, USA \and Univ. Grenoble Alpes, CNRS, IPAG, F-38000 Grenoble, France \and Observatorio Astronómico Nacional (OAN-IGN), Apartado 112, 28803 Alcalá de Henares, Madrid, Spain \and Department of Physics and Astronomy, University of Exeter, Stocker Road, Exeter EX4 4QL, UK \and Department of Space, Earth and Environment, Chalmers University of Technology, Onsala Space Observatory, 43992 Onsala, Sweden}

	\date{Received 09 December 2025 / Accepted 11 May 2029}

	\abstract{Post-asymptotic giant branch (post-AGB) binaries are surrounded by
	stable dusty discs. Despite extensive evidence for disc-binary interactions and
	complex morphologies, systematic studies of the inner disc rim morphology and
	dynamics are lacking.}
	{We image the detailed inner rim morphology for a sample of 'full' post-AGB circumbinary discs observed using near-infrared interferometry. At resolutions down to $\sim\!1\!-\!2\,\mathrm{mas}$ -- corresponding to $\sim \!1\!-\!10\,\mathrm{AU}$ -- we aim to reveal potential substructures that may trace perturbations from the binary or embedded substellar companions, or that arise from hydrodynamical instabilities.}
	{We developed a systematic image reconstruction workflow using a combination of the SPARCO and ORGANIC algorithms. This was applied to VLTI/PIONIER data of eight diverse post-AGB binaries obtained within the INSPIRING ESO large programme, providing high-fidelity images of dust continuum emission at the inner rim. Extensive tests were applied to assess the robustness of the recovered image features.}
	{The dusty disc rim is well-resolved for all targets. Only one of the images can be fully accounted for by simple radiative transfer effects due to disc inclination, while in several others indications of potential substructures are detected. Strikingly, four exhibit robust azimuthal brightness enhancements at locations not expected from inclination effects alone. These can indicate strong radiative or dynamical responses to the binary, or vortices formed via hydrodynamical instabilities. One target -- IW~Car -- displays an even more puzzling morphology, showing a single large-scale outer flux arc and several small-scale arcs closer to the binary, possibly revealing accretion streams onto the binary, a misaligned innermost disc, or a spiral feature located in the disc or in an outflow.}
	{This work presents the first homogeneous interferometric imaging survey of the inner regions of post-AGB circumbinary discs, enabling direct comparison of inner rim morphologies across a representative sample. The inner disc regions are highly diverse and dynamic, harbouring numerous candidate substructures. Multi-wavelength and time-series imaging will be essential in constraining the extent, motion and wavelength-dependence of these features and uncovering their physical origins.}


	\maketitle
	\nolinenumbers

	\section{Introduction}
	Dusty circumstellar discs are observed in systems at various phases of stellar
	evolution. One such type of system is the post-asymptotic giant branch (post-AGB)
	binary \citep[for a recent review, see][and references therein]{VanWinckel2025}.
	The post-AGB phase represents a brief period of contraction after the envelope
	of a low-to-intermediate mass star has been stripped on the AGB. The star
	remains at a constant luminosity of the order of
	$\sim\!10^{3}\!-\!10^{4}\,\mathrm{L_{\odot}}$ \citep[][]{VanWinckel2003},
	heating up until it reaches the white dwarf phase in
	$\sim\!10^{4}\!-\!10^{5}\,\mathrm{yr}$ \citep[][]{MillerBertolami2016}. A
	large subset of post-AGB systems -- about one third -- show spectral energy
	distributions (SEDs) with broad infrared (IR) excesses, indicative of stable dusty
	circumbinary discs \citep[e.g.][]{deRuyter2006, Kamath2014, Kamath2015, Kluska2022}.
	The binarity of these disc sources has been confirmed through extensive radial
	velocity (RV) monitoring, showing that the companions are likely on the main
	sequence \citep[][]{Oomen2018}. Due to the strong link between IR excess and
	binarity, it is thought that the disc is formed from envelope material ejected
	during a poorly understood phase of binary interaction on the AGB \citep[][]{VanWinckel2019}.
	Recently, \citet{Kluska2022} presented an SED catalogue of 85 systems,
	encompassing all disc-bearing Galactic post-AGB binaries known up to that date.
	Based on these SEDs, the authors showed that the majority of the dust discs
	can be considered part of the 'full' disc category, defined as discs with inner
	rims close to the dust sublimation radius and a corresponding excess starting in
	the near-infrared (NIR). In addition, they also identified a sizeable population
	of 'transition' discs, typified by larger inner cavities -- confirmed using optical
	interferometry by \citet{Corporaal2023b} -- and excess starting in the mid-infrared
	(MIR).

	The stability of post-AGB circumbinary discs is confirmed through the detection
	of Keplerian velocity fields in single-dish and interferometric (sub-)$\mathrm{mm}$
	observations of CO lines \citep[][]{Bujarrabal2013a, Bujarrabal2013b, Bujarrabal2015, Bujarrabal2017, Bujarrabal2018, GallardoCava2021},
	revealing gas discs with outer radii of $\sim \! 100 - \!1000\,\mathrm{AU}$ (angular
	sizes $\lesssim \! 1.5 \!- \!2 \, \mathrm{\arcsec}$ at typical system
	distances of $\gtrsim \! 1 \,\mathrm{kpc}$). IR spectroscopy, $\mathrm{mm}$ photometry
	and $\mathrm{mm}$ interferometry reveal significant silicate grain crystallinity
	and grain growth \citep[e.g.][]{Gielen2008, Sahai2011, Scicluna2020, Bujarrabal2023}
	-- in some cases up to at least $\mathrm{mm}$ size -- showing that dust
	processing has proceeded to an advanced stage. The hot inner disc rim, at a
	typical size of several $\mathrm{AU}$ ($\sim \! 1 \! - \! 10 \, \mathrm{mas}$),
	has been resolved in thermal dust emission using optical interferometry \citep[e.g.][]{Kluska2019}.
	The outer disc regions are instead resolved in scattered light using single-telescope
	adaptive optics imagers \citep[e.g.][]{Andrych2023, Andrych2025}. Together with
	extensive radiative transfer (RT) modelling of the inner rim from optical interferometric
	data \citep[][]{Hillen2014, Hillen2015, Kluska2018, Corporaal2023a}, these observations
	have generally shown that post-AGB circumbinary discs display high degrees of similarity
	to passively irradiated protoplanetary discs (PPDs) around young stellar objects
	(YSOs).

	Post-AGB circumbinary discs are not isolated from the rest of their system.
	While the influence of the central binary on the circumbinary disc morphology
	remains poorly constrained, there is strong observational evidence that the disc
	significantly affects the central binary. First, systematic depletion of dust-forming
	elements in the post-AGB primary's atmosphere, a phenomenon called 'refractory
	depletion', can be explained through dust trapping and subsequent re-accretion
	of cleaned gas from the circumbinary disc \citep[e.g.][]{Giridhar1998, VanWinckel2003, Oomen2019, Mohorian2024, Mohorian2025}.
	Second, spectroscopic time-series reveal strong blue-shifted absorption in the
	Balmer lines, caused by a magnetically launched jet from a circumcompanion accretion
	disc that is itself likely fed by re-accretion from the larger circumbinary disc
	\citep[e.g.][]{Gorlova2012, Bollen2022, Verhamme2024, DePrins2024}. This circumcompanion
	disc can be detected as a point source in optical interferometric observations
	\citep[][]{Hillen2016, Anugu2023}. In addition, \citet{Kluska2022} and \citet{Mohorian2025}
	found a correlation between transition discs and refractory depletion strength,
	leading the former to suggest that giant planets may be carving the transition
	disc cavities -- as has been similarly proposed for PPDs around YSOs \citep[e.g.][]{Kama2015}
	-- and raise the possibility of second-generation planet formation. Together,
	these findings demonstrate that the inner regions of post-AGB circumbinary
	discs must be viewed as highly dynamic environments, shaped by ongoing interactions.

	Given their general similarities, it is natural to compare post-AGB discs with
	their younger PPD counterparts. Around PPDs, high-resolution thermal (sub)-$\mathrm{mm}$
	and optical scattered-light imaging has revealed a wealth of substructures -- spirals,
	arcs, rings, gaps, shadows, and crescents \citep[e.g.][]{Andrews2018, Garufi2024}.
	These features have become central to studies of planet formation, as they are
	often interpreted as signposts of embedded planets or as outcomes of instabilities
	that may themselves promote planet formation \citep[for a detailed review, see][]{Bae2023}.
	Indeed, previous efforts to resolve the outer regions of 11 post-AGB
	circumbinary discs with scattered light imaging \citep[][]{Ertel2019, Andrych2023, Andrych2024, Andrych2025}
	have also shown diverse candidate substructures, including arc-, spiral- and ring-like
	features. Conversely, the detailed morphology and presence of substructures near
	the inner post-AGB disc rim, where the effects of the binary are expected to
	be strongest, have been investigated to a much lesser extent.

	While the inner rim has been resolved for larger samples using optical interferometry,
	this was typically through sparse observations suitable only for limited
	geometric modelling \citep[][]{Hillen2017, Kluska2019, Corporaal2023b}. Given
	the expected complexity of the inner disc, information is preferably extracted
	using image reconstruction instead. Indeed, non-parametric image reconstructions
	have proven to be a powerful tool to reveal unexpected features in the inner regions
	of various systems -- including PPDs \citep[e.g.][]{Kluska2016_MWC158, Kluska2020a, Ibrahim2023, Setterholm2025},
	stellar outflows \citep[e.g.][]{Weigelt2016,Planquart2024} and stellar surfaces
	\citep[e.g.][]{Drevon2024} -- whose presence would often be hard to infer using
	parametrised geometric models. Imaging of post-AGB inner circumbinary disc
	rims has nevertheless been limited to only two single-object studies, targeting
	IRAS~08544-4431 and HD~101584 \citep[][]{Hillen2016, Kluska2020b}, respectively
	revealing a perturbed azimuthal brightness distribution and a secondary ring --
	the latter possibly marking a secondary condensation front in a disc wind. These
	findings clearly deviated from the classical view of a smooth, inclined
	circular rim centred on a single star, where the emission is simply expected to
	be brightest and symmetric along the far side of the projected rim due to self-shadowing
	of the near side \citep[e.g.][]{Hofmann2022}.

	In this study, we aim to systematically investigate the morphological complexity
	of post-AGB circumbinary disc inner rims. To this end, we present the first systematic
	NIR interferometric imaging survey for a diverse sample of targets, using
	observations taken with VLTI/PIONIER. Probing thermal emission from the hot inner
	rim dust, we focus on the detection of any asymmetric features possibly
	indicative of substructures induced by the central binary, substellar companions,
	or hydrodynamical instabilities. The article is structured as follows. We
	describe our targets in Sect.\ \ref{sect:target_systems} and the observations and
	data in Sect.\ \ref{sect:observational_details}. Our image reconstruction workflow
	is developed in Sect.\ \ref{sect:methodology}, and then applied to our targets
	in Sect.\ \ref{sect:results}, where we describe the detected features. During
	this, we take into account multiple tests meant to assess the reliability of the
	image features. The possible origins of the rim positions and detected
	features are discussed in Sect.\ \ref{sect:discussion}. Finally, we summarise our
	results and present our conclusions in Sect.\ \ref{sect:conclusions}.

	\section{Target systems}
	\label{sect:target_systems}

	\begin{table*}
		\caption{Properties of the target sample and their observations.}
		\label{table:target_stars_and_observations}
		\centering
		\renewcommand{\arraystretch}{1.5}
		\resizebox{\textwidth}{!}{
		\begin{tabular}{lccccccccccc}
			\hline
			\hline
			Name                     & IRAS ID    & $P_{\mathrm{orb}}$ & $a_{1}\sin{i_{\mathrm{bin}}}$ & $f(m)$               & $d_{\mathrm{Gaia}}$       & $L_{*}$                      & $T_{\mathrm{eff}}^{\mathrm{SED}}$ & $d_{\mathrm{prim}}$ & RVb & $\Delta t_{\mathrm{obs}}/P_{\mathrm{orb}}$ & Refs.\        \\[-0.8ex]
			                         &            & $(\mathrm{d})$     & $\mathrm{(AU)}$               & $\mathrm{(M_\odot)}$ & $\mathrm{(kpc)}$          & $\mathrm{(L_{\odot})}$       & $\mathrm{(K)}$                    &                     & ph. & $\mathrm{(\%)}$                            &               \\
			\hline
			\object{AI~Sco}          & 17530-3348 & $977^{\dagger}$    & --                            & --                   & $2.9^{+0.5}_{-0.3}$       & $5600^{+1800}_{-1400}$       & $4500\pm250$                      & $-2.13$             & y   & $3.3$                                      & 2, 3. 4       \\
			\object{EN~TrA}          & 14524-6838 & $1488 \pm 9$       & $2.11 \pm 0.09$               & $0.57 \pm 0.07$      & $2.7^{+0.2}_{-0.3}$       & $2800^{+2500}_{-400}$        & $5500\pm250$                      & $-2.86$             & n   & $1.9$                                      & 1, 3, 4       \\
			\object{HD~95767}        & 11000-6153 & $1990 \pm 60$      & $2.14 \pm 0.16$               & $0.33 \pm 0.07$      & $4.4^{+0.3}_{-0.3}$       & $14\,800_{-2500}^{+15\,500}$ & $7100\pm250$                      & $-3.19$             & y   & $2.0$                                      & 1, 3, 4, 7    \\
			\object{HD~108015}       & 12222-4652 & $906 \pm 6$        & $0.28 \pm 0.02$               & $0.0036 \pm 0.0009$  & $5.1^{+0.8}_{-0.6}$       & $19\,800_{-4400}^{+5300}$    & $6850\pm250$                      & $-3.16$             & n   & $5.0$                                      & 1, 3, 4       \\
			\object{HR~4049}         & 10158-2844 & $430.6 \pm 0.1$    & $0.627 \pm 0.010$             & $0.177 \pm 0.008$    & $1.4^{+0.3}_{-0.2}$       & $12\,600_{-2000}^{+8\,200}$  & $7200\pm250$                      & $-3.14$             & y   & $3.5$                                      & 1, 2, 3, 4, 5 \\
			\object{IRAS~15469-5311} & --         & $390.2 \pm 0.7$    & $0.438 \pm 0.015$             & $0.074 \pm 0.008$    & $3.5^{+0.4}_{-0.3}$       & $16\,400^{+11\,900}_{-1400}$ & $7000\pm250$                      & $-3.17$             & n   & $7.7$                                      & 1, 3, 4       \\
			\object{IW~Car}          & 09256-6324 & $1449^{\dagger}$   & --                            & --                   & $1.310^{+0.060}_{-0.070}$ & $5500^{+4800}_{-1800}$       & $6500\pm250$                      & $-3.17$             & y   & $5.6$                                      & 2, 3, 4       \\
			\object{PS~Gem}          & 07008+1050 & $1288.6 \pm 0.3$   & $1.507 \pm 0.034$             & $0.274 \pm 0.019$    & $1.8^{+0.1}_{-0.1}$       & $3300^{+1100}_{-100}$        & $6100\pm250$                      & $-3.03$             & y   & $8.7$                                      & 1, 3, 4, 6    \\
			\hline
		\end{tabular}
		\renewcommand{\arraystretch}{1.0}
		} \tablefoot{The binary period $P_{\mathrm{orb}}$, projected primary semi-major axis $a_{1}\sin{i_{\mathrm{bin}}}$ and mass function $f(m)$ are derived from SB1 fitting by (1). When RV monitoring is unavailable -- as marked by $\dagger$ -- $P_{\mathrm{orb}}$ is instead taken to be the period of the detected RVb phenomenon by (2). System distances $d_{\mathrm{Gaia}}$ are based on the Gaia DR3 'geometric' estimates by (3). Primary luminosities $L_{*}$ and effective temperatures $T_{\mathrm{eff}}^{\mathrm{SED}}$ (formal uncertainty of $250\,\mathrm{K}$) are derived from the SED fits by (4). The errors on $L_{*}$ result from uncertainty on the total reddening. The primary's $H$-band spectral index $d_{\mathrm{prim}}$ (defined at $\lambda_{0}= 1.65 \, \mathrm{\mu m}$), is derived from a power-law fit to the de-reddened model spectrum. We also note detections of the RVb phenomenon, as found by (2), (5), (6) and (7). $\Delta t_{\mathrm{obs}}/P_{\mathrm{orb}}$ denotes the ratio between the timespan that the observations were taken in and the binary period.}
		\tablebib{(1)~\citet{Oomen2018}; (2)~\citet{KissBodi2017}; (3)~\citet{bailer_jones2021}; (4)~\citet{Kluska2022}; (5)~\citet{Waelkens1991a}; (6)~\citet{VanWinckel1999}; (7)~\citet{Kiss2007}.}
	\end{table*}

	In this study, we present NIR $H$-band interferometric imaging observations of
	eight full disc post-AGB binaries. All observations were obtained within the 'INterferometric
	Survey of Post-AGB binaries Interacting with their RING' large programme at
	the European Southern Observatory (INSPIRING, ESO 1103.D-0797, PI: J.\ Kluska).
	The data were taken using the Precision Integrated-Optics Near-infrared
	Imaging ExpeRiment \citep[PIONIER,][]{LeBouquin2011}, mounted on the Very
	Large Telescope Interferometer \citep[VLTI,][]{Haguenauer2010}. The target
	selection is described below. Further details of the data and observations are
	given in Sect.\ \ref{sect:observational_details}.

	The INSPIRING targets were selected out of the wider sample of $\sim\!85$
	known Galactic disc-bearing post-AGB binaries \citep[][]{Kluska2022}. This was
	based on the following selection criteria:

	\textit{Criterion 1:} The target is observable with both PIONIER and GRAVITY --
	with analysis of the GRAVITY snapshot data being deferred to future work -- on
	the VLTI auxiliary telescopes (ATs), satisfying constraints on correlated flux
	and sky position.

	\textit{Criterion 2:} An estimated binary orbital period of $P_{\mathrm{orb}}>
	1 \, \mathrm{yr}$ (Table \ref{table:target_stars_and_observations}),
	permitting the necessary AT array configuration changes for imaging with
	minimal orbital motion during the observing campaign (see Sect.\ \ref{sect:observational_details}).

	\textit{Criterion 3:} Clearly resolved circumstellar emission in prior PIONIER
	snapshots \citep{Kluska2019}, as shown by a long-baseline plateau in the
	squared visibilities ($V2$, Sect.\ \ref{sect:observational_details}).

	Applying these criteria yielded 11 targets. In this work, we analysed eight of
	these with clear disc contributions in their PIONIER data and excluded three --
	IRAS~08544-4431, HR~4226, and U~Mon -- from our analysis. Briefly, IRAS~08544$-$4431
	is reserved for a dedicated PIONIER time-series imaging; HR~4226 shows no clear
	$H$-band disc signal in PIONIER -- though the outer disc is resolved with the
	Spectro-Polarimetic High contrast imager for Exoplanets REsearch \citep[SPHERE;][]{Andrych2025}
	-- and its small RV amplitudes together with a relatively low effective
	temperature ($T_{\rm eff}\! \approx\! 4000\,\mathrm{K}$) cast doubt on its
	post-AGB classification \citep[][]{Andrych2025}. Hence, we defer a separate,
	multi-wavelength study; and for U~Mon the PIONIER observables seemingly only probe
	the binary. The inner rim is likely further out and colder, and hence undetectable
	in NIR, consistent with U Mon's SED categorisation as a possible transition
	disc system (\citealt[][]{Kluska2022}, with the outer disc being resolved with
	SPHERE by \citealt{Andrych2023}).

	Based on their SEDs, the eight systems analysed in this work are all
	classified as full discs by \citet{Kluska2022}, i.e.\ with inner rims close to
	the sublimation radius. This is a selection bias due to the natural tendency
	of this class of objects to have strong NIR excesses, and thus for the disc signal
	to be resolvable within PIONIER's sensitivity limits. Nevertheless, they form a
	diverse sample of post-AGB binaries, covering a broad range in orbital period,
	stellar effective temperature and luminosity, refractory depletion level and
	previous inclination angle estimates. The main target properties relevant to our
	analysis and interpretation are summarised in Table \ref{table:target_stars_and_observations}.
	For completeness, we give extensive descriptions of pertinent results from previous
	literature on these targets in Appendix \ref{sect:appendix_target_individual_descriptions}.

	\section{Data and observations}
	\label{sect:observational_details} For our eight targets (Sect.\ \ref{sect:target_systems}),
	we obtained VLTI/PIONIER \citep{LeBouquin2011} observations in the $H$ band
	using the ATs in various configurations. A grism disperses the light into six spectral
	channels ($R\!\sim\!30$; $1.50-\!1.80\,\mu\mathrm{m}$). Each PIONIER integration
	delivers six baseline squared visibility ($V2$) and four -- of which three are
	independent -- triplet closure phase ($\phi3$) measurements for every spectral
	channel. While the former are sensitive to the radial intensity dependence of
	the source along different directions, the latter constrain its point-asymmetry.
	Each target was assigned six K-type giant calibrators at on-sky separations of
	$< 5^{\circ}$ found using SearchCal\footnote{\url{https://www.jmmc.fr/searchcal}}
	\citep[][]{Chelli2016}, excluding objects with uncertain luminosity class,
	known variability, or multiplicity. Every two science observations were
	bracketed -- in a CAL-SCI-CAL-SCI-CAL sequence -- by observations of three
	different calibrators.

	To enable imaging, the observations were taken using various AT configurations
	(baselines spanning $\sim\!10-\!130\,\mathrm{m}$) to properly fill the spatial
	frequency plane, also called the $(u, v)$ plane. We retrieved all reduced and calibrated
	INSPIRING observations from the Jean-Marie Mariotti Center Optical Interferometry
	Database (OiDB)\footnote{\url{http://oidb.jmmc.fr}}. The data were reduced using
	the standard pndrs\footnote{\url{http://www.jmmc.fr/pionier}} data reduction
	package \citep[][]{LeBouquin2011}. We then filtered the observations based on
	any severe problems with fringe tracking, weather conditions, etc. Observations
	showing a less-than-optimal ESO quality control grade (i.e.\ grade B or C)
	were manually checked. Any such observations that still presented good quality
	data on $>4$ baseline pairs were kept. Some observations were graded poor due
	to $(u, v)$-plane overlap with previous observations, while still showing excellent
	data quality.

	To minimise the effects of orbital motion and variable illumination on the inner
	disc, we carefully selected observations within the smallest possible time
	window that still provided adequate $(u,v)$ coverage for imaging. This strategy
	avoids merging widely separated epochs, which could otherwise introduce time-dependent
	signatures. The fractional time-span of the selected observations relative to the
	orbital period, $\Delta t_{\rm obs}/P_{\rm orb}$, is typically only a few per cent
	(Table~\ref{table:target_stars_and_observations}), ensuring that orbital
	smearing has a limited impact on the reconstructed morphologies.

\begin{figure*}
		\centering
		\includegraphics[width=1.0\linewidth]{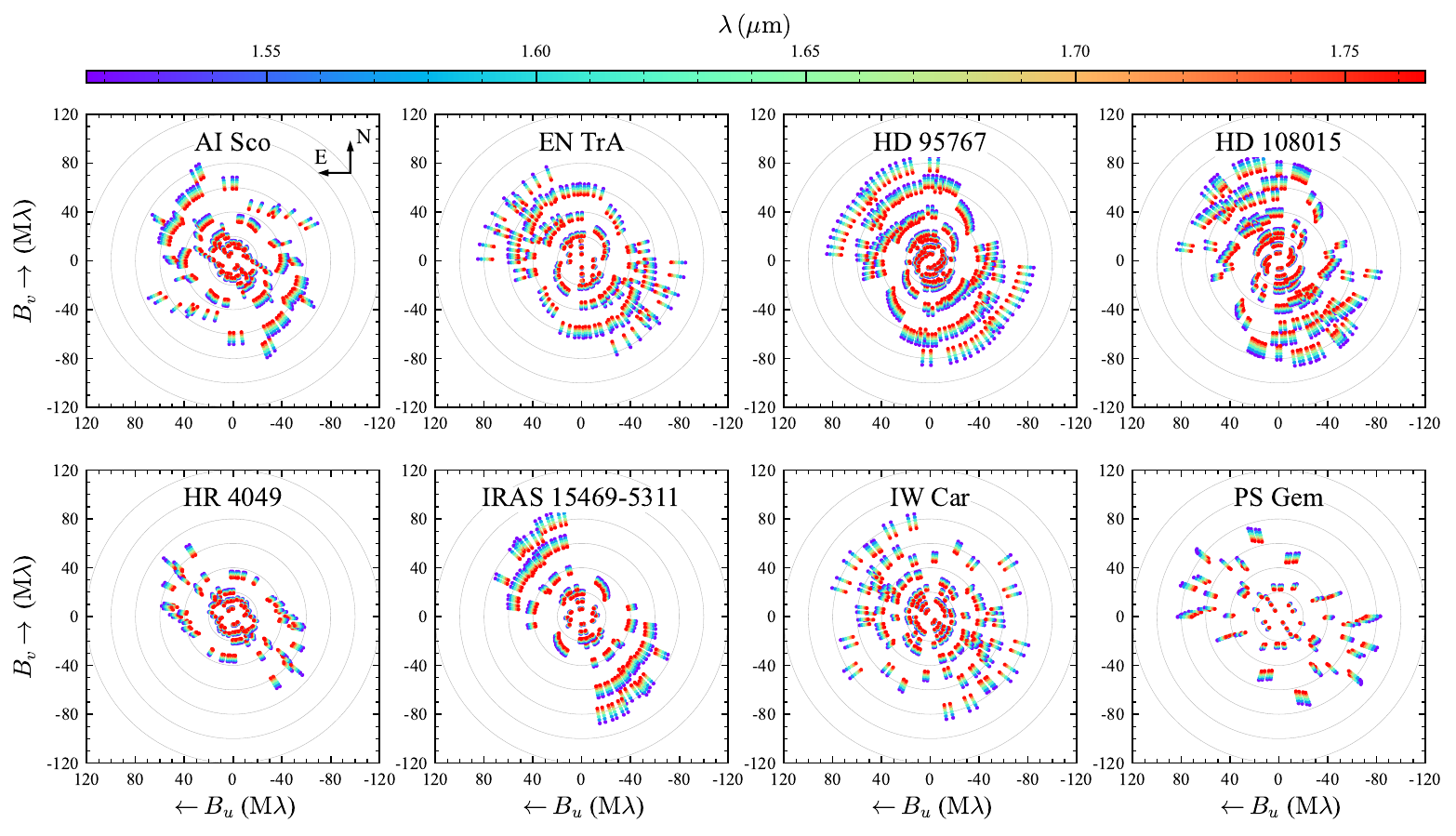}
		\caption{$(u, v)$ coverage for the targets' selected data. Observations are
		denoted by scattered circles, coloured by their wavelength.}
		\label{fig:uv_coverage}
	\end{figure*}

	Logs of the selected observations including the observational conditions and
	calibrators are available online on
	\href{https://doi.org/10.5281/zenodo.20155600}{Zenodo}. The $(u, v)$ coverages,
	dirty beams (i.e.\ the point spread functions, the centre of which forms the formal
	resolution element called the 'beam size') and corresponding $V 2$ and $\phi3$
	data are shown in Figs.\ \ref{fig:uv_coverage}, \ref{fig:dirty_beams}, \ref{fig:observables_v2}
	\& \ref{fig:observables_t3}, respectively.

	\section{Methodology}
	\label{sect:methodology} Given that the inner disc regions are expected to be morphologically
	complex, the principal aim of the analysis is to produce non-parametric image
	reconstructions of the circumstellar environment. Our imaging workflow makes use
	of the Semi-Parametric Approach for Reconstruction of Chromatic Objects (SPARCO)
	formalism, which we describe in Sect.\ \ref{sect:visibility_formalism}. To
	provide initial estimates on the different SPARCO parameters (flux ratios,
	chromaticities and positions), we perform a simplified geometric modelling of
	our targets in Sect.\ \ref{sect:geometric_modelling}. Lastly, our image reconstruction
	workflow itself is described in Sect.\ \ref{sect:img_rec}.

	\subsection{The SPARCO visibility formalism}
	\label{sect:visibility_formalism} In the $H$ band, the primary star presents itself
	as a bright central emission, while the disc emission is much more diffuse. In
	addition, the data show strong chromaticity, caused by the relative flux change
	between the primary and disc -- revealed by the strong $V2$ slopes across
	wavelength channels in Fig.\ \ref{fig:observables_v2}. We addressed these
	challenges by adopting the SPARCO approach \citep[][]{Kluska2014}. This
	approach defines the total source visibility $\tilde{V}(u, v)$ as a flux-weighted
	combination of the central stars and the circumstellar environment, following
	\begin{eqnarray}
		\label{eq:visibility_model} \tilde{V}_{\mathrm{tot}}(u, v,\lambda) & = &
		\frac{\sum_{j}f_{j}(\lambda) \tilde{V}_{j}(u,v)}{\sum_{j}f_{j}(\lambda)}, \\
		f_{j}(\lambda) & = & f_{j,0}\left(\lambda/\lambda_0\right)^{d_{j}},\\ \sum_{j}
		f_{j,0} & = & 1,
	\end{eqnarray}
	where $j$ represents a component -- in our case the primary star, secondary star
	or disc rim -- and $f_{j}$ its spectrum. Following \citealt{Kluska2019}, we defined
	the spectrum by a spectral index $d_{j}$ and flux fraction $f_{j,0}$ at reference
	wavelength $\lambda_{0}\! =\!1.65\,\mathrm{\mu m}$. This assumes constant
	component morphologies across wavelength channels, which is reasonable for the
	narrow $H$ band. The essence of SPARCO is then to approximate the central
	stars in Eq.\ \ref{eq:visibility_model} using closed-form geometric components,
	while applying an image reconstruction algorithm to recover just the disc rim
	$\tilde{V}_{\mathrm{rim}}(u,v)$. This greatly reduces the required pixel dynamic
	range and allows us to use data from all wavelength channels simultaneously,
	significantly improving image fidelity and the capability to spatially
	separate the central stars from the disc rim.

	Throughout the remainder of this work, we refer to the central star geometric parameters
	as the 'SPARCO parameters'. We modelled the primary as a uniform disc centred
	at the origin and the secondary as a point source. The SPARCO parameters are then
	the primary's flux fraction $f_{\mathrm{prim,0}}$, uniform diameter
	$\mathrm{UD}_{\mathrm{prim}}$ and spectral index $d_{\mathrm{prim}}$, the disc
	spectral index $d_{\mathrm{rim}}$, and the secondary's flux fraction
	$f_{\mathrm{sec,0}}$ and position
	$(\Delta \alpha_{\mathrm{sec}}, \Delta \delta_{\mathrm{sec}})$. Throughout our
	analysis, we fixed $d_{\mathrm{prim}}$ to the value derived from the SED fit (Table
	\ref{table:target_stars_and_observations}). This is because our data are only sensitive
	to relative flux differences between components. The SPARCO parameter values
	need fine-tuning in order to properly reveal the circumstellar environment
	\citep[cf.][]{Kluska2014, Kluska2016_MWC158, Kluska2020a}.

	We combined SPARCO with an image reconstruction algorithm and a SPARCO parameter
	optimisation routine into a consistent image reconstruction workflow, which we
	describe in Sect.\ \ref{sect:img_rec}. To proceed with image reconstruction, however,
	we first required preliminary estimates on the SPARCO parameters, which we obtained
	via a simplified geometric modelling of the binary and inner rim.

	\subsection{Initial SPARCO estimates via geometric modelling}
	\label{sect:geometric_modelling} To provide a first estimate of the SPARCO parameters,
	we fitted simple geometric models to the data using the PMOIRED\footnote{\url{https://github.com/amerand/PMOIRED}}
	package \citep{Merand2022}. PMOIRED allows the user to flexibly define the source
	morphology as a combination of parametrised components. We adopted the
	visibility model of Eq.\ \ref{eq:visibility_model}. This step of the analysis
	is designed to provide initial estimates of the SPARCO parameters (Sect.\ \ref{sect:visibility_formalism}),
	as these provide crucial input for our imaging workflow.

	PMOIRED fits parameters by minimising the reduced $\chi^{2}_{\nu}$ using a
	Levenberg-Marquardt algorithm \citep[][]{Marquardt1963}. To avoid local minima,
	we opted to gradually increase our model complexity, using the best fit from
	the previous model as a starting point for the next -- similar to the strategy
	adopted by \citealt{Kluska2019}. We proceeded through the following stages:

	\textit{Step 1: primary and symmetric power law disc}. Assuming a maximum primary
	stellar diameter of $300 \, \mathrm{R_\odot}$ \citep[e.g.][]{Bollen2022}, we fitted
	the primary using a uniform disc of diameter $\mathrm{UD}_{\mathrm{prim}}$, on
	the condition that the corresponding angular size of the maximum stellar
	diameter is above half of the interferometric beam size. If lower, we modelled
	the primary as a point source. The inner disc rim was modelled as an azimuthally
	symmetric disc with a radial power law surface brightness. It is inclined with
	inclination $i_{\mathrm{rim}}$ and oriented by position angle
	$\mathrm{PA}_{\mathrm{rim}}$, defined anti-clockwise from north to east. We
	note that the derived inclination assumes a circular rim, where the only source
	of ellipticity is its on-sky projection. The rim is further defined by its
	flux fraction $f_{\mathrm{disc,0}}$, inner diameter $\theta_{\mathrm{rim}}$ (interior
	to which the intensity is zero), outer diameter $\theta_{\mathrm{out}}$, radial
	brightness power index $\rho_{\mathrm{rim}}$ and spectral index $d_{\mathrm{rim}}$.
	$\theta_{\mathrm{out}}$ is fixed to an angular size corresponding to
	$300 \, \mathrm{AU}$. Our data only probe the inner disc, so its value has
	little effect on the results. Both components were centred at $(0,0)$.

	In addition, we added a grey over-resolved background of flux
	$f_{\mathrm{bkg,0}}$, on the condition that it improves the short baseline $V2$
	residuals and causes a decrease of the Bayesian information criterion \citep[BIC;][]{Schwarz1978}
	$\Delta \mathrm{BIC}< -100$. We adopted this conservative BIC threshold -- a
	more common choice is $\Delta \mathrm{BIC}\!<\!-10$ -- because the errors on PIONIER
	data can be underestimated (as is generally the case for interferometric
	instruments, e.g.\ \citealt{Monnier2012, GRAVITY_collab2019}), and more lenient
	thresholds often resulted in overfitting.

	\textit{Step 2: primary with offset and azimuthally modulated disc}. The disc centre
	was offset from the primary by $(\Delta \alpha_{\mathrm{rim}},\Delta \delta_{\mathrm{rim}}
	)$. In another fit, the disc brightness was azimuthally modulated using a first
	order harmonic, as described by its amplitude $A_{\mathrm{mod,1}}$ and
	position angle relative to the $\mathrm{PA}$ of the rim
	$\psi_{\mathrm{mod,1}}$. Each complexity increase was again only accepted if $\Delta
	\mathrm{BIC}\!< \!-100$.

	\textit{Step 3: binary with offset and azimuthally modulated disc}. A grey point-source
	secondary companion was added to the fit (chromatic companions did not improve
	the $\mathrm{BIC}$). It is defined by its flux fraction $f_{\mathrm{sec,0}}$
	and offset from the primary
	$(\Delta \alpha_{\mathrm{sec}},\Delta \delta_{\mathrm{sec}})$ (initially set to
	the disc centre found in step 2). We stress that the secondary's flux is likely
	not photospheric, instead originating from jet-launching circumcompanion
	accretion discs \citep[e.g.][]{Hillen2016, Bollen2022, Verhamme2024, DePrins2024}.
	We added this component only if $\Delta \mathrm{BIC}\!<\!- 100$.

	\textit{Step 4: correction of the secondary detection via preliminary reconstructions}.
	Due to our simplistic model of the inner rim (including only a first order azimuthal
	modulation), step 3 can be highly degenerate and prone to placing a secondary at
	unphysical positions, mimicking rim morphology features. As a result, we
	included a feedback step from the image reconstructions in order to validate
	the secondary detections, and, if necessary, repeated step 3 with a corrected initial
	secondary position. We provide a further description of this feedback step in
	Sect.\ \ref{sect:validation_geometric_secondary_via_imaging}.

	Parameter uncertainties on the final geometric models were obtained via bootstrapping
	\citep[][]{Efron1979}, using $20 \, 000$ samples. The only target that we
	could not fit a geometric model to is IW~Car, as its complex morphology proved
	unable to be reproduced with a single-disc model (see also Fig.\ \ref{fig:organic_imgs_last_four}).
	We tested two-disc models for IW~Car, but these resulted in ill-conditioned
	solutions with large parameter degeneracies, providing no meaningful constraints
	as the PMOIRED minimisation engine became immediately trapped in local minima.
	We therefore restricted our analysis to single-disc models for compatibility
	across the sample. The parameters of our final geometric model fits are found in
	Table \ref{table:geom_params}. The corresponding model disc images are
	presented in Fig.\ \ref{fig:pmoired_images}, though we note that the images
	are not used as initial starting points for the image reconstructions.

	\subsection{Image reconstruction workflow}
	\label{sect:img_rec} Image reconstruction allows us to represent the source
	morphology in a non-parametric way, providing a powerful tool for discovering unexpected
	features whose presence could be difficult to decidedly infer using parametric
	models. However, sparse $(u,v)$ coverages and non-linearity between source morphology
	and observables make it an ill-posed, non-convex inverse problem \citep[][]{thiebaut2017}.
	As a result, the loss function to be minimised is typically formulated in a Bayesian
	way as
	\begin{equation}
		\label{eq:img_rec_tot_loss}\mathcal{L}_{\mathrm{tot}}= \mathcal{L}_{\mathrm{data}}
		+ \mu \mathcal{L}_{\mathrm{reg}},
	\end{equation}
	consisting of both a data loss $\mathcal{L}_{\mathrm{data}}$ and a prior-like regularisation
	function $\mathcal{L}_{\mathrm{reg}}$. Many different images can fit the data
	well, leading to low $\mathcal{L}_{\mathrm{data}}$ despite not being astrophysically
	valid. The role of $\mathcal{L}_{\mathrm{reg}}$ is to reign in the
	reconstruction, guiding it toward images with certain qualitative properties.
	Its relative numerical weight in the reconstruction is set via the
	regularisation weight $\mu$.

	\subsubsection{ORGANIC image reconstruction}
	\label{sect:organic_image_reconstruction}

	\label{sect:organic_imgrec} For our image reconstructions, we utilised the
	ORGANIC\footnote{\url{https://github.com/DePrinsT/organic}} algorithm \citep[][]{Claes2020}.
	It uses a Generative Adversarial Network \citep[GAN,][]{Goodfellow2014} -- a type
	of convolutional neural network (CNN) -- to encode RT models as physically
	informed priors for image reconstruction. Specifically, ORGANIC is trained on
	an extensive grid of $H$- and $K$-band thermal dust emission models of symmetric
	disc rims, simulated using the MCMax\footnote{\url{https://github.com/michielmin/MCMax3D}}
	RT code \citep[][]{Min2009}. We briefly describe the ORGANIC algorithm below,
	but refer interested readers to \citet{Claes2020} for more details.

	The GAN architecture consists of a 'generator' and a 'discriminator'. The
	generator takes a size $100$ latent Gaussian noise vector and maps it to a
	$128\!\times\! 128$ image. The discriminator takes a $128\!\times\!128$ image and
	maps it to a scalar score $\hat s \in [0, 1]$. The overall ORGANIC algorithm
	consists of the following two steps:

	\textit{Step 1: initial training}. The generator and discriminator are
	iteratively and alternately trained. In each training step, the generator
	produces a batch of 'fake' images using a set of random input noise vectors. These
	are presented to the discriminator together with a batch of 'real' images from
	the RT training set. In alternating fashion, the generator is trained to deceive
	the discriminator into evaluating its images as real (aiming at a
	discriminator score $\hat s\!\sim\!1$ for its images), while the discriminator
	is trained to discern the 'fake' generator images from the 'real' RT ones (aiming
	for $\hat s \!\sim\!0$ and $\hat s\!\sim\!1$ for the generated and RT images,
	respectively). When properly trained, this adversarial game converges to a state
	where the generator produces an image distribution similar to the RT models \citep[][]{Goodfellow2014}.

	\textit{Step 2: image reconstruction.} The GAN can then be confronted with any
	set of $V2$ and $\phi3$ data. With the discriminator fixed, the generator starts
	from a random latent noise vector. This essentially corresponds to drawing a
	random initial image similar to the RT training images. The generator is then further
	trained to minimise Eq.\ \ref{eq:img_rec_tot_loss} by directly updating the
	weights in its convolutional kernels. The loss terms are now
	\begin{eqnarray}
		\label{eq:organic_loss} \mathcal{L}_{\mathrm{data}} & = & \frac{l_{V2}+ l_{\phi3}}{N_{V2}+
		N_{\phi3}}, \\ l_{V2} & = & \sum_{i=1}^{N_{V2}} \frac{(V2_{\mathrm{{obs,i}}}-
		V2_{\mathrm{{img,i}}})^{2}}{\sigma_{V2,i}^{2}} = \chi^2_{V2}, \\ l_{\phi3} &
		= & \sum_{i=1}^{N_{\phi3}} \frac{2 \left(1 - \cos{(\phi3_\mathrm{{obs,i}} - \phi3_\mathrm{{img,i}}})\right)}{\sigma_{\phi3.i}^{2}}
		,\\ \mathcal{L}_{\mathrm{reg}} & = & -\ln{\hat s_{\mathrm{img}}},
	\end{eqnarray}
	with the subscripts 'img' and 'obs' respectively denoting the generated image and
	the observed data. $N_{V2}$ and $N_{\phi3}$ are the number of observed $V2$
	and $\phi3$ points, and $l_{V2}$ and $l_{\phi3}$ the corresponding data losses.
	We assumed a Von Mises distribution for $\phi3$ in order to account for its
	angular nature. The goal of $\mathcal{L}_{\mathrm{reg}}$ is to make the final images
	look more like the RT training models. However, most of the regulatory properties
	of ORGANIC are implicitly inherited from the CNN architecture and the initial training.
	We demonstrate and discuss this in Appendix
	\ref{sect:appendix_regularization_weight}.

	To account for the random nature of the starting vector, ORGANIC repeats this step
	for $N_{\mathrm{restart}}$ different random generator inputs. We set
	$N_{\mathrm{restart}\!}=\!50$. This results in an image cube, the median of
	which is adopted as the final image. The standard deviation along the cube for
	each pixel $j$, $\sigma_{\mathrm{restart},j}$, then represents the uncertainty
	induced by the choice of latent vector. Regardless, some of the $N_{\mathrm{restart}}$
	images in the cube are badly converged, ending up in poor local minima. To account
	for this, we added a procedure using principal component analysis \citep[PCA; e.g.][]{Gewers2021}
	and k-means clustering \citep[][]{Lloyd1982} to filter out these images from the
	cube. We applied PCA to the image cube, retaining 15 components to reduce dimensionality
	(accounting for $\sim\!70- 80\%$ of the inter-image variance). We then applied
	k-means clustering with three clusters on the PCA subspace, discarding the cluster
	with the worst median $\mathcal{L}_{\mathrm{tot}}$. This proved highly
	effective in singling out poorly converged images.

	\begin{figure}
		\centering
		\includegraphics[width=1.0\linewidth]{
			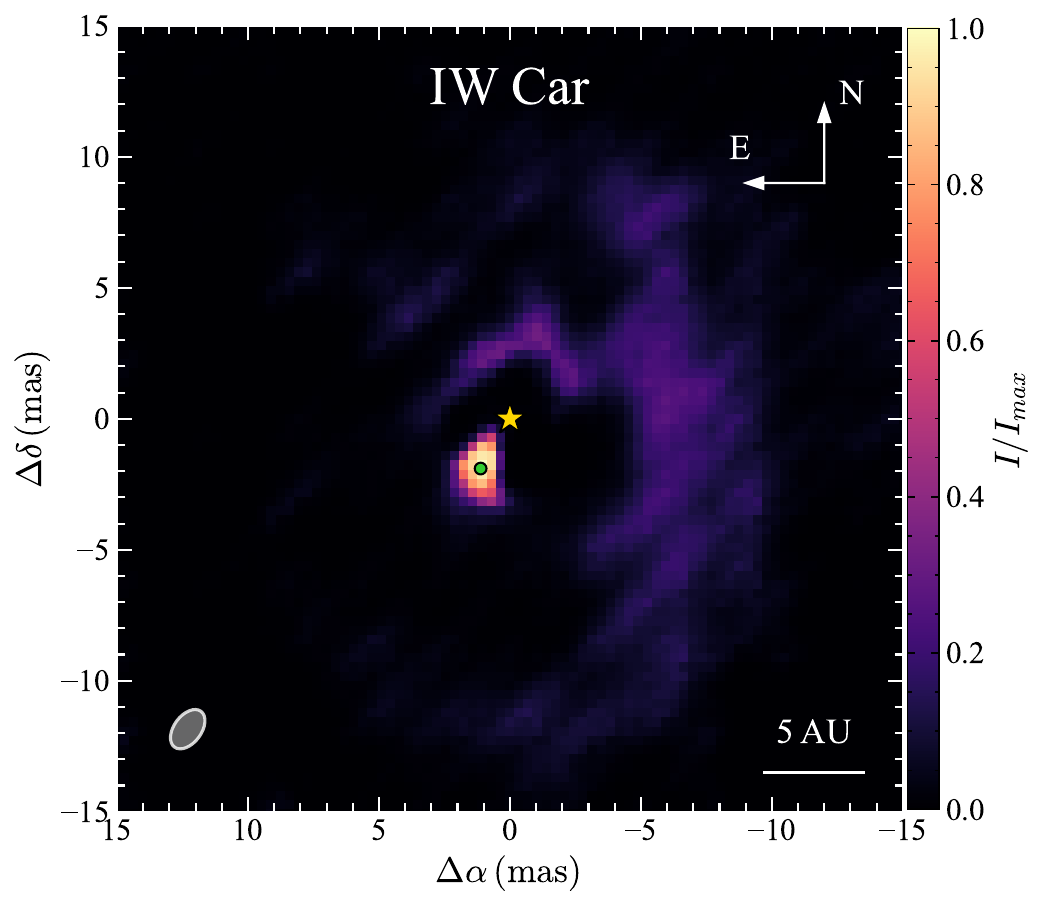
		}
		\caption{Peak-normalised IW~Car ORGANIC image without a SPARCO secondary. We
		identify a strong unresolved flux peak as flux from the secondary. The
		primary is marked by a gold star, while a green dot marks the centre of an elliptical
		Gaussian fit to the peak. The interferometric beam size and a
		$5 \, \mathrm{AU}$ scale bar are shown in the bottom.}
		\label{fig:IW_Car_secondary_detection}
	\end{figure}

	\subsubsection{SPARCO optimisation}
	\label{sect:sparco_optimisation} We applied ORGANIC in combination with the SPARCO
	approach. Our binaries can have up to six free SPARCO parameters: $f_{\mathrm{prim,0}}$,
	$\mathrm{UD}_{\mathrm{prim}}$, $d_{\mathrm{rim}}$, $f_{\mathrm{sec,0}}$, $\Delta
	\alpha_{\mathrm{sec}}$ and $\Delta \delta_{\mathrm{sec}}$ (Sect.\ \ref{sect:visibility_formalism}).
	This made the computational cost of grid searches \citep[e.g.][]{Kluska2016_MWC158, Kluska2020a}
	prohibitive. We instead made use of the BADS\footnote{Its Python interface, pyBADS
	\citep{Singh2024}, is available at \url{https://github.com/acerbilab/pybads}}
	optimisation algorithm \citep[][]{Acerbi2017}, running ORGANIC for each
	considered set of SPARCO parameters and choosing the one with the lowest value
	of $\mathcal{L}_{\mathrm{tot}}$ (Eq.\ \ref{eq:img_rec_tot_loss}) on the resulting
	image. BADS alternates between a broader systematic grid search stage and a
	fast, detailed local search using a Gaussian process surrogate for $\mathcal{L}
	_{\mathrm{tot}}$ \citep[][]{Rasmussen2006}. It is highly efficient in the required
	amount of ORGANIC runs and can readily escape shallow local minima.

	We took the initial SPARCO parameter values from the corresponding final geometric
	models in Table \ref{table:geom_params} (after validation and correction of
	the secondary position from preliminary imaging; see Sect.\ \ref{sect:validation_geometric_secondary_via_imaging}).
	Initial tests showed that BADS+ORGANIC had the tendency to place the secondary
	at non-physical positions and heavily reduce its flux contribution, instead placing
	a slightly smeared flux in the image reconstruction at the true secondary position.
	This is likely because the binary motion causes the secondary to better fit
	the data as a slightly smeared Gaussian instead of a point source. As a result,
	we fixed $(\Delta \alpha_{\mathrm{sec}},\Delta \delta_{\mathrm{sec}})$ to
	their final geometric model values (Table \ref{table:geom_params}), leaving only
	$f_{\mathrm{prim,0}}$, $\mathrm{UD}_{\mathrm{prim}}$, $d_{\mathrm{rim}}$ and $f
	_{\mathrm{sec,0}}$ free.

	Since we could not fit a geometric model for IW~Car, the secondary position
	was instead fully inferred from a Gaussian fit to the strong unresolved flux
	peak in a BADS+ORGANIC run omitting the secondary star SPARCO component, which
	is identified as emission from the secondary \citep[cf.][]{Hillen2016}. This resulted
	in $(\Delta \alpha_{\mathrm{sec}},\Delta \delta_{\mathrm{sec}})\!=\!(1.12,-1.90
	)\, \mathrm{mas}$ (Fig.\ \ref{fig:IW_Car_secondary_detection}). The other initial
	values were set to the fit values of \citet{Kluska2019}, derived from previous
	PIONIER snapshot data.

	\subsubsection{Selection of the regularisation weight and pixel scale}
	As optimisation of $\mu$ via the usual L-curve method \citep[e.g.][]{Renard2011}
	is prohibited due to the current ORGANIC architecture (see the discussion in Appendix
	\ref{sect:appendix_regularization_weight}), we performed manual exploration in
	a reasonable range ($\mu = 0.1 \! - \!2$). Recovering the same main features
	for different values (see also Fig.\ \ref{fig:organic_mu_effect}), we fixed
	$\mu = 0.2$ during our imaging workflow for all targets. This value generally provided
	a good balance between suppression of reconstruction artefacts while avoiding excessive
	over-regularisation that tends to make the rims too sharp.

	Starting from images with a pixel scale corresponding to Nyquist sampling (two
	pixels fitting in the beam size) we gradually increased the sampling factor.
	We applied the BADS optimisation each time, until all discs were deemed properly
	resolved with $\sim\!20-30$ pixels across the inner rim diameter. During this,
	we took care to ensure the $128\!\times\!128$ pixel field-of-view (set by the
	architecture of ORGANIC) contained all of the rim's flux.

	\subsubsection{Correction of the secondary detection}
	\label{sect:validation_geometric_secondary_via_imaging} Due to the simplistic geometric
	model used for the disc rim, step 3 in the geometric modelling (Sect.\ \ref{sect:geometric_modelling})
	can initially fit faulty secondary positions. As a result, we first validated the
	secondary detection by performing preliminary BADS+ORGANIC runs where the secondary
	was omitted from the SPARCO parameters (i.e.\ $f_{\mathrm{sec}}\!=\!0$). In the
	resulting image reconstructions, the secondary emission, if significant,
	appears as a clear unresolved flux contribution (as seen in Fig.\ \ref{fig:IW_Car_secondary_detection}
	for IW~Car). If the model secondary position proved to be mimicking features
	of the disc rim, we reran the geometric fitting step 3 with a corrected initial
	position to procure the final geometric models. For HD~108015 the geometric
	companion detection was fully spurious, and the final geometric model was instead
	taken from step 2.

	\subsubsection{Uncertainty estimation}
	\label{sect:uncertainty_estimation} After optimising the SPARCO parameters
	using BADS, we performed a final bootstrap procedure with $500$ resampled datasets,
	applying ORGANIC to each and storing the resulting final median image in a
	cube. The standard deviation along this cube was then adopted as the uncertainty
	induced by the data, $\sigma_{\mathrm{boot},j}$. This was added in quadrature to
	the restart-induced error, $\sigma_{\mathrm{restart},j}$, of the image reconstructed
	on the original dataset to provide the total error for each pixel $j$, i.e.\ $\sigma
	_{tot,j}\!=\!\sqrt{\smash[b]{\sigma_{\mathrm{restart},j}^2 + \sigma_{\mathrm{boot},j}^2}}$.
	This total error was used to calculate any significance contours.

	\subsubsection{Workflow and image feature validation}
	We implemented an extensive suite of tests to validate our imaging workflow
	and assess the robustness of features detected in the final images.

	In Appendix~\ref{sect:appendix_geom_models_and_synth_reconstructions}, we applied
	our full imaging workflow on synthetic datasets, derived from the PMOIRED
	model images, to identify potential biases in the inner rim parameters and the
	general position of azimuthal brightness enhancements. Appendix~\ref{sect:appendix_squeeze_reconstructions_appendix}
	presents image reconstructions performed with the more classical SQUEEZE
	reconstruction package \citep[][]{Baron2010}. The physics-informed
	regularisation capabilities of ORGANIC (Appendix \ref{sect:organic_implicit_regularisation})
	allowed us to reconstruct coherent images for HR~4049 and PS~Gem, where SQUEEZE
	failed as it heavily overfitted the data due to the sparser $(u,v)$ coverage.
	Otherwise, SQUEEZE recovers the same general features as our ORGANIC images,
	confirming the convergence of our ORGANIC workflow. Lastly, Appendix~\ref{sect:appendix_robustness_against_uv_coverage}
	assesses the robustness of the image features against the specifics of the
	$(u, v)$ coverage and dirty beam artefacts. In what follows, we explicitly
	address the robustness of image features revealed by our tests when needed.

	\section{Results}
	\label{sect:results}

	\begin{table*}
		\caption{ORGANIC image reconstruction parameters and image-derived inner rim
		parameters.}
		\label{table:img_rec_params}
		\setlength{\tabcolsep}{1.5mm}
		\centering
		\resizebox{\textwidth}{!}{
		\renewcommand{\arraystretch}{1.7}
		\begin{tabular}{lccccccccccccccccc}
			\hline
			\hline
			Target                     & SF\tablefootmark{\,a} & $\mathrm{PS}$\tablefootmark{\,a} & $\mathrm{FOV}$\tablefootmark{\,a} & $\mathcal{L}_{\mathrm{data}}$ & $\mathcal{L}_{\mathrm{reg}}$ & $f_{\mathrm{prim},0}$ & $\mathrm{UD}_{\mathrm{prim}}$ & $f_{\mathrm{sec},0}$ & $\Delta \alpha_{\mathrm{sec}}$\tablefootmark{\,b} & $\Delta \delta_{\mathrm{sec}}$\tablefootmark{\,b} & $d_{\mathrm{rim}}$ & $\theta_{\mathrm{rim}}$ & $i_{\mathrm{rim}}$   & $\mathrm{PA}_{\mathrm{rim}}$ & $\Delta \alpha_{\mathrm{rim}}$ & $\Delta \delta_{\mathrm{rim}}$ & $r_{\mathrm{hl}}$      \\[-0.8ex]
			                           &                       & $\mathrm{(mas)}$                 & $\mathrm{(mas)}$                  &                               &                              & $(\%)$                & $\mathrm{(mas)}$              & $(\%)$               & $(\mathrm{mas})$               & $(\mathrm{mas})$               &                    & $(\mathrm{mas})$        & $(^{\circ})$         & $(^{\circ})$                 & $(\mathrm{mas})$               & $(\mathrm{mas})$               & $(\mathrm{mas})$       \\
			\hline
			AI~Sco\tablefootmark{\,c}  & $5$                   & $0.230$                          & $29.5$                            & $2.09$                        & $0.94$                       & $64.5$                & --                            & $2.76$               & $-1.08$                        & $-1.38$                        & $-0.59$            & $4.8^{+0.8}_{-0.5}$     & $46^{+10}_{-14}$     & $135^{+7}_{-7}$              & $0.51^{+0.13}_{-0.15}$         & $-1.4^{+0.3}_{-0.2}$           & $2.89^{+0.12}_{-0.23}$ \\
			EN~TrA                     & $4$                   & $0.265$                          & $33.9$                            & $1.63$                        & $0.91$                       & $72.1$                & --                            & $5.37$               & $-0.70$                        & $-1.65$                        & $0.30$             & $6.1^{+0.3}_{-0.2}$     & $50^{+3}_{-4}$       & $37^{+5}_{-7}$               & $-1.02^{+0.09}_{-0.09}$        & $-0.89^{+0.17}_{-0.21}$        & $3.99^{+0.13}_{-0.27}$ \\
			HD~95767                   & $4$                   & $0.263$                          & $33.7$                            & $0.96$                        & $0.97$                       & $44.1$                & --                            & $24.75$              & $-0.72$                        & $-0.51$                        & $2.36$             & $6.01^{+0.05}_{-0.05}$  & $23.7_{-0.5}^{+1.0}$ & $7_{-5}^{+8}$                & $0.03_{-0.03}^{+0.03}$         & $-0.51_{-0.03}^{+0.03}$        & $3.43^{+0.13}_{-0.13}$ \\
			HD~108015                  & $3$                   & $0.327$                          & $41.9$                            & $0.99$                        & $0.93$                       & $56.0$                & --                            & --                   & --                             & --                             & $1.45$             & $6.28_{-0.06}^{+0.06}$  & $30_{-3}^{+3}$       & $81_{-5}^{+5}$               & $-0.57_{-0.04}^{+0.04}$        & $0.58_{-0.06}^{+0.06}$         & $3.78^{+0.16}_{-0.16}$ \\
			HR~4049\tablefootmark{\,c} & $2$                   & $0.635$                          & $81.3$                            & $0.97$                        & $0.92$                       & $65.4$                & $0.72$                        & $0.24$               & $-1.70$                        & $-0.69$                        & $2.39$             & $17.2_{-1.2}^{+1.5}$    & $48_{-4}^{+4}$       & $95_{-8}^{+8}$               & $-1.5_{-0.8}^{+0.6}$           & $-0.4_{-0.3}^{+0.3}$           & $16.5^{+3.6}_{-1.9}$   \\
			IRAS~15469-5311            & $3$                   & $0.294$                          & $37.6$                            & $0.82$                        & $0.89$                       & $52.9$                & --                            & --                   & --                             & --                             & $1.84$             & $8.99_{-0.09}^{+0.09}$  & $35_{-2}^{+2}$       & $51_{-3}^{+3}$               & $0.23_{-0.06}^{+0.06}$         & $-0.29_{-0.05}^{+0.05}$        & $5.3^{+0.3}_{-0.2}$    \\
			IW~Car\tablefootmark{\,d}  & $3$                   & $0.350$                          & $44.8$                            & $1.52$                        & $0.95$                       & $63.6$                & $0.50$                        & $1.99$               & $1.12$                         & $-1.90$                        & $0.89$             & $6.9_{-0.5}^{+0.5}$     & $48_{-5}^{+5}$       & $144_{-7}^{+8}$              & $0.15_{-0.18}^{+0.19}$         & $-0.1_{-0.2}^{+0.2}$           & $11.7^{+1.0}_{-0.4}$   \\
			                           &                       &                                  &                                   &                               &                              &                       &                               &                      &                                &                                &                    & $20.3_{-0.8}^{+0.9}$    & $57_{-2}^{+2}$       & $176_{-3}^{+3}$              & $-1.0_{-0.2}^{+0.3}$           & $-1.4_{-0.4}^{+0.4}$           & $11.9^{+1.0}_{-0.3}$   \\
			PS~Gem\tablefootmark{\,c}  & $3$                   & $0.311$                          & $39.8$                            & $0.71$                        & $0.88$                       & $87.0$                & $0.37$                        & --                   & --                             & --                             & $0.61$             & $9.0_{-0.3}^{+0.4}$     & $28_{-3}^{+5}$       & $161_{-12}^{+7}$             & $0.027_{-0.15}^{+0.15}$        & $-0.4_{-0.2}^{+0.2}$           & $7.5^{+0.8}_{-0.8}$    \\
			\hline
		\end{tabular}
		}
		\renewcommand{\arraystretch}{1}
		\tablefoot{\tablefoottext{a}{$\mathrm{SF}$, $\mathrm{PS}$ \& $\mathrm{FOV}$ are the image sampling factor (i.e.\ how many pixels fit in the beam size), pixel scale and field-of-view.} \tablefoottext{b}{The secondary positions $(\Delta \alpha_{\mathrm{sec}}, \Delta \delta_{\mathrm{sec}})$ are taken from the final geometric models (Table \ref{table:geom_params}; except for IW~Car, where it follows the fitted Gaussian position in Fig.\ \ref{fig:IW_Car_secondary_detection}).} \tablefoottext{c}{HR~4049 and possibly PS Gem likely have overestimated $r_{\mathrm{hl}}$ values, since their images account for an over-resolved flux component (see Fig.\ \ref{fig:radial_brightness_profiles}). Similarly, $r_{\mathrm{hl}}$ could be slightly underestimated for AI~Sco, since the flux close to the primary is systematically offset southwards (see Sect.\ \ref{sect:AI_Sco_detected_features}).} \tablefoottext{d}{For IW~Car, we find both inner flux arcs and a large outer flux arc (see Sect.\ \ref{sect:IW_Car_detected_features}), and we provide fits for both.}}
	\end{table*}

\begin{figure*}[t]
		\centering

		\centering
		\begin{subfigure}
			{0.5\textwidth}
			\includegraphics[width=\linewidth]{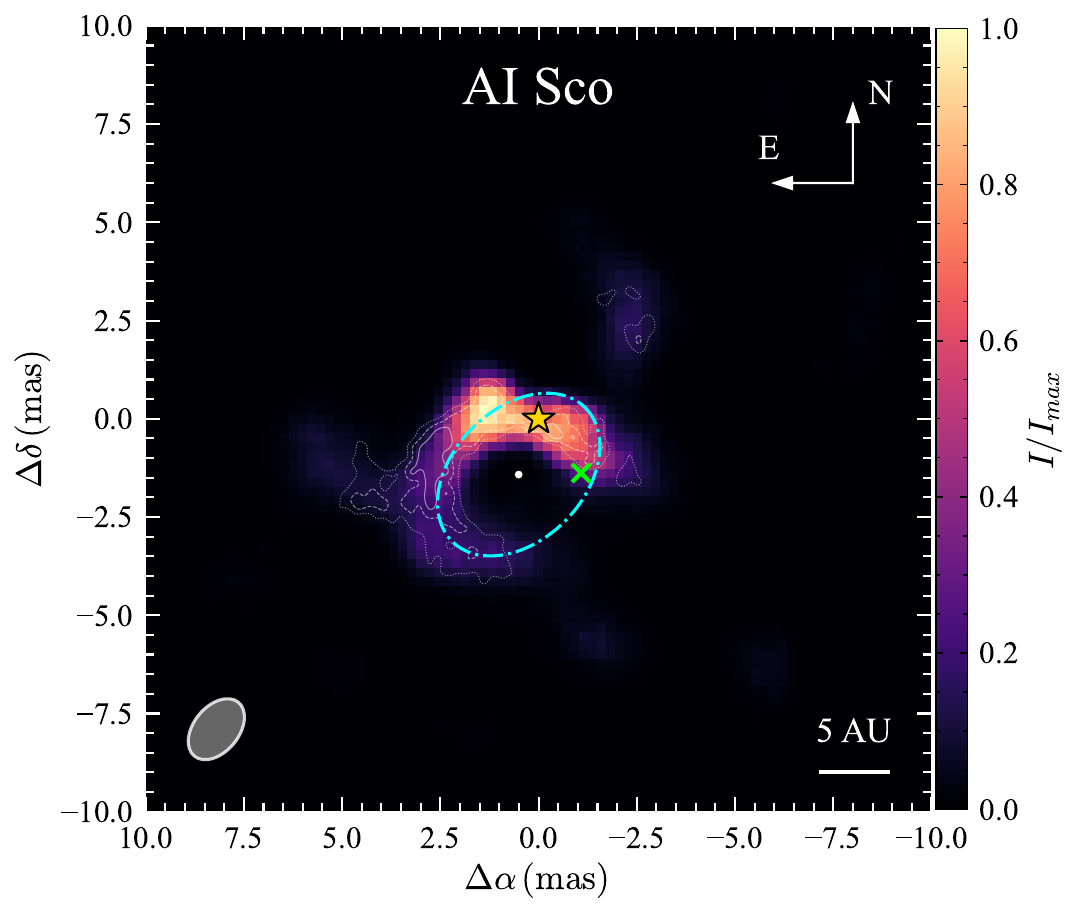}
		\end{subfigure}\hfil
		\begin{subfigure}
			{0.5\textwidth}
			\includegraphics[width=\linewidth]{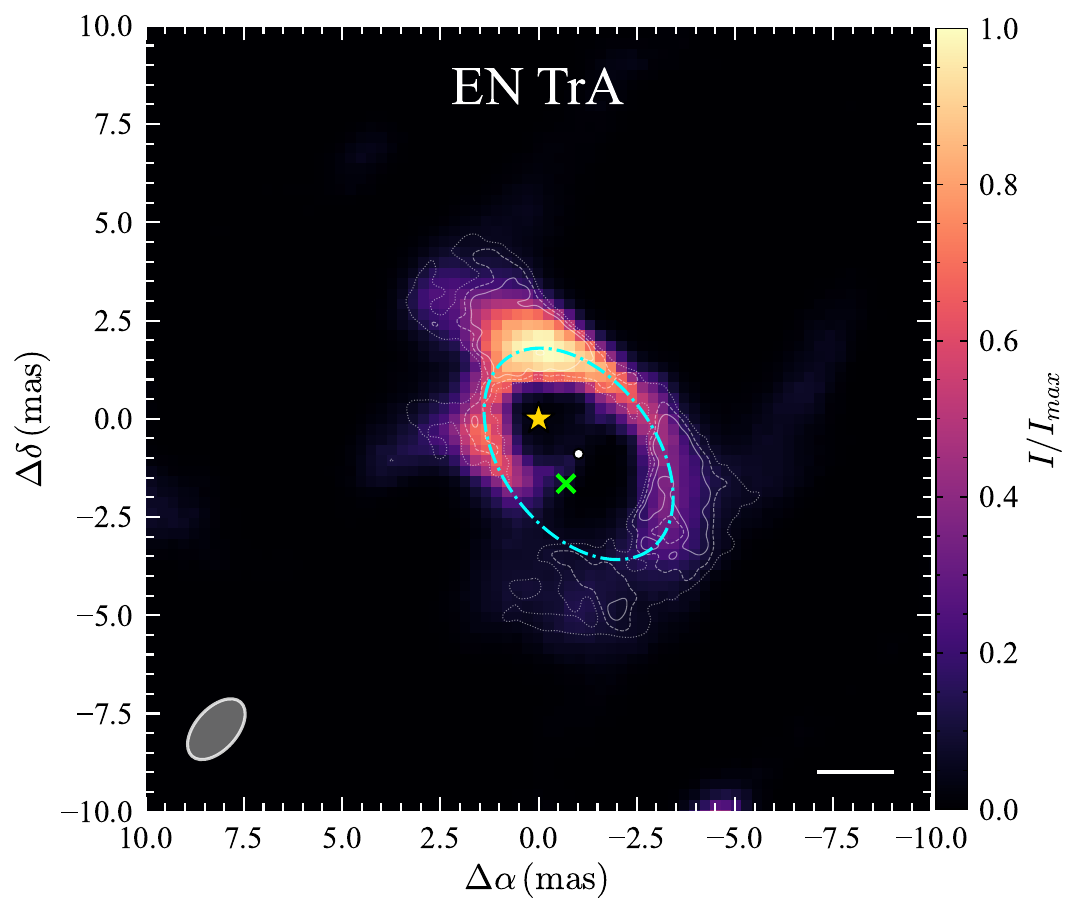}
		\end{subfigure}\hfil

		\vspace{-1ex}

		\begin{subfigure}
			{0.5\textwidth}
			\includegraphics[width=\linewidth]{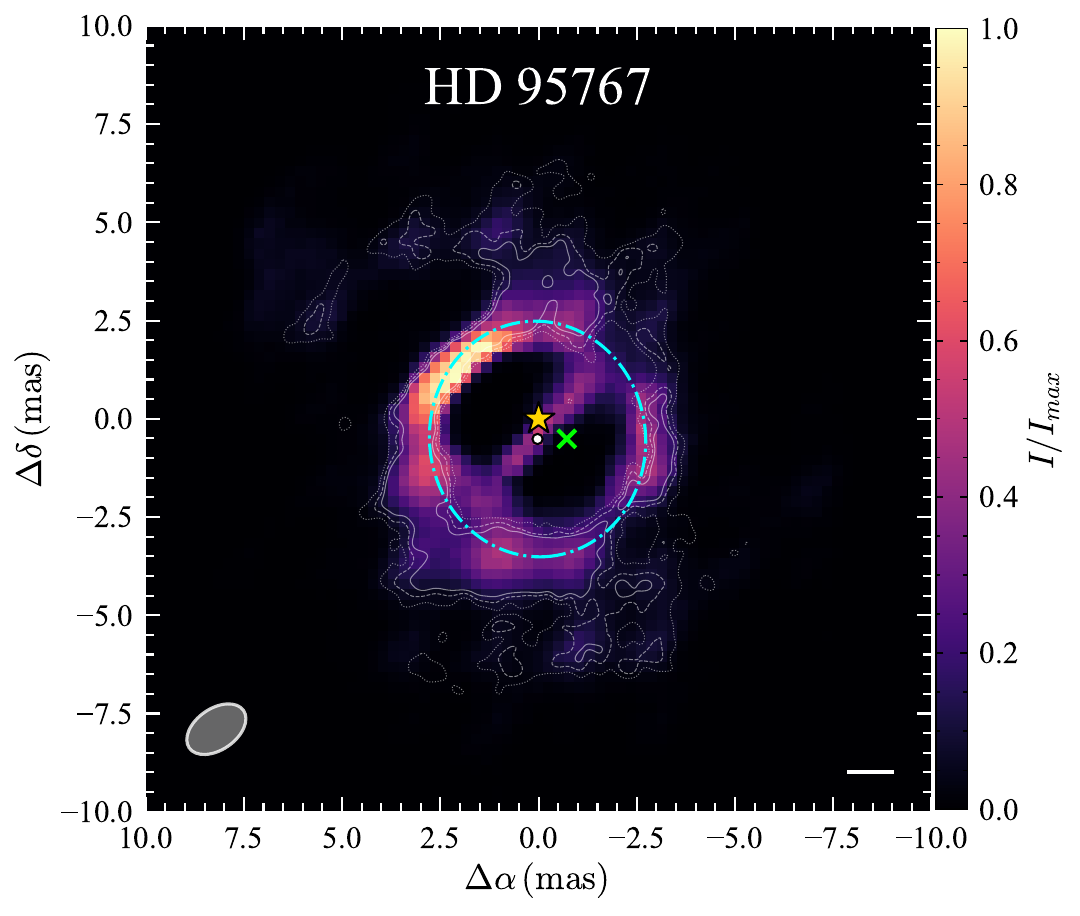}
		\end{subfigure}\hfil
		\begin{subfigure}
			{0.5\textwidth}
			\includegraphics[width=\linewidth]{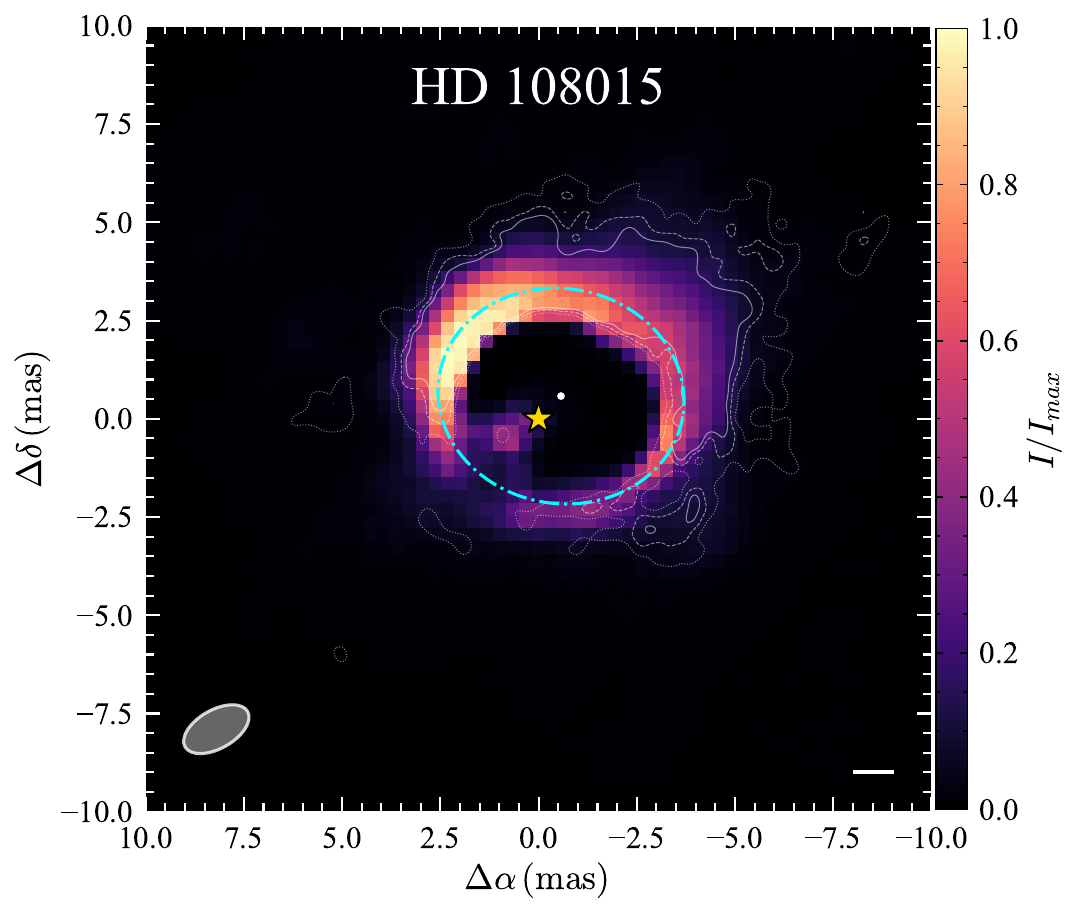}
		\end{subfigure}\hfil

		\caption{Peak-normalised ORGANIC images for our first four targets (using the
		image parameters in Table \ref{table:img_rec_params}). The SPARCO primary and
		secondary (if detected) are denoted by a golden star and green cross,
		respectively. The median inner rim fit and the position of its centre (see
		Table.\ \ref{table:img_rec_params}) are denoted by a dash-dotted cyan
		ellipse and a white dot, respectively. $3$, $4$ and $5\sigma$ contours are given
		as dotted, dashed and solid white lines. The interferometric beam size and a
		$5 \, \mathrm{AU}$ scale bar are shown in the bottom.}
		\label{fig:organic_imgs_first_four}
	\end{figure*}

\begin{figure*}[t]
		\centering

		\centering
		\begin{subfigure}
			{0.5\textwidth}
			\includegraphics[width=\linewidth]{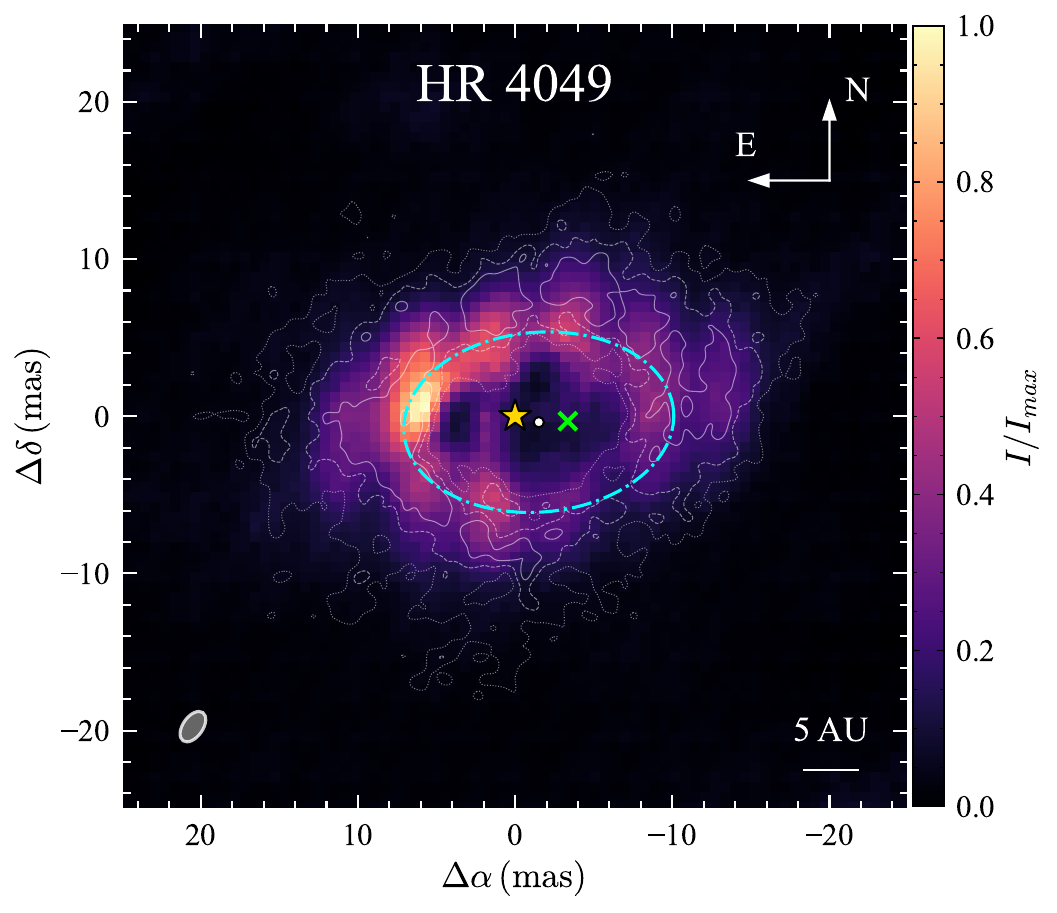}
		\end{subfigure}\hfil
		\begin{subfigure}
			{0.5\textwidth}
			\includegraphics[width=\linewidth]{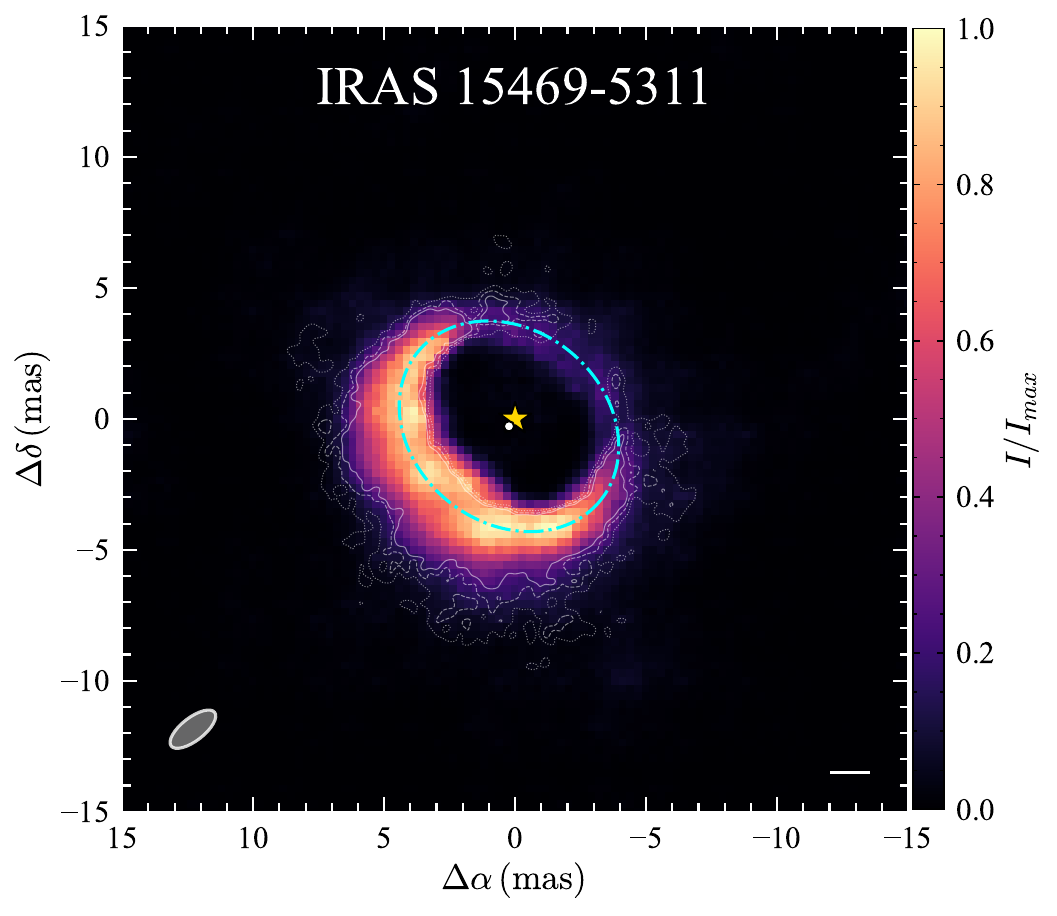}
		\end{subfigure}\hfil

		\vspace{-1ex}

		\begin{subfigure}
			{0.5\textwidth}
			\includegraphics[width=\linewidth]{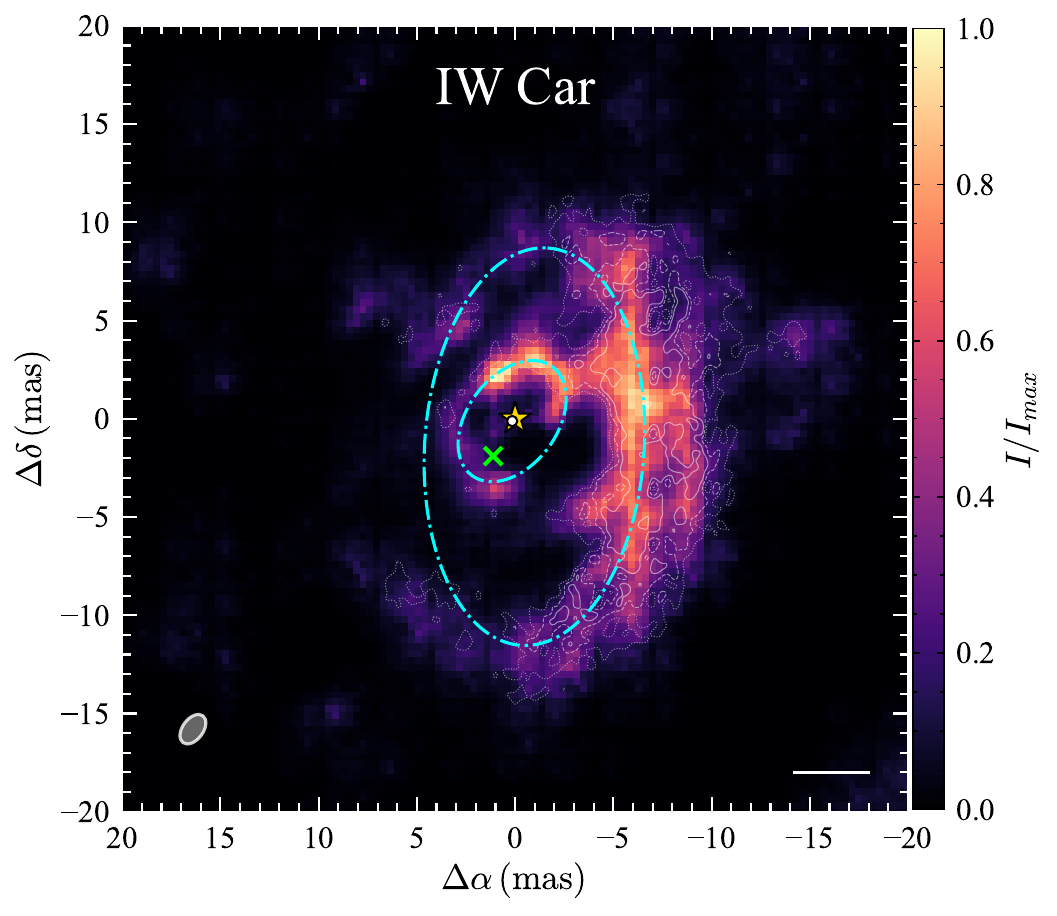}
		\end{subfigure}\hfil
		\begin{subfigure}
			{0.5\textwidth}
			\includegraphics[width=\linewidth]{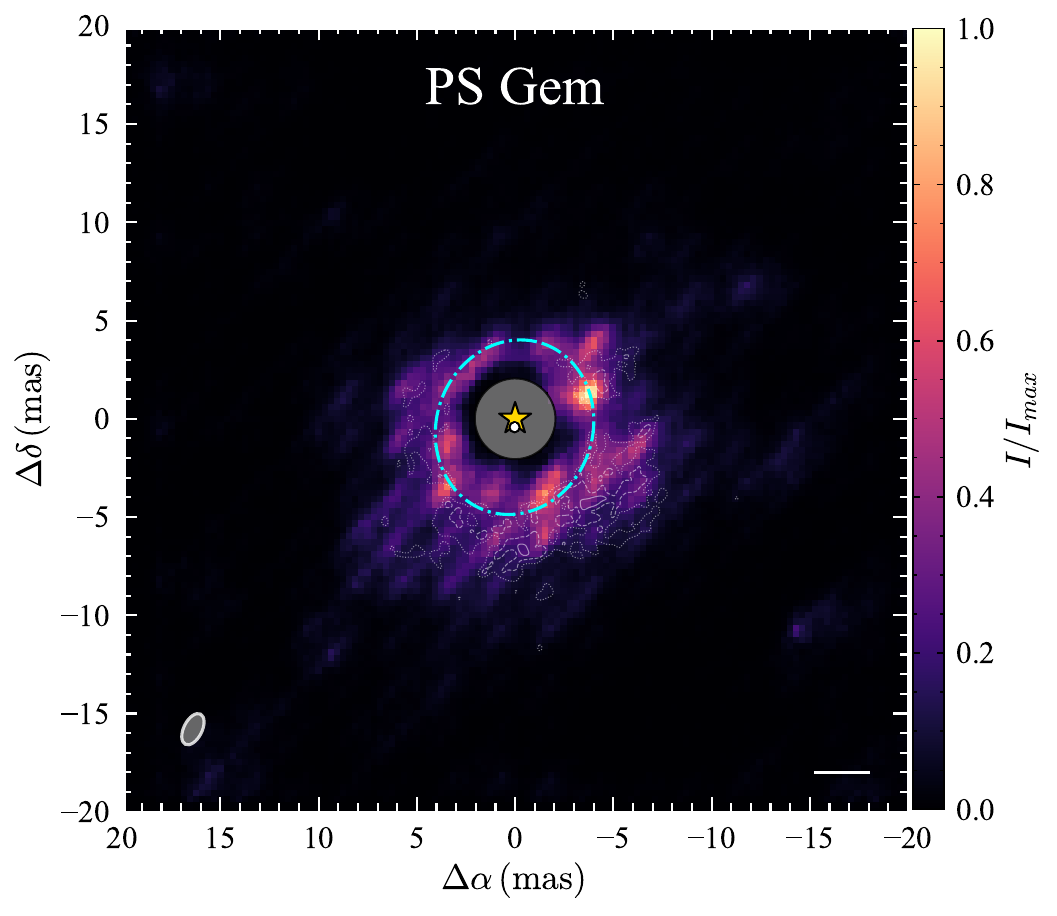}
		\end{subfigure}\hfil

		\caption{Analogous to Fig.\ \ref{fig:organic_imgs_first_four}, but for our
		last four targets. For IW~Car, we show inner rim fits for both the inner and
		outer flux arcs (see Sect.\ \ref{sect:IW_Car_detected_features}). For the
		sake of clarity, the rim fit centre is only given for the innermost rim fit.
		For PS~Gem, the strong flux smearing from the primary (see Sect.\ \ref{sect:PS_Gem_detected_features})
		has been masked out, as denoted by a grey circle, to better reveal the
		circumstellar medium.}
		\label{fig:organic_imgs_last_four}
	\end{figure*}

	We applied the image reconstruction workflow developed in Sect.\ \ref{sect:methodology}
	to our eight targets (Sect.\ \ref{sect:target_systems}), providing high fidelity,
	non-parametric images of their circumstellar environment. The parameters for our
	final ORGANIC images are summarised in Table \ref{table:img_rec_params}, while
	the reconstructed images themselves are shown in Figs.\ \ref{fig:organic_imgs_first_four}
	\& \ref{fig:organic_imgs_last_four}. These image reconstructions and their noise levels are available online on \href{https://doi.org/10.5281/zenodo.20155600}{Zenodo}. The dusty inner disc rim is well-resolved
	for all targets. We first present a fitting routine to extract the overall inner
	rim properties and radial surface brightness profiles in Sect.\ \ref{sect:rim_fitting_and_brightness_profile_calculation},
	and briefly describe the derived rim inclinations in the context of the 'RVb
	phenomenon' -- long-term secular brightness modulations following
	$P_{\mathrm{orb}}$ (e.g. \citealt{Manick2017}) -- in Sect.\ \ref{sect:rvb_phenomenon}.
	We then describe the detected morphological features in Sect.\ \ref{sect:detected_features}.
	A comprehensive discussion of the possible underlying physical mechanisms is presented
	in Sect. \ref{sect:discussion}.

	\begin{figure}[t]
		\centering
		\includegraphics[width=1.0\linewidth]{
			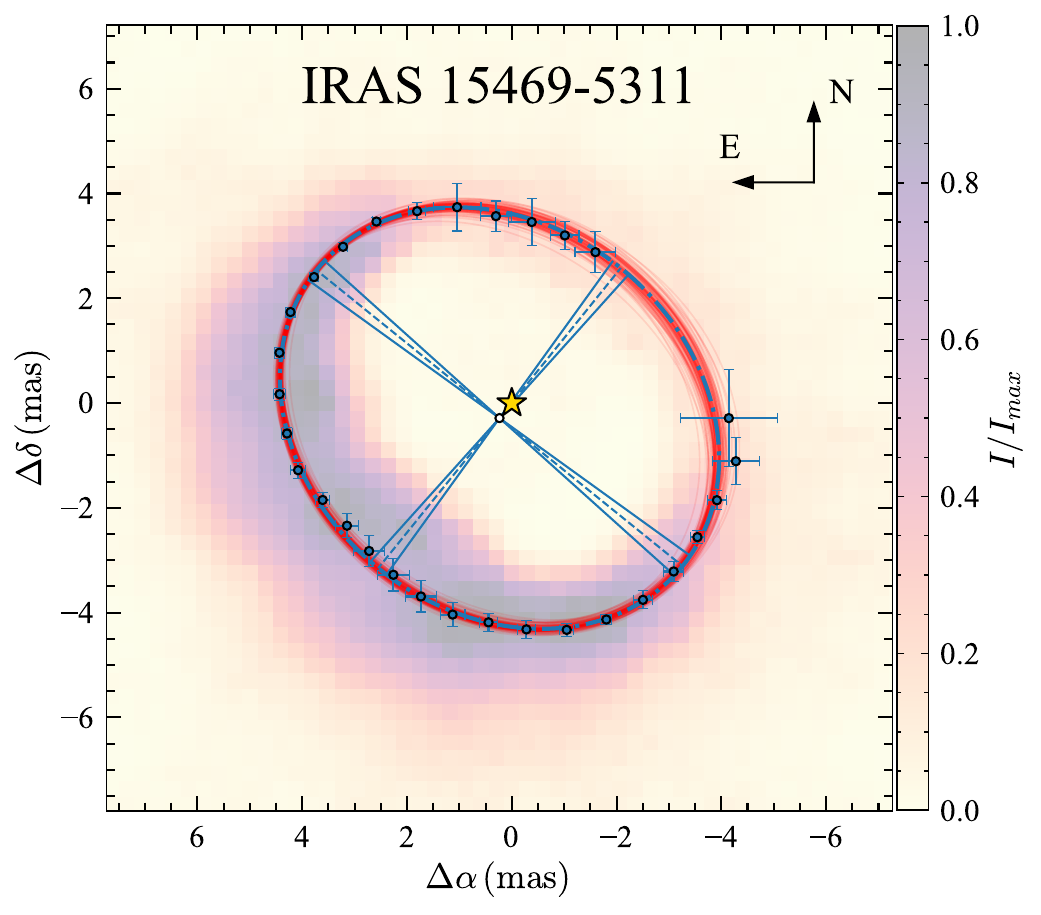
		}
		\caption{Example of the inner rim fitting procedure for IRAS~15469-5311.
		Blue errorbars denote the rim positions along different radial rays. Red ellipses
		show 100 random samples drawn from the rim's MCMC chains, with the median
		ellipse (Table \ref{table:img_rec_params}) shown in dot-dashed blue. Lines drawn
		from the median rim centre (white dot) indicate the direction and error of the
		projected rim's minor and major axes.}
		\label{fig:organic_inner_rim_fit_example}
	\end{figure}

	\subsection{Rim fitting and brightness profile calculation}
	\label{sect:rim_fitting_and_brightness_profile_calculation} We developed an ellipse
	fitting procedure to retrieve the inner rim parameters directly from the images.
	Radial rays were drawn from the primary every $10^{\mathrm{\circ}}$
	azimuthally, including only rim sections of higher per-pixel significance ($\geq
	\!3 \sigma$
	for all targets except PS Gem, for which we use $\geq\!2 \sigma$). A Gaussian was
	matched to each ray's radial profile to constrain the radial position of peak brightness.
	This was repeated for every image in the ORGANIC cubes to determine the rim
	positions and their uncertainties, taken as the median and standard deviation across
	the cube. For AI~Sco, we placed a $0.9\,\mathrm{mas}$ diameter circular mask over
	the primary during this procedure, since the image flux close to its position
	is likely biased slightly southwards (see Sect.\ \ref{sect:AI_Sco_detected_features}).

	An ellipse was then fit to the positions in order to derive the rim parameters
	and their uncertainties. For this, we used the Markov Chain Monte Carlo (MCMC)
	package emcee\footnote{\url{https://github.com/dfm/emcee}} \citep[][]{Foreman-Mackey2013}
	with 32 walkers performing 100~000 steps each. An example of this procedure is
	shown in Fig.\ \ref{fig:organic_inner_rim_fit_example}. The resulting inner
	rims are displayed on Figs.\ \ref{fig:organic_imgs_first_four} \& \ref{fig:organic_imgs_last_four},
	and their parameters summarised in Table \ref{table:img_rec_params}. We stress
	that the emission does not probe the mid-plane, but originates from (sub-)$\mathrm{\mu
	m}$
	grains lifted to layers around $1\!-\!2$ gas scale heights higher up \citep[e.g.][]{Hillen2014, Kluska2018}.
	On-sky projection can then skew the centre of emission ($\Delta \alpha_{\mathrm{rim}}$,
	$\Delta \delta_{\mathrm{rim}}$) along the projected rim's minor axis, away
	from the centre of mass \citep[e.g.][]{deBoer2016}. We note that in case the
	inner rim is sharp and wall-like, the limited beam size of the observations causes
	the imaged morphology of the rim to be slightly washed out. This can cause our
	rim fitting method to induce a systematic positive bias of typically
	$\lesssim 1 \, \mathrm{mas}$ ($\lesssim 2 \, \mathrm{mas}$ for HR~4049 \& PS~Gem)
	in $\theta_{\mathrm{in}}$ and a negative bias of
	$\lesssim 10^{\mathrm{\circ}}$ in $i_{\mathrm{rim}}$ relative to the true
	values (see Sect.\ \ref{sect:geom_model_synthetic_reconstructions}). However, at
	least for rims set by sublimation physics, there is little reason to suspect
	the rim would be sharp in the first place (see Sect.\ \ref{sect:inner_rim_comparison_to_sublimation}).

	Using concentric elliptical apertures following the rim orientations (Table
	\ref{table:img_rec_params}), we then calculated radial surface brightness
	profiles. This was again repeated for every image in the image cubes to provide
	$1\sigma$ uncertainties. The resulting profiles are shown in Fig.\ \ref{fig:radial_brightness_profiles}.
	For IW~Car, where we find both inner flux arcs and a large outer flux arc (see
	Sect.\ \ref{sect:IW_Car_detected_features}), the rim fitting procedure was applied
	to both features. The IW~Car brightness profile calculation was based on the
	orientation of the rim fit to the inner arcs.

	\subsection{Apparent rim inclinations and the RVb phenomenon}
	\label{sect:rvb_phenomenon} Using PIONIER snapshots, \citet{Kluska2019}
	performed geometric model fitting of the inner rim and central binary for a
	large sample of 23 post-AGB circumbinary discs. They found that the apparent inclination
	of rims observable with PIONIER ranges between $i_{\mathrm{rim}}\!\sim\!10-65^{\circ}$.
	A significant subset of post-AGB binaries also displays a long-term brightness
	modulation following the binary period known as the RVb phenomenon. It is commonly
	thought to be induced by periodic obscuration along the binary orbit as the line-of-sight
	to the post-AGB primary grazes the inner rim, and is thus considered to be an
	indicator of high rim inclination \citep[e.g.][]{VanWinckel1999, Manick2017}.
	Nevertheless, \citet{Kluska2019} only found high apparent inclinations ($i_{\mathrm{rim}}
	\gtrsim 60^{\circ}$) for three out of nine of their RVb targets, but the authors
	advised caution when considering their inclination values given the generally poor
	$(u,v)$ coverage.

	The RVb phenomenon has also been firmly established in five out of eight of our
	targets: AI~Sco, HD~95767, HR~4049, IW~car \& PS~Gem (Table
	\ref{table:target_stars_and_observations}). Using our images obtained under
	greatly superior $(u,v)$ coverage than the snapshots in \citet{Kluska2019}, we
	find moderate apparent inclinations for AI~Sco, HR~4049 \& IW~Car ($45^{\circ}\lesssim
	\! i_{\mathrm{rim}}\! \lesssim 60^{\circ}$; Table \ref{table:img_rec_params}).
	Moreover, we find low apparent inclinations for HD~95767 and PS~Gem ($i_{\mathrm{rim}}
	\lesssim 30^{\circ}$; Table \ref{table:img_rec_params}), though we note their
	RVb-related $V$-band brightness semi-amplitudes are relatively weak at
	$\Delta V \lesssim 0.1 \, \mathrm{mag}$ \citep[][]{Kiss2007, VanWinckel1999}.
	These results corroborate the findings of \citet[][]{Kluska2019}, showing that
	the RVb phenomenon can indeed still occur in the presence of moderate-to-low
	inclination disc rims. This implies very high disc scale heights in case the conventional
	disc obscuration hypothesis holds true, and that the exact link between disc
	rim inclination and the presence of the RVb phenomenon is not yet clear.

	\subsection{Detected features}
	\label{sect:detected_features} Below, we describe the detected morphological features
	in our images, and, when possible, note on connections with features detected
	in previous studies on the same systems. Especially pertinent in this respect again
	are the PIONIER snapshot geometric fitting results of \citet{Kluska2019}. Since
	this included all our targets, this enables comparison across a multi-year
	timebase. We note that this comparison should nevertheless be approached cautiously,
	as \citet{Kluska2019} used sparse $(u,v)$ coverages and relatively simplistic inner
	rim models up to only second order in azimuthal modulation. Potential
	reconstruction artefacts revealed by our tests in Appendices~\ref{sect:appendix_geom_models_and_synth_reconstructions},
	\ref{sect:appendix_squeeze_reconstructions_appendix} \& \ref{sect:appendix_robustness_against_uv_coverage}
	are addressed on a case-by-case basis.

	\subsubsection{AI~Sco}
	\label{sect:AI_Sco_detected_features} Simple RT inclination effects on a smooth,
	circular rim, centred on a single star, dictate that the NIR emission
	originates from the far side of the rim relative to the observer. The near side
	is namely obscured due to self-shadowing, resulting in emission that is
	antisymmetric relative to the projected rim's major axis and symmetric
	relative to the minor one \citep[e.g.][]{Hofmann2022}. While the north-eastern
	side of AI~Sco's rim is generally brighter than the south-western -- likely marking
	the former as the far side -- there is also a strong brightness enhancement
	towards the primary's position and the rim's major axis (Fig.\ \ref{fig:organic_imgs_first_four}).
	The flux in the immediate vicinity of the primary is also slightly offset
	southward (by about $\!\sim \!1\, \mathrm{mas}$) from where it's expected to
	be according to the rim fit. As shown in Appendix \ref{sect:appendix_geom_models_and_synth_reconstructions},
	this is a bias likely induced by the lack of long baseline observations
	towards the north and north-east (Fig.\ \ref{fig:uv_coverage}) in combination with
	the close approach of the primary to the rim. Nevertheless, this morphology cannot
	be accounted for solely with inclination effects.

	The brightness enhancement next to the primary dominates and heavily biases our
	geometric model towards it, which fails to fully reproduce the fainter 'tail'
	of emission towards the south-east, resulting in a poor goodness-of-fit at $\chi
	^{2}_{\nu}\!\sim\! 5$ (Fig.\ \ref{fig:pmoired_images} \& Table \ref{table:geom_params}).
	We generally caution that in cases like AI~Sco, where the azimuthal brightness
	profile has sharp features, our simple geometric models with only one order of
	azimuthal modulation can fail to capture the full complexity and present faulty
	rim orientations (cf.\ AI~Sco's values in Tables \ref{table:geom_params} \& \ref{table:img_rec_params}).
	Nevertheless, the \citet{Kluska2019} model for AI~Sco (data taken
	$\sim\!5 \,\mathrm{yr}$ before ours) similarly found a strong brightness enhancement
	in the direction of the primary (see the model images in their Fig.\ C.1).
	This suggests that the bright enhancement could be long-lived, lasting at least
	several years, and following the primary.

	\subsubsection{EN~TrA}
	\label{sect:EN_TrA_detected_features} While the inner rim is significantly
	brighter along the north-western side compared to the south-eastern -- suggesting
	the former represents the far side of the inclined rim -- the image cannot be
	fully explained by just inclination effects (Fig.\ \ref{fig:organic_imgs_first_four}).
	The rim has a brightness enhancement shifted towards its major axis. This enhancement
	aligns with the direction of the primary, with both tending towards the north-east
	side of the rim. The \citet{Kluska2019} geometric model fit (data taken $\sim\!
	7\,\mathrm{yr}$ before ours) found a brightness enhancement almost symmetric
	along the north-western side. While this positioning was compatible with disc inclination
	effects, it also aligned with the direction of the primary at that time. As for
	AI~Sco, the brightness enhancement in our image could last for multiple years
	and follow the primary.

	\subsubsection{HD~95767}
	\label{sect:HD95767_detected_features} The inner rim shows a strong brightness
	enhancement to the north-east, shifted towards its major axis. It could
	tentatively be aligned with the direction of the primary. The geometric model of
	\citet{Kluska2019} (data taken $\sim\!7.5\,\mathrm{yr}$ before ours) found the
	disc to be brightest in the south-east, aligned in the direction of the primary.
	While suggestive, this should be approached with caution. As shown in Appendix
	\ref{sect:appendix_robustness_against_uv_coverage}, the brightness enhancement
	is highly sensitive to the $(u,v)$ coverage, likely due the combination of a
	small over-\ or under-subtraction of the SPARCO secondary together with a particularly
	high secondary flux fraction ($f_{\mathrm{sec,0}}\!\sim \!25\%$, Table \ref{table:img_rec_params}).
	Similarly, the small amount of lower per-pixel significance emission in the
	inner rim cavity is highly likely to be a dirty beam artefact related to the
	specifics of the $(u,v)$ coverage (see Sect.\ \ref{sect:identifying_beam_artefacts}).

	The inner rim emission seems to faintly bulge towards the north and south,
	with the former curving into an arch-like shape. Although this feature is intriguing,
	we similarly urge caution in its interpretation, given the high sensitivity of
	the brightness enhancement and the fact that the dirty beam exhibits a negative
	flux arch at a comparable location (Fig.\ \ref{fig:dirty_beams}). Due to these
	complications, we exclude HD~95767 from our discussion on potential
	substructure formation mechanisms further on in Sect.\ \ref{sect:origins_of_structures}.

	\subsubsection{HD~108015}
	\label{sect:HD108015_detected_features} The inner rim is generally brighter along
	the northern side compared to the southern, suggesting that the former
	represents the far side of the inclined rim. There is a strong brightness enhancement
	shifted towards the major axis in the north-east (though there is some
	uncertainty up to $\sim\!10^{\circ}$ in its position angle; see Appendix \ref{sect:appendix_squeeze_reconstructions_appendix}).
	The primary motion in HD~108015 is not expected to be resolved much at all.
	Using the spectroscopic value of the primary's projected semi-major axis, $a_{1}
	\sin{i_{\mathrm{bin}}}$ (Table \ref{table:geom_params}) and assuming co-planarity
	between the binary (i.e.\ $i_{\mathrm{bin}}\!=\!i_{\mathrm{rim}}$; Table
	\ref{table:img_rec_params}) -- a reasonable assumption given that the disc is
	thought to form from the binary -- we estimate $a_{1}\!\sim\! 0.1\,\mathrm{mas}$.
	Hence, the offset between the projected rim centre and primary position is in this
	case mostly a projection effect due to the fact that the $H$-band continuum is
	emitted from higher-up regions in the disc (Sect.\ \ref{sect:rim_fitting_and_brightness_profile_calculation}).
	The small amount of flux interior of the cavity towards the south-east is likely
	an artefact related to the $(u,v)$ coverage (see Appendix
	\ref{sect:appendix_robustness_against_uv_coverage}).

	It is thus not clear towards which direction of the disc rim the primary is offset.
	\citet{Kluska2019} (data taken $\sim\!7.5\,\mathrm{yr}$ before ours) instead found
	a strong brightness enhancement in the south-western sector of the inner rim,
	approximately opposite of the enhancement we find. If this feature were to be following
	the primary, given the binary period of $906\pm6 \, \mathrm{d}$ and the fact
	that the orbit is circular \citep[][]{Oomen2018}, we would have expected it to
	have moved by only $33 \pm 7\mathrm{^\circ}$ (calculated between the mean MJD of
	each observing epoch). Hence, this indicates that the enhancement found in our
	image is likely moving at its own angular speed, independent of the binary orbit.

	\subsubsection{HR~4049}
	\label{sect:HR4049_detected_features} The rim is generally brighter along the northern
	side relative to the southern, marking the former as its far side. A strong brightness
	enhancement is seen along the eastern side of the rim's major axis, aligned
	with the position of the primary. The imaged rim has a slightly choppy appearance,
	likely caused by the sparser $(u,v)$ coverage along the larger baselines (Fig.\ \ref{fig:uv_coverage}).
	While the western brightness enhancement seems aligned with the primary
	position, the geometric model by \citet{Kluska2019} (data taken $\sim\!5\,\mathrm{yr}$
	before ours) instead shows the main brightness enhancement to also be oriented
	towards the west, while the primary was oriented east towards a secondary brightness
	enhancement. As for HD~108015, this tentatively suggests the main brightness enhancement
	could be long-lived and moving at its own pace, though we do not exclude the
	possibility that the brightness enhancement is following the primary instead.

	\subsubsection{IRAS~15469-53110}
	\label{sect:IRAS154569-5311_detected_features} The rim is brighter along the
	south-eastern side, marking it as the far side of the inner rim. Furthermore, the
	rim is highly anti-symmetric along its projected major axis, and
	simultaneously symmetric along the minor axis. This indicates that the rim morphology,
	at least at the resolution probed by our observations, is consistent with that
	of a smooth, inclined circular rim with a centred star, without any other substructure
	or perturbation.

	\subsubsection{IW~Car}
	\label{sect:IW_Car_detected_features} IW~Car's inner disc regions display a highly
	complex morphology. While a broad outer flux arc is seen to the west, smaller flux
	arcs are seen closer to the inner binary. These are generally of slightly
	lower per-pixel significance ($\leq 3\sigma$), yet represent a significant fraction
	of the total image flux at $12. 4 \pm 1.6\,\%$ ($\sim\!4.3 \pm 0.5\, \%$ of
	the total flux in $H$). We note that the exact details of the inner arcs' azimuthal
	profile are not fully well-constrained (see Appendix
	\ref{sect:appendix_squeeze_reconstructions_appendix}).

\begin{figure*}
		\sidecaption
		\includegraphics[width=12cm]{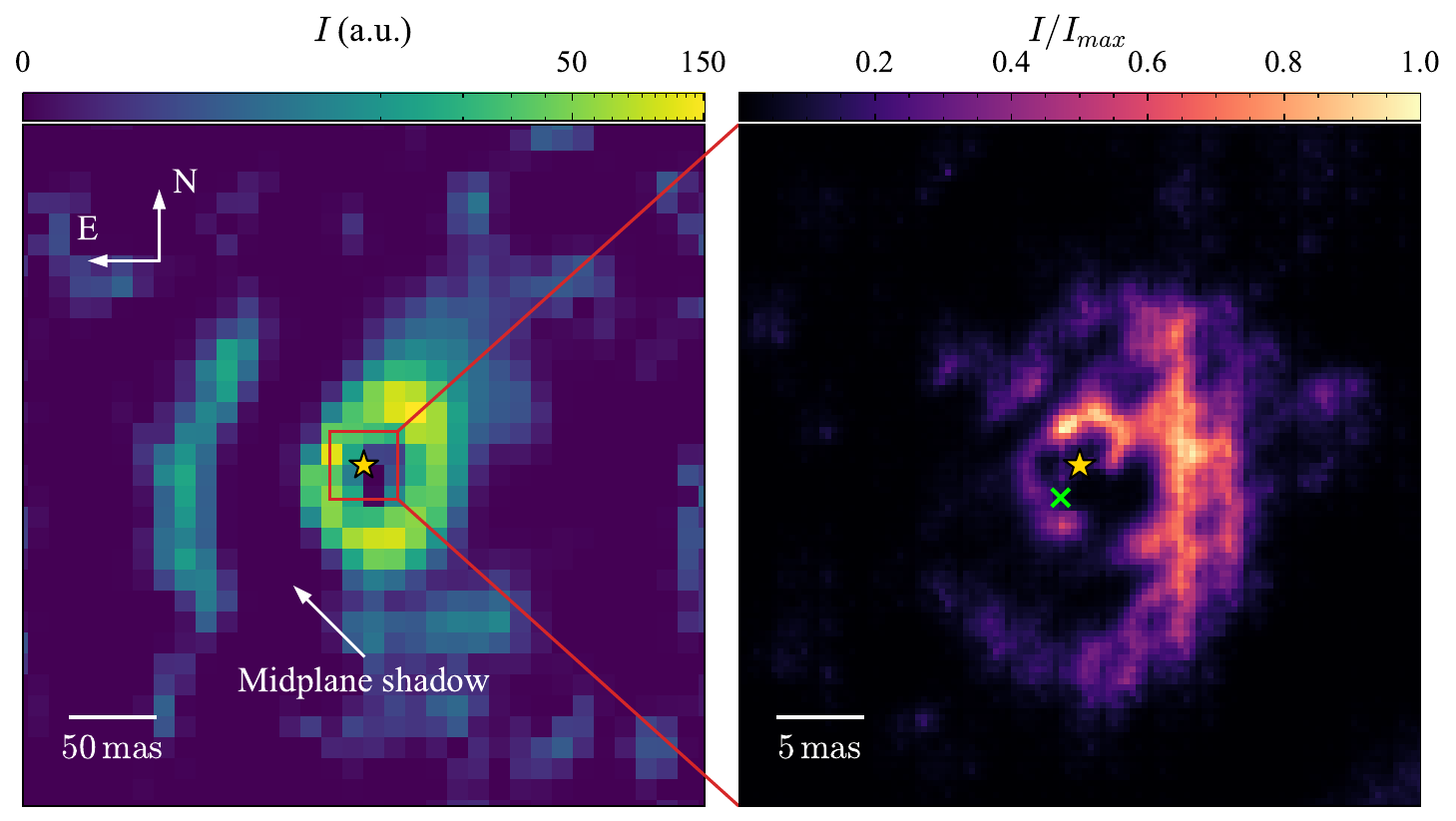}
		\caption{SPHERE-PIONIER comparison for IW~Car. \textit{Left:} SPHERE/IRDIS
		image \citep[polarised $Q_{\phi}$ intensity, adapted from][]{Andrych2023},
		probing $H$-band starlight scattered on the outer disc. The image is presented
		on an $\mathrm{arcsinh}$ scale to highlight faint features. The central dark
		hole is a data reduction artefact due to over-subtraction of the central
		unresolved polarised flux \citep{Andrych2023}. We mark the midplane's shadow
		and the position of the primary. The partial arc to the east likely traces light
		scattered from the side of the disc facing away from the observer \citep{Andrych2023}.
		\textit{Right:} zoomed inset with our ORGANIC PIONIER image, probing $H$-band
		thermal emission in the inner disc regions. The position of the secondary is
		marked by a green cross.}
		\label{fig:IW_Car_SPHERE_VS_PIONIER}
	\end{figure*}

	It is not a priori clear whether the outer arc or inner arcs, or perhaps both,
	represent a bona fide rim structure. As a result, we applied our rim fitting procedure
	(Sect.\ \ref{sect:rim_fitting_and_brightness_profile_calculation}) to each
	separately. If both do represent rims, their orientation is likely misaligned,
	with a difference in inclination and position angle of $\sim \!8 \pm 5\mathrm{^\circ}$
	and $33\pm8\mathrm{^\circ}$ ($i = 48.5 \pm 5 \mathrm{^\circ}$ \&
	$\mathrm{PA}= 144_{-7}^{+8}\mathrm{^\circ}$ for the inner arcs and
	$i = 57\pm 2\mathrm{^\circ}$ \& $\mathrm{PA}= 176 \pm 3 \mathrm{^\circ}$ for
	the outer arc; Table \ref{table:img_rec_params}). Both these orientations are relatively
	consistent with the orientation of the outer disc found by \citet{Andrych2023}
	(Fig.\ \ref{fig:IW_Car_SPHERE_VS_PIONIER}), using SPHERE/IRDIS $H$-band
	polarimetric scattered light imaging ($i = 41.7^{+6.5}_{-7.7}\mathrm{^\circ}$ \&
	$\mathrm{PA}= 161^{+12}_{-10}\mathrm{^\circ}$; taken $\sim\!1\,\mathrm{yr}$
	before our data). This comparison is however complicated by the fact that the
	outer disc orientation could be biased by artefacts induced during the SPHERE
	data reduction. Specifically, the inner 'holes' due to subtraction of the
	unresolved polarised flux \citep{Andrych2023}. It is not clear which of the
	features seen with PIONIER traces the actual inner rim of the outer disc seen
	by SPHERE.

	The geometric fit by \citet{Kluska2019} (data taken $\sim\!5 \,\mathrm{yr}$
	before ours) recovered the outer flux arc, though it could not capture the
	inner arcs and the full complexity of the inner disc (as shown by their poorer
	goodness-of-fit at $\chi^{2}_{\nu}\!= \!3.9$).

	\subsubsection{PS~Gem}
	\label{sect:PS_Gem_detected_features} PS Gem's image reconstruction shows a very
	strong flux smearing from the primary (masked in Fig.\ \ref{fig:organic_imgs_last_four}),
	with a contribution of $\sim\!15\pm 3 \,\%$ to the image flux ($\sim\!1.9 \pm 0
	.3\, \%$
	of the total $H$-band flux). This is likely due to the primary's orbital
	movement over the timespan of the observations ($\Delta t_{\mathrm{obs}}/P_{\mathrm{orb}}
	\!=\!8.7 \,\%$; Table \ref{table:target_stars_and_observations}), causing it
	to be better represented as a slightly smeared Gaussian instead of a uniform
	disc. Its effect is exacerbated by the low flux contribution of the disc at $\sim
	\!13\,\%$ (causing only weak non-zero closure phase signal; Fig.\ \ref{fig:observables_t3})
	and the strong dirty beam secondary lobes caused by the sparser $(u,v)$ coverage
	(Figs.\ \ref{fig:uv_coverage} \& \ref{fig:dirty_beams}). This causes the retrieved
	inner rim azimuthal morphology to be very choppy and mostly of low per-pixel significance
	($\leq 3\sigma$), though the radial intensity profile is still well-constrained
	(Fig.\ \ref{fig:radial_brightness_profiles}). While we tentatively discern a
	bright spot in the west and a broader brightness enhancement in the south of
	the rim (the latter also being retrieved by our geometric model; Fig.\ \ref{fig:pmoired_images}),
	we forgo including them in our discussion on the origins of potential substructures
	in Sect.\ \ref{sect:origins_of_structures} due to these complications.

	Overall, five out of eight systems show robust, non-axisymmetric brightness
	features inconsistent with solely inclination effects, indicating that inner rims
	of post-AGB circumbinary discs are often perturbed.

	\section{Discussion}
	\label{sect:discussion} In this section, we first examine the mechanisms that
	determine the inner rim position (Sect.\ \ref{sect:mechanism_inner_rim_position}).
	We then turn to the possible origins of the robustly detected substructure
	candidates, drawing heavily on the extensive literature available for protoplanetary
	discs (Sect.\ \ref{sect:origins_of_structures}).

	\subsection{Mechanism behind the inner rim position}
	\label{sect:mechanism_inner_rim_position} We investigate our detected inner
	rims in the context of two possible mechanisms setting the rim position: dust sublimation
	and binary truncation.

	\subsubsection{Dust sublimation}
	\label{sect:inner_rim_comparison_to_sublimation} All our targets exhibit SEDs representative
	of full discs \citep{Kluska2022}, suggesting their energetics are consistent with
	an inner rim near the dust sublimation radius. To validate this scenario, we confront
	the derived inner rim radii ($\theta_{\mathrm{rim}}/ 2$; Table \ref{table:img_rec_params})
	with the expected dust sublimation radius \citep[e.g.][]{Lazareff2017}, which
	follows
	\begin{equation}
		\label{eq:sublimation_radius}R_{\mathrm{sub}}= \frac{1}{2}\left(\frac{C_{\mathrm{bw}}}{\epsilon}
		\right)^{1/2}\left(\frac{L_{*}}{4\pi \sigma_{\mathrm{SB}}T_{\mathrm{sub}}^{4}}
		\right)^{1/2},
	\end{equation}
	with $C_{\mathrm{bw}}$ the backwarming coefficient, $\epsilon$ the grain cooling
	efficiency, $L_{*}$ the primary's luminosity (Table \ref{table:target_stars_and_observations}),
	$\sigma_{\mathrm{SB}}$ the Stefan-Boltzmann constant and $T_{\mathrm{sub}}$ the
	sublimation temperature. Since $R_{\mathrm{sub}}$ and $L_{*}^{1/2}$ are
	proportional to the target distance, the angular size of $R_{\mathrm{sub}}$ is
	robust against any errors on the parallax. This includes those induced by the
	binary motion, as our targets were not marked as astrometric binaries in Gaia DR3.

	\citet{Kluska2019} found a typical inner rim temperature of
	$\sim\!1100\!-\!140 0 \, \mathrm{K}$. Meanwhile, \citet{Corporaal2023a} similarly
	found a rim temperature of $\sim\!1250\, \mathrm{K}$ for the prototypical full
	disc IRAS~08544-4431 through RT modelling. Given these values, we assume a conservative
	range of $T_{\mathrm{sub}}\!=\!1000\!-\!1500 \, \mathrm{K}$. The cooling
	efficiency reaches its highest value, $\epsilon = 1$, for large grains ($a_{\mathrm{grain}
	}\!\gtrsim \!\lambda_{\mathrm{NIR}}/ (2\pi)$; \citealt{Lazareff2017}). Given that
	grain growth in post-AGB discs is known to easily proceed up to $\mathrm{\mu m}$
	sizes and beyond \citep[e.g.][]{Gielen2008, Scicluna2020, Bujarrabal2023}, and
	the fact that large grains significantly help cool smaller grains even at moderate
	abundances \citep[][]{Kama2009}, we assume $\epsilon\!\approx\!1$. The
	backwarming coefficient can take on values of $C_{\mathrm{bw}}\!=\!1\!-\!4$. The
	lower limit corresponds to optically thin dust destruction, while the upper
	corresponds to an instantaneously optically thick, wall-like rim \citep[][]{Kama2009}.

\begin{figure*}
		\centering

		\centering
		\begin{subfigure}
			{0.246\textwidth}
			\includegraphics[width=\linewidth]{
				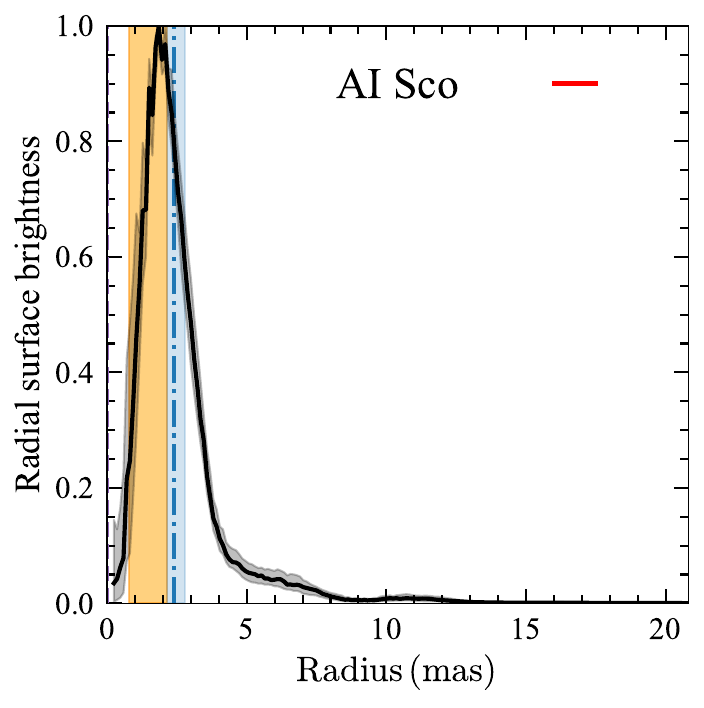
			}
		\end{subfigure}
		\begin{subfigure}
			{0.246\textwidth}
			\includegraphics[width=\linewidth]{
				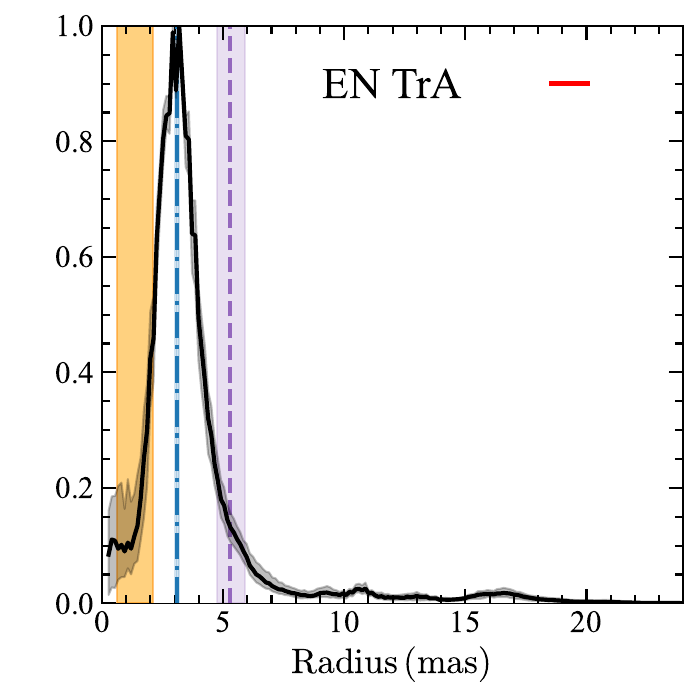
			}
		\end{subfigure}
		\begin{subfigure}
			{0.246\textwidth}
			\includegraphics[width=\linewidth]{
				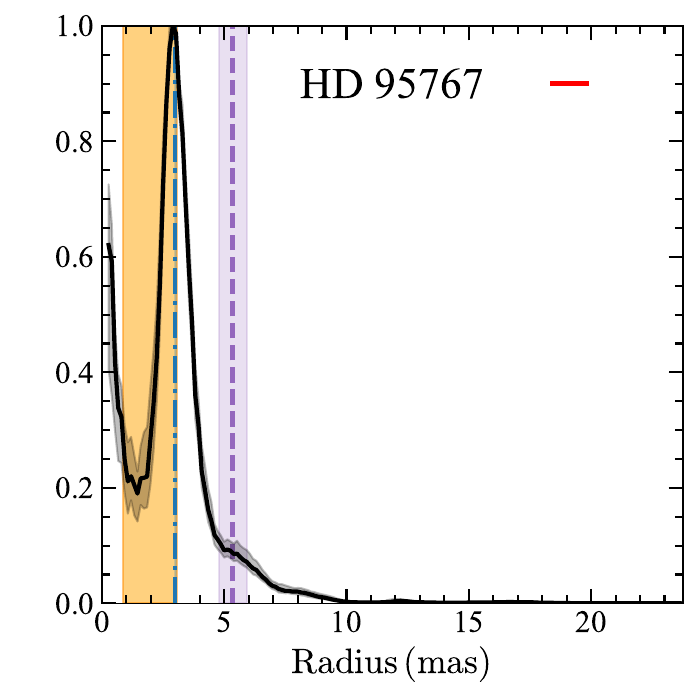
			}
		\end{subfigure}
		\begin{subfigure}
			{0.246\textwidth}
			\includegraphics[width=\linewidth]{
				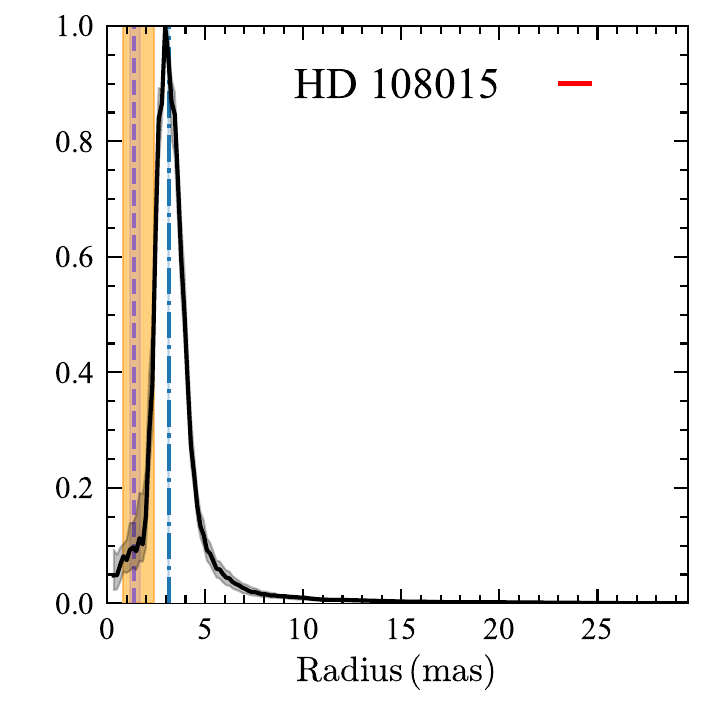
			}
		\end{subfigure}

		\vspace{-2.45ex}

		\begin{subfigure}
			{0.246\textwidth}
			\includegraphics[width=\linewidth]{
				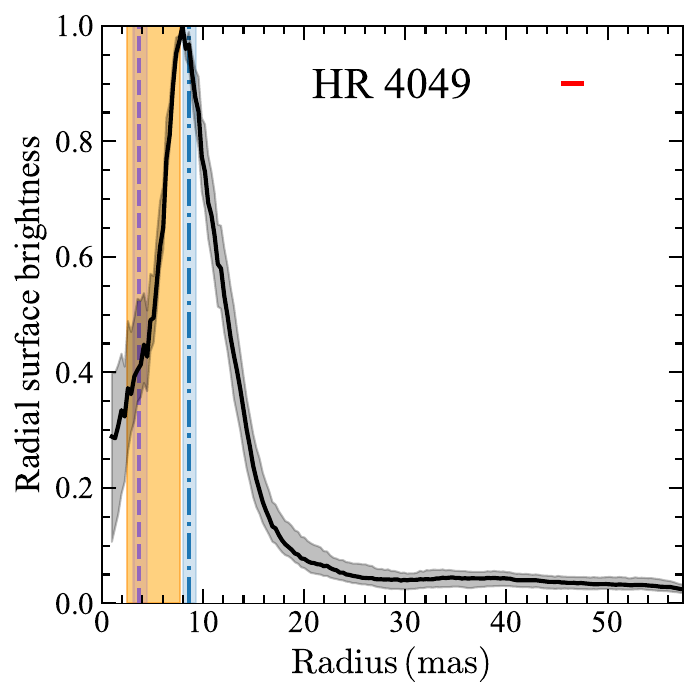
			}
		\end{subfigure}
		\begin{subfigure}
			{0.246\textwidth}
			\includegraphics[width=\linewidth]{
				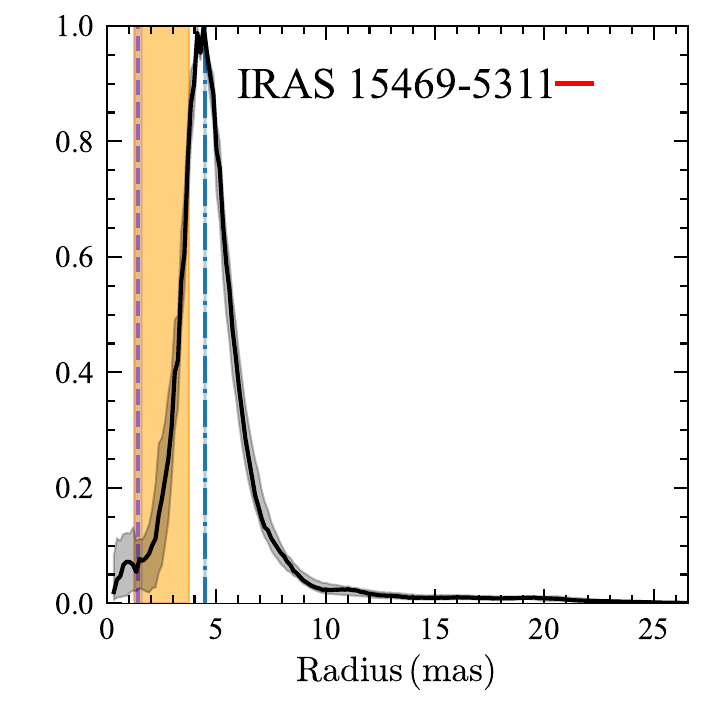
			}
		\end{subfigure}
		\begin{subfigure}
			{0.246\textwidth}
			\includegraphics[width=\linewidth]{
				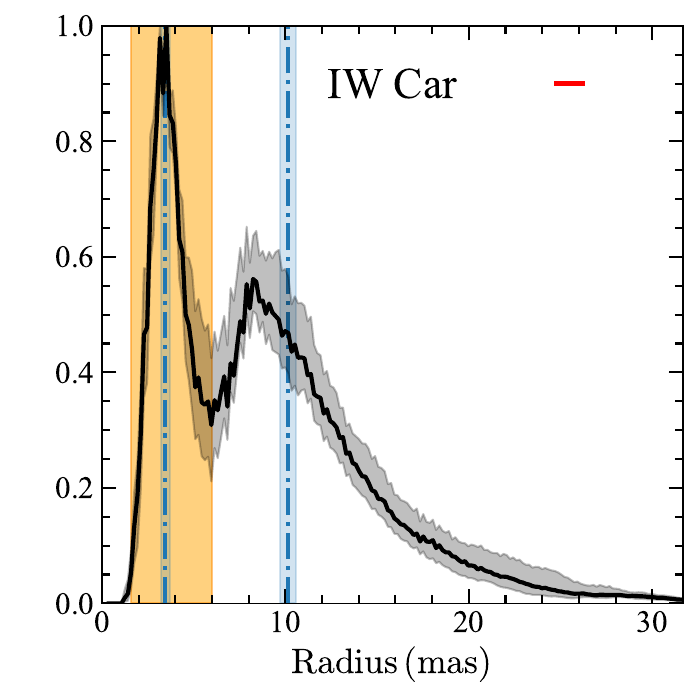
			}
		\end{subfigure}
		\begin{subfigure}
			{0.246\textwidth}
			\includegraphics[width=\linewidth]{
				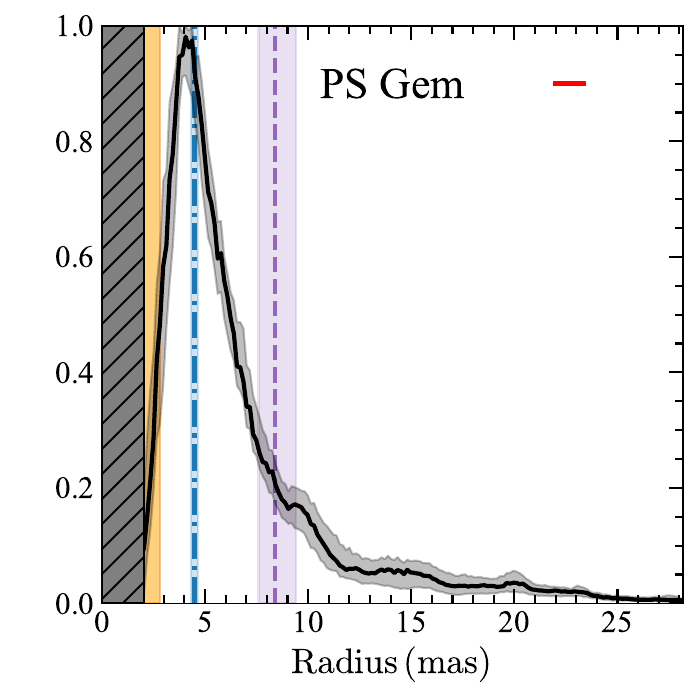
			}
		\end{subfigure}

		\caption{Peak-normalised ORGANIC radial surface brightness profiles. The profiles
		and their $1 \sigma$ contours are given in black, with red bars denoting the
		azimuthally averaged beam size. An orange strip represents the range of sublimation
		radii for $C_{\mathrm{bw}}= 1$ (Sect.\ \ref{sect:inner_rim_comparison_to_sublimation}).
		The image-derived inner rim radius ($\theta_{\mathrm{rim}}/2$; Table \ref{table:img_rec_params})
		is given as dot-dashed blue line, and for targets where RV monitoring is available
		(Table \ref{table:target_stars_and_observations}), a dashed purple line
		provides an upper limit to the binary truncation radius $R_{\mathrm{trunc}}$
		(Sect.\ \ref{sect:dynamical_truncation}). Shaded regions indicate $1 \sigma$
		errors for both. For IW~Car, we display inner rim radii for both the inner
		and outer flux arcs (see Sect.\ \ref{sect:IW_Car_detected_features}), with
		the brightness profile calculation based on the orientation of the inner
		arcs. For PS~Gem, a mask was placed in front of the primary to mitigate the
		effects of strong flux smearing (Sect.\ \ref{sect:PS_Gem_detected_features}),
		as denoted by a hatched strip.}
		\label{fig:radial_brightness_profiles}
	\end{figure*}

	We show the expected range of dust sublimation radii for
	$C_{\mathrm{bw}}\!=\!1$ (also accounting for the $1\sigma$ uncertainties on $L_{*}$)
	in Fig.\ \ref{fig:radial_brightness_profiles}. Disregarding IW~Car, where it is
	not clear whether the innermost circumstellar flux actually traces the inner rim
	(see Sect.\ \ref{sect:IW_Car_structure_origin_discussion}), the image-derived rim
	radii lie at or just beyond ($\sim\!1\text{--}2\, \mathrm{mas}$) the outer edge
	of the expected $C_{\mathrm{bw}}= 1$ sublimation radii. However, varying the backwarming
	up to $C_{\mathrm{bw}}\!=\!4$ can raise the $R_{\mathrm{sub}}$ estimates by factors
	of up to two, which can readily provide an excellent match to every target. We
	conclude that the inner rim positions are compatible with dust sublimation,
	though most rims are likely not fully optically thin. We also stress that sublimation
	rims are not well-defined at a single radius. Simulations of PPD sublimation
	fronts \citep[e.g.][]{Kama2009, Flock2016, Flock2017} instead often display broad,
	wedge-shaped geometries (extending from $C_{\mathrm{bw}}\!=\!1$ to $C_{\mathrm{bw}}
	\!\approx\! 4$). Unfortunately, the sharpness of the rim is hard to discern
	from imaging alone, and RT modelling of the SEDs and interferometry \citep[cf.][]{Hillen2014, Kluska2018}
	would be needed to constrain it.

	\subsubsection{Dynamical truncation by the binary}
	\label{sect:dynamical_truncation} Alternatively, the location of the inner rim
	could be set by binary truncation, where resonance torques and non-resonant interactions
	can carve a cavity in the disc \citep[e.g.][]{Artymowicz1994, Ragusa2020}. Detailed
	3D hydrodynamical simulations of circumbinary discs show that binaries can carve
	cavities with truncation radii of $R_{\mathrm{trunc}}\lesssim 5a$, with $a$
	the total binary semi-major axis \citep[though the value decreases with
	decreasing binary mass ratio and orbital eccentricity, and increasing rim scale
	height;][]{Hirsh2020}. While we lack full orbital solutions for our targets,
	we can still set a conservative upper limit on the truncation radius when RV
	monitoring is available. We do not attempt to use the detected secondary positions,
	since single astrometric points do not provide strong enough constraints on
	the orbital parameters, and dedicated astrometric modelling would be required for
	this purpose \citep[e.g.][]{Anugu2023}.

	First, the spectroscopic mass function of the secondary, $f(m)$, can be rewritten
	as
	\begin{equation}
		\label{eq:mass_function_rewritten}f(m) = \frac{M_{\mathrm{sec}}^{3}}{(M_{\mathrm{prim}}+
		M_{\mathrm{sec}})^{2}}\sin^{3}i_{\mathrm{bin}}= \frac{M_{\mathrm{prim}}}{a^{2}(a
		- a_{1})}(a_{1}\sin i_{\mathrm{bin}})^{3}.
	\end{equation}
	As we did for HD~108015 in Sect.\ \ref{sect:HD108015_detected_features}, we assume
	co-planarity between the disc rim and the binary (i.e.\ $i_{\mathrm{bin}}\!=\!i
	_{\mathrm{rim}}$; Table \ref{table:img_rec_params}). Following \citealt{Oomen2018},
	we assume the post-AGB primary to follow the mass distribution of single white
	dwarfs in the Galactic field. Specifically, we adopt the value of
	$M_{\mathrm{prim}}= 0.62\pm0.11 \,\mathrm{M_\odot}$ from \citealt{Tremblay2017}.
	If the values of $a_{1}\sin{i_{\mathrm{bin}}}$ and $f(m)$ are available from
	RV monitoring (Table \ref{table:target_stars_and_observations}), Eq.\ \ref{eq:mass_function_rewritten}
	can readily be solved for the total semi-major axis $a$. The upper limit for the
	binary truncation radius inferred from this, $R_{\mathrm{trunc}}\lesssim 5a$, is
	displayed for our targets with previous RV monitoring on Fig.\ \ref{fig:radial_brightness_profiles}.
	We find that the inner rims of HD~108015, HR~4049 and IRAS~15469-5311 are
	highly unlikely to be set through tidal truncation, and are mostly unaffected by
	torque from the binary. Without tighter constraints on the binary orbital
	properties and rim scale height, we cannot make such assumptions for the other
	targets, and the inner rim might be set either by binary truncation or by
	sublimation physics.

	\subsection{Origins of the detected potential substructures}
	\label{sect:origins_of_structures} We identified several robust brightness
	asymmetries from our images in Sect.\ \ref{sect:detected_features}. Most
	strikingly, we found a number of brightness enhancements along the inner rim,
	not symmetric with the rim minor axis and inconsistent with inclination
	effects. For IW~Car, the image reconstruction revealed an even more puzzling
	morphology, finding inner flux arcs seemingly misaligned with a large, outer
	flux arc to the west. We now address several physical mechanisms that could lie
	at the origin of these features, as well as possible ways to discern them. We stress
	that while we are sensitive to thermal emission from small grains well-coupled
	to the gas, the emission can be optically thick \citep[e.g.][]{Hillen2014, Kluska2019, Corporaal2023a}.
	As a result, relating the brightness contrasts of features seen in the images (Figs.\ \ref{fig:organic_imgs_first_four}
	\& \ref{fig:organic_imgs_last_four}) to gas density enhancements predicted by
	mechanisms mentioned below is outside the scope of this work.

	\subsubsection{Thermal response to the primary's orbit}
	\label{sect:struct_origin_primary_irradiation} Comparing with the geometric
	PIONIER snapshot modelling study of \citet{Kluska2019}, we found that the
	azimuthal brightness enhancement in AI~Sco, EN~TrA \& HR~4049 (Sects.\ \ref{sect:AI_Sco_detected_features},
	\ref{sect:EN_TrA_detected_features} \& \ref{sect:HD95767_detected_features}) might
	be long-lived and following the movement of the primary star. We propose that
	the irradiation of the primary can preferentially heat parts of the disc rim
	it is closest to, locally increasing the temperature, scale height and
	brightness. Indeed, a similar mechanism was previously invoked to explain the
	MIR variability of the circumbinary PPD CoKu~TAU/4, showing that the thermal response
	of the inner rim is almost instant relative to the binary motion for typical density
	and temperature values \citep{Nagel2010}. This was also cautiously considered by
	\citet[][]{Kluska2018} as an explanation for the rim brightness enhancement in
	the post-AGB disc IRAS~08544-4431.

	If this mechanism is indeed active for post-AGB circumbinary discs, these
	brightness enhancements should follow the primary's position along its orbit. While
	comparison of our images with the modelling results of \citet{Kluska2019} is
	strongly suggestive of this, the complexity of the retrieved morphologies necessitates
	multi-epoch imaging to fully confirm this scenario.

	\subsubsection{Dynamical interaction with the binary}
	\label{sect:struct_origin_binary_dynamical} \citet{Ragusa2020} performed
	comprehensive 3D hydrodynamical simulations for co-planar binary-truncated
	discs, showing that the inner rim becomes eccentric -- even for circular
	binaries -- at mass ratios $q\geq0.05$. This criterion is easily fulfilled by
	our targets \citep[][]{Oomen2018}. The simulations show two simultaneously
	occurring types of substructure. First, a long-lived crescent-like gas density
	enhancement at the eccentric rim, caused by a 'traffic jam', as the orbits of
	the gas are squeezed due to gradients in disc eccentricity \citep[][]{Ataiee2013}.
	Second, a co-moving overdensity for mass ratios $q\gtrsim 0.2$, possibly
	formed by gas streams launched across the cavity \citep[e.g.][]{Shi2012, Ragusa2017, Calcino2019}.
	Under our currently limited constraints on the central binary orbit and disc
	rim scale height, these substructure formation mechanisms could potentially be
	active in all our targets except HD~108015, HR~4049 and IRAS~15469-5311 (where
	the rim is unlikely to be affected by binary torque; Sect.\ \ref{sect:dynamical_truncation}).
	AI~Sco is a particularly enticing candidate for these mechanisms, since the
	binary stars closely approach the rim, and the apparent rim centre is heavily displaced
	in the direction of the projected major axis relative to the binary,
	potentially indicating an eccentric disc cavity (Fig.\ \ref{fig:organic_imgs_first_four}).

	These binary-induced substructures are readily discernible from both each other
	and a thermal response to the primary through their expected motion. Features
	caused by thermal response should orbit on a timescale of $P_{\mathrm{orb}}$ (Table
	\ref{table:target_stars_and_observations}), while the binary-induced traffic jam
	and co-moving overdensity are, respectively, precessing on a timescale of $\sim
	\!100-\!1000\,P_{\mathrm{orb}}$ and orbiting with the local Keplerian velocity
	\citep[][]{Ragusa2020}. This clear separation in expected motion provides a powerful
	way of discriminating between thermal and dynamical origins.

	\subsubsection{Hydrodynamical instability}
	\label{sect:struct_origin_instabilities}Another possible mechanism causing
	asymmetric structure in inner disc rims is hydrodynamical instability. Specifically,
	the Rossby wave instability \citep[RWI; e.g.][]{Lovelace1999, Meheut2013, Bae2023}.
	RWI occurs when there is a local maximum in the vortensity of the disc,
	typically induced by an initial bump in the gas pressure. If local viscosity is
	low, the disc can form a single, long-lived vortex, orbiting at approximately
	the local Keplerian velocity \citep[][]{Lyra2012, Meheut2012}. While there are
	currently no constraints on the vortensity or viscosity in our target discs --
	there being only a single constraint on viscosity via RT modelling of the full
	post-AGB disc IRAS 08544-4431 \citep[][]{Kluska2018,Corporaal2023a} -- RWI provides
	a physically motivated framework for producing long-lived azimuthal asymmetries
	at the rim moving at timescales different from the binary. The brightness
	enhancements we find for HD~108015 and HR~4049 are possibly moving separately
	from the primary (Sects.\ \ref{sect:HD108015_detected_features} \&
	\ref{sect:HR4049_detected_features}), and are likely unaffected by binary
	torque (Sect.\ \ref{sect:dynamical_truncation}). This makes them consistent
	with a possible vortex scenario.

	Several ways of inducing such initial pressure bumps are considered for PPDs. One
	such scenario is the carving of gaps and cavities by massive planets \citep[e.g.][]{Pinilla2012}.
	Nevertheless, despite their status as potential second-generation planet
	formation sites \citep[][]{Kluska2022}, no candidate planet detections in post-AGB
	discs have been made so far. A scenario that does not invoke the presence of planets
	are turbulent 'dead zones'. A dead zone pressure bump occurs in discs where
	accretion is mediated by the magnetorotational instability \citep[MRI;][]{Balbus1998},
	as gas transitions from ionised to non-ionised and MRI turbulence becomes
	quiescent \citep[e.g.][]{Flock2016, Flock2019}. Previous work on PPDs showed
	that thermal ionisation alone is sufficient to produce strong ionisation
	transitions around $\sim\!1000\,\mathrm{K}$ \citep[e.g.][]{Desch2015, Jankovic2021},
	positioning the resulting pressure bump and RWI vortex close to the silicate
	sublimation rim \citep[][]{Flock2017}.

	If RWI operates in the inner rim, the bright region should orbit at the local
	Keplerian velocity. It would be readily discernible with multi-epoch imaging from
	the other formation mechanisms considered here, except the binary-induced co-moving
	overdensities. Both types of feature namely move at similar angular speeds,
	and resolving their subtle kinematic differences (a lack of vorticity in the binary-induced
	overdensities) is beyond the capabilities of current facilities.

	\subsubsection{IW~Car: a morphological puzzle}
	\label{sect:IW_Car_structure_origin_discussion} The inner disc of IW~Car shows
	a highly complex morphology (Sect.\ \ref{sect:IW_Car_detected_features}). A
	large flux arc is seen to the west, with additional smaller inner flux arcs found
	closer to the binary (Fig.\ \ref{fig:organic_imgs_last_four}).

	The large western flux arc might trace the inner rim of the outer disc probed
	by SPHERE (Fig.\ \ref{fig:IW_Car_SPHERE_VS_PIONIER}). The inner arcs could
	then probe accretion streams from the western rim onto the binary, with the emission
	originating from either hot, optically thick gas \citep[e.g.][]{Kraus2008}, or
	specific refractory dust grains that can survive at higher temperatures inward
	of the main rim \citep[such as corundum or iron;][]{Kama2009}. Alternatively, if
	both the western arc and the inner arcs represent rim-like structures, the two
	are likely misaligned relative to each other (Sect.\ \ref{sect:IW_Car_detected_features}
	\& Table \ref{table:img_rec_params}). This could be accounted for by a warped
	innermost disc \citep{Facchini2013}, which was recently put forward to explain
	morphological differences between visual and NIR images of the outer disc of
	the post-AGB binary IRAS~08544-4431 \citep[][]{Andrych2024}. Such warps can be
	induced by initial inclination misalignments between the binary orbit and the
	disc \citep[e.g.][]{Young2023}, or by the presence of an embedded giant planet
	on an inclined orbit \citep[][]{Nealon2018}.

	In case the inner arcs trace the actual inner rim of the outer disc observed with
	SPHERE (Fig.\ \ref{fig:IW_Car_SPHERE_VS_PIONIER}), the large western arc possibly
	traces a loosely wound spiral-like feature. This can explain why the western
	flux arc seemingly attaches to the inner flux arcs north-west of the primary (Fig.\ \ref{fig:organic_imgs_last_four}).
	Spiral arms can be induced in the disc by the presence of an embedded giant
	planet, in which case they co-move with the orbit of their perturber \citep[e.g.][]{Dong2015, Ren2020}.
	They may also be induced in an outflow lofted above the disc, similar to the
	putative secondary condensation front previously found for the full post-AGB
	disc HD~101584 \citep{Kluska2020b}. Gravitational interaction with the binary
	is also a possibility, though the resulting spirals are typically more flocculent
	and highly wound \citep[e.g.][]{Price2018}.

	Additional imaging epochs are needed to disentangle these interpretations. Such
	observations could confirm if the inner arcs are transient accretion streams or
	represent a stable rim structure, and whether the western flux arc is static
	and indicative of a rim structure, or moving and indicative of a dynamical spiral.

	\section{Summary and conclusions}
	\label{sect:conclusions} This study aimed to reveal the morphological complexity
	of the inner regions in a varied sample of post-AGB circumbinary discs by spatially
	resolving them. To this end, we successfully introduced a consistent image reconstruction
	workflow, combining the SPARCO approach with the ORGANIC reconstruction and
	BADS optimisation algorithms. We applied this workflow to VLTI/PIONIER observations
	of eight diverse full disc targets, obtained under the INSPIRING ESO large
	programme, providing the first homogeneous, high fidelity imaging survey of
	the inner disc regions.

	Our images clearly resolve the inner disc regions for all targets. The dusty inner
	rim sizes are generally compatible with dust sublimation, under the condition
	that the rims are not always fully optically thin. The possibility that the rim
	is set by binary tidal truncation instead cannot be excluded except for three targets
	-- HD~108015, HR~4049 \& IRAS~15469-5311. Further constraints on the binary properties
	-- for example combining RV monitoring with relative interferometric
	astrometry -- ideally in addition with constraints on the inner rim height from
	RT modelling, would clarify whether discs could be binary-truncated on a
	system-by-system basis.

	The images reveal a high degree of morphological complexity, displaying asymmetric
	azimuthal brightness enhancements that are significantly shifted way from the minor
	projected rim axis. After extensive testing of the image robustness, we demonstrate
	that the overall structure and orientation of these brightness enhancements is
	reliable in four targets -- though finer details of the azimuthal profile, such
	as small-scale clumpiness, can be affected by artefacts and specifics of the
	imaging routine (Appendices \ref{sect:appendix_geom_models_and_synth_reconstructions}
	\& \ref{sect:appendix_squeeze_reconstructions_appendix}). These enhancements are
	inconsistent with brightness asymmetry induced by simple RT effects of an inclined,
	smooth circular rim surrounding a single star. Only one target's image -- IRAS~15469-5311
	-- can be fully accounted for by invoking just inclination effects. Comparison
	with the results of the PIONIER snapshot geometric modelling study by \citet{Kluska2019},
	providing a multi-year timebase with respect to our observations, tentatively suggests
	these brightness enhancements could be long-lived and moving. They likely
	either follow the position of the primary star or move at an angular speed
	independent of the binary. Nevertheless, given the poor $(u,v)$ coverage in
	the \citet{Kluska2019} study, follow-up imaging is needed to fully confirm
	these scenarios. One target -- IW~Car -- shows a large outer flux arc to the west,
	as well as smaller inner flux arcs close to the central binary.

	Drawing on the extensive literature available for PPDs, various substructure formation
	mechanisms able to explain these features are considered. A thermal response
	of the inner rim to the variable irradiation of the primary as it orbits
	provides a natural explanation for brightness enhancements following the
	primary. In cases where the inner rim is set by binary truncation, slowly
	precessing, crescent-like features as well as co-moving overdensities can be produced.
	Co-moving vortices induced by RWI also form a strong contender for features
	moving independent of the binary. These could occur at gaps and cavities carved
	by massive planets or at magnetic dead zones close to the sublimation radius.
	Given the different expected rates of angular movement for these mechanisms, time-series
	imaging would provide a strong means of both fully establishing their presence
	and discerning amongst them. Only RWI vortices and binary-induced
	overdensities would be hard to disentangle, since both substructures co-move at
	the local Keplerian velocity. For IW~Car the inner flux arcs could represent
	either accretion streams onto the central stars or a slightly warped innermost
	disc, with the large western arc denoting the disc rim. Alternatively, the inner
	arcs might probe the main disc rim, and the western arc a spiral-like feature
	located in the disc or in an outflow. The extent to which these features can affect
	the general disc structure and spatial distribution of gas and dust -- like
	the ubiquitous refractory depletion, potentially suspected to be induced by dust
	trapping substructures \citep[e.g.][]{Oomen2019, Kluska2022, Mohorian2024} -- remains
	to be explored.

	Future time-resolved and multi-wavelength imaging campaigns, including MIR
	observations with VLTI/MATISSE and (sub)-mm observations with ALMA, will allow
	us to probe more radially extended and vertically deeper disc regions, as well
	as produce the first inner rim images for transition disc targets. This will provide
	the temporal and spectral leverage needed to fully unveil the dynamics of these
	complex environments. Dedicated modelling of substructure formation scenarios
	would then be called for in order to fully uncover their implications with
	regards to refractory depletion, possible second-generation planet formation or
	changes to the orbital architecture of any surviving first-generation planets.

	\section*{Data availability}
	Observation logs for the selected VLTI/PIONIER observations and the final ORGANIC image reconstructions are publicly available
	on Zenodo via \url{https://doi.org/10.5281/zenodo.20155600}. The observations
	themselves are publicly available as Phase 3 reduced data products through the
	ESO Science Archive. More information about accessing the Phase 3 products can
	be found via \url{https://doi.eso.org/10.18727/archive/113}.
	\begin{acknowledgements}
		We thank the anonymous referee for their thorough comments and suggestions,
		which helped us improve the quality of the manuscript. TDP acknowledges support
		of the Research Foundation - Flanders (FWO) under grant 11P6I24N. DK, HVW,
		and KA acknowledge the support of the Australian Research Council through
		Discovery Project DP240101150. JA and VB acknowledge the financial support of
		the project CRISPNESS, grant PID2023-146056NB-C21, funded by by MICIU/AEI/10.13039/501100011033
		and by ERDF/EU. SK acknowledges support from an ERC Consolidator Grant (Grant
		Agreement ID 101003096) and STFC Small Award (ST/Y002695/1). This research has
		made use of the SIMBAD database, operated at CDS, Strasbourg, France. This work
		was performed on the OzSTAR national facility at Swinburne University of Technology.
		The OzSTAR program receives funding in part from the Astronomy National
		Collaborative Research Infrastructure Strategy (NCRIS) allocation provided
		by the Australian Government, and from the Victorian Higher Education State Investment
		Fund (VHESIF) provided by the Victorian Government. This research has made
		use of the Jean-Marie Mariotti Center OiDB service. This research has made use
		of the PIONIER data reduction package of the Jean-Marie Mariotti Center.
		This research has made use of the Jean-Marie Mariotti Center SearchCal
		service, which involves the JSDC and JMDC catalogues.
	\end{acknowledgements}

	\bibliographystyle{aa.bst}
	\bibliography{refs.bib}

	\begin{appendix}
		\nolinenumbers

		\section{Individual target descriptions}
		\label{sect:appendix_target_individual_descriptions}

		Below, we provide a more extensive discussion of our selected targets in the
		context of the existing literature, focusing on known properties and previous
		studies that provide insights on the structure and morphological complexity of
		the circumbinary disc.

		\subsection{AI~Sco}
		AI~Sco's primary is non- to weakly-depleted, as marked by its low zinc over titanium
		abundance of $\mathrm{[Zn/Ti]}\!\sim\!0.3 \, \mathrm{dex}$ \citep[][]{Giridhar2005}.
		It is an RV Tauri type pulsator, and clearly shows the 'RVb' phenomenon
		\citep[][]{Kiss2007}. This binary-associated effect describes a long-term
		modulation of photometric brightness superimposed on the shorter period
		variations caused by RV Tau-like pulsations. The modulation periods typically
		match with spectroscopically derived orbital periods, and the primary star
		is often faintest and reddest at inferior conjunction. This supports the usual
		interpretation that the effect is caused by variable extinction along the line-of-sight
		(LOS) to the primary as it grazes the inner rim of the dusty circumbinary disc
		\citep[e.g.][]{Kiss2007, Manick2017}. Detection of the RVb phenomenon is hence
		though to imply a high inclination for the circumbinary dust disc -- though
		\citep[][]{Kluska2019} found that the RVb phenomenon can occur for moderately
		inclined systems. While originally coined for RV Tauri stars, the term is broadly
		used for any post-AGB system showing a similar long-term brightness modulation,
		regardless of whether the object is an RV Tauri pulsator or not.

		\citet{Kluska2019} performed geometric modelling on a PIONIER snapshot survey
		of disc-bearing post-AGB binaries using models of varying complexity. The
		most complex models include a binary star, an up to second order azimuthally
		modulated Gaussian inner disc rim and an over-resolved background flux. This
		resulted in a best-fit model for AI~Sco with a strong secondary detection at
		a flux fraction of $f_{\mathrm{sec,0}}= 24. 7^{+6.5}_{-6.9}\, \mathrm{\%}$ (defined
		at reference wavelength $\lambda_{0}= 1.65 \, \mathrm{\mu m}$), a weak background
		flux at $f_{\mathrm{bg,0}}= 1.1 \pm 0.4 \, \mathrm{\%}$ and an inner rim with
		diameter $\theta_{\mathrm{rim}}= 4.6 \pm 0.2 \, \mathrm{mas}$, inclination
		$i_{\mathrm{rim}}={47.3^{+2.3}_{-2.6}}\mathrm{^\circ}$ and position angle
		$\mathrm{PA}_{\mathrm{rim}}= 179{^{+3}_{-3}}\mathrm{^\circ}$. An inclined circular
		inner rim appears as an ellipse due to its projection on the plane of the sky.
		For a symmetrically and passively irradiated smooth disc, the primary $H$-band
		brightness enhancement of the inner rim lies along the long side of the ellipse,
		symmetric with respect to the projected minor axis, due to self-absorption of
		the near side's emission \citep[e.g.][]{Hofmann2022}. The main brightness enhancement
		of AI~Sco's inner rim in the \citet{Kluska2019} model instead lies along the
		short side of the projected rim, seemingly in phase with the binary position
		of the primary (see their Fig.\ C.1 for model images of the disc rims).

		\subsection{EN~TrA}
		EN~TrA's primary is mildly depleted at $\mathrm{[Zn/Ti]}\!\sim\!0.6 \, \mathrm{dex}$
		\citep[][]{VanWinckel1997}. It is an RV Tau pulsator, but no firm detection
		of the RVb phenomenon has been established yet \citep[][]{VanWinckel2009}.
		The best-fit model by \citet{Kluska2019} shows a significant contribution from
		the secondary $f_{\mathrm{sec,0}}= 7.8^{+1.7}_{-1.2}\, \mathrm{\%}$, a weak
		over-resolved background
		$f_{\mathrm{bg,0}}= 2.5_{-1.0}^{+0.9}\, \mathrm{\%}$, and an inner rim with diameter
		$\theta_{\mathrm{rim}}= 7.7^{+1.7}_{-1.6}\, \mathrm{mas}$ and orientation of
		$i_{\mathrm{rim}}= 52.7^{+7.8}_{-9.1}\mathrm{^\circ}$ and $\mathrm{PA}_{\mathrm{rim}}
		= 13^{+9}_{-7}\mathrm{^\circ}$. A strong azimuthal brightness enhancement of
		the model disc appears aligned with the long side of the projected ellipse, which
		is also in phase with the position of the primary.

		\subsection{HD~95767}
		HD~95767 has a non-depleted primary at $\mathrm{[Zn/Ti]}\!\sim\!0.0 \, \mathrm{dex}$
		\citep[][]{VanWinckel1997}. \citet{Kiss2007} marked it as a low-amplitude
		semi-regular or multiply periodic variable and firmly detected the RVb
		phenomenon, implying a possibly high inclination for the circumbinary disc.
		\citet{Kluska2019} fitted a model with a strong secondary companion flux at $f
		_{\mathrm{sec,0}}= 12.3^{+2.3}_{-1.2}\, \mathrm{\%}$, a weak background at
		$f_{\mathrm{bg,0}}= 1.1^{+0.8}_{-0.7}\, \mathrm{\%}$, and, contrary to what
		is expected from the clear presence of the RVb phenomenon, a pole-on inner rim
		at a diameter of $\theta_{\mathrm{rim}}= 6.8^{+0.4}_{-0.5}\, \mathrm{mas}$.
		The main brightness enhancement along the rim seemed in phase with the
		primary's position.

		The polar orientation derived by \citet{Kluska2019} is likely an artefact from
		their extremely sparse $(u,v)$ coverage, covering only the small AT
		baselines. Our imaging results instead find a significantly non-zero inclination
		(Table \ref{table:img_rec_params}).

		\subsection{HD~108015}
		HD~108015's primary is non-depleted at $\mathrm{[Zn/Ti]\!\sim\!0.1 \,
		\mathrm{dex}}$ \citep[][]{VanWinckel1997}. It was marked as a low-amplitude semi-regular
		or multiply periodic variable without an RVb phenomenon by \citet{Kiss2007}.
		The best-fit model of \citet{Kluska2019} shows no detection of the secondary,
		a significant background flux at $f_{bg, 0}= 8.5 \pm 0.4 \, \mathrm{\%}$ and
		an inner rim with diameter $\theta_{\mathrm{rim}}= 6.7 \pm 0.1 \, \mathrm{mas}$
		and a low-inclination orientation of $i_{\mathrm{rim}}= 12.9^{+4.7}_{-5.9}\mathrm{^\circ}$
		and $\mathrm{PA}_{\mathrm{rim}}= 40^{+16}_{-19}\mathrm{^\circ}$. The main
		brightness enhancement along the rim is out of phase with the primary's
		position relative to the rim centre (with the primary's projected position likely
		being mostly a projection effect; see Sect.\ \ref{sect:HD108015_detected_features}),
		and is instead shifted counter-clockwise. Geometric modelling of snapshot
		VLTI/MATISSE data by \citet{Corporaal2023b} using a model consisting of the primary
		and a power-law surface brightness disc (and modelling only the $V2$ data) resulted
		in a much smaller inner rim disc diameter of $\theta_{\mathrm{rim}}= 3.4 \pm
		0.6 / 3. 6\pm0.2\, \mathrm{mas}$ in $L$/$N$ band respectively. The fitted
		orientation of the disc was
		$i_{\mathrm{rim}}= 35 \pm 11/32 \pm 2\mathrm{^\circ}$ and
		$\mathrm{PA}_{\mathrm{rim}}= 45 \pm 13 / 15 \pm 2\mathrm{^\circ}$, which seems
		inconsistent with the lower inclination found by \citet{Kluska2019}.

		These inconsistencies between geometric modelling studies are likely caused
		by their sparse $(u,v)$ coverages, causing strong parameter degeneracies between
		the rim size, orientation, and azimuthal and radial extent of the emission (this
		is why the rim orientation is best fixed during geometric modelling if
		constraints from other sources are available, as in e.g.\ \citealt{Setterholm2025}).
		This is compounded by the use of qualitatively different disc radial intensity
		profiles (Gaussian profiles in \citealt{Kluska2019} versus radial power laws
		in \citealt{Corporaal2023b}). Our imaging results (Table \ref{table:img_rec_params})
		are more consistent with the rim size found by \citet{Kluska2019}, but also
		more consistent with the inclination found by \citet{Corporaal2023a}. We
		note that this does not invalidate the findings of either study, as each was
		internally consistent in their geometric modelling approach, and their main
		results were based on general distributions of the fitted parameters (and,
		in the case of \citealt{Corporaal2023a}, the relative comparison between
		different SED categories).

		\subsection{HR~4049}
		HR~4049 is an extremely depleted system, hallmarked by its remarkably low
		metallicity $\mathrm{[Fe/H]}\!\sim\!-4.8 \, \mathrm{dex}$ and high zinc over
		iron ratio $\mathrm{[Zn/Fe]}\!\sim\!3.5 \, \mathrm{dex}$ \citep[][]{VanWinckel1995, Oomen2019}.
		Photometric studies identified the primary as non-pulsating, but the RVb phenomenon
		was firmly detected \citep[][]{Waelkens1991a, Kiss2007}. \citet{Kluska2019}
		fitted a model with a fairly weak secondary flux fraction at
		$f_{\mathrm{sec,0}}= 2.5^{+1.1}_{-0.8}\, \mathrm{\%}$ and strong background
		contribution of $f_{\mathrm{bg,0}}= 17.0 \pm 0.7 \, \mathrm{\%}$. The fitter
		inner rim presented a diameter of $\theta_{\mathrm{rim}}= 16.4 \pm 0.5 \, \mathrm{mas}$
		and orientation $i_{\mathrm{rim}}= 49.3^{+3.2}_{-3.3}\mathrm{^\circ}$ and $\mathrm{PA}
		_{\mathrm{rim}}= 63^{+7}_{-6}\mathrm{^\circ}$. While a secondary brightness
		enhancement was in phase with the primary position, the main brightness enhancement
		appeared on the opposite side along the major axis of the projected disc.

		The surface of the outer disc regions was imaged in the SPHERE/IRDIS $H$-band
		polarimetric imaging study of eight disc-bearing post-AGB binaries by \citet{Andrych2023}.
		The outer discs in this study appear as a bright central 'ring', marked by a
		low intensity cavity in the centre. This cavity is an artefact of the image reduction,
		caused by the subtraction of the central unresolved polarised flux. The disc
		orientation derived from this ring was deemed unreliable, as it provides a
		low inclination ($i\!\sim\!16.8^{\circ}$) incompatible with the robust
		detection of the RVb phenomenon and the results of \citet{Kluska2019} (as
		well as our imaging results; Table \ref{table:img_rec_params}). The circumbinary
		disc otherwise showed no potential substructures on the scales probed by IRDIS.

		\subsection{IRAS~15469-5311}
		IRAS~15469-5311's primary is strongly depleted with $\mathrm{[Zn/Ti]}\!\sim \!
		1.8 \, \mathrm{dex}$ \citep[][]{Maas2005}. While the primary pulsates, the
		pulsation pattern seems rather erratic, and not consistent with a classical RV
		Tau variable \citep[][]{Kiss2007}. No clear photometric detection of an RVb phenomenon
		has been published yet. The \citet{Kluska2019} PIONIER snapshot model fit consists
		of a binary star, a significant over-resolved background at
		$f_{\mathrm{bg,0}}= 12.8^{+0.5}_{-0.6}\, \mathrm{\%}$ and an inner rim with
		diameter $\theta_{\mathrm{rim}}= 10.4 \pm 0.3 \, \mathrm{mas}$ and
		orientation $i_{\mathrm{rim}}={53.5^{+1.8}_{-2.2}}\mathrm{^\circ}$ and
		$\mathrm{PA}_{\mathrm{rim}}= 64{^{+2}_{-2}}\mathrm{^\circ}$. The secondary detection
		was highly tentative, with a flux fraction of only
		$f_{\mathrm{sec,0}}= 0.1^{+0.1}_{-0.1}\, \mathrm{\%}$. The primary brightness
		enhancement lies along the long side of the projected rim, and is not in
		phase with the primary's position, consistent with the expected RT effects of
		a symmetrically illuminated, smooth inclined disc rim. The MATISSE snapshot modelling
		by \citet{Corporaal2023b} resulted in a disc with an orientation of
		$i_{\mathrm{rim}}= 47 \pm 5\mathrm{^\circ}$ and
		$\mathrm{PA}_{\mathrm{rim}}= 50 \pm 4\mathrm{^\circ}$. The inner rim diameter
		was found to be much smaller than the \citet{Kluska2019} value, at $\theta_{\mathrm{rim}}
		= 3.4 \pm 0.4 \, \mathrm{mas}$. As for HD~108015, this is most probably caused
		by sparse $(u,v)$ coverages and the resulting strong parameter degeneracies.

		The outer disc surface was polarimetrically imaged by \citet{Andrych2023}, revealing
		a bright central ring at an orientation of $i_{\mathrm{rim}}={19.6^{+18}_{-16}}
		\mathrm{^\circ}$ and $\mathrm{PA}_{\mathrm{rim}}={139^{+27}_{-30}}\mathrm{^\circ}$.
		This orientation differs from that of the disc inner rim found by \citet{Kluska2019}
		and \citet{Corporaal2023b}, again possibly due to data reduction artefacts affecting
		the central ring. These SPHERE observations also showed a secondary ring-like
		substructure offset from the central ring (see their Fig.\ 5). Interpreting
		these as part of one continuous disc, \citet{Andrych2023} derived a disc
		scale height of $\sim\!190 \, \mathrm{AU}$ at a separation of
		$\sim\!1100 \, \mathrm{AU}$.

		\subsection{IW~Car}
		IW~Car is strongly depleted at $\mathrm{[Zn/Ti]}\!\sim\!2.1 \, \mathrm{dex}$
		\citep[][]{Giridhar1994}. It is an RV Tau pulsator with a firmly detected
		RVb phenomenon \citep[][]{KissBodi2017}. \citet{Kluska2019} fitted an inner rim
		model with a binary $f_{\mathrm{sec,0}}= 3.5 \pm 0.1 \, \mathrm{\%}$ and a significant
		over-resolved component at $f_{\mathrm{bg,0}}= 10.5 \pm 0.3 \, \mathrm{\%}$.
		The fitted inner rim had a large diameter of
		$\theta_{\mathrm{rim}}= 23.0^{+0.7}_{-0.6}\, \mathrm{mas}$ and was
		remarkably broad (Gaussian full-width-half-maximum (FWHM) equal to $0.71 \pm
		0.02 \, \theta_{\mathrm{rim}}$). The rim was oriented at $i_{\mathrm{rim}}= 4
		4.6^{+1.7}_{-1.8}\mathrm{^\circ}$ and $\mathrm{PA}_{\mathrm{rim}}= 155 \pm 2\mathrm{^\circ}$.
		\citet{Corporaal2023b} fitted a smaller inner rim using VLTI/MATISSE at $\theta
		_{\mathrm{rim}}= 4.4 \pm 0.4/4. 8 \pm 0. 2 \, \mathrm{mas}$ in $L$/$N$ band respectively
		(likely probing the inner flux arcs we resolve in our imaging; Sect.\ \ref{sect:IW_Car_detected_features}),
		though the model generally fails to fit the large baseline data in the $L$
		band.

		IW~Car was included in the SPHERE/IRDIS study of \citet{Andrych2023}, where
		the system clearly presented a bright arc to the east of the usual central
		ring, separated by a dark shadow (see their Fig.\ 5). This was interpreted as
		scattered light from the two sides of the disc, separated by the optically
		thick near-side mid-plane, leading to a disc scale height estimate of $\sim\!
		190 \, \mathrm{AU}$. The central ring orientation was found to be, $i_{\mathrm{rim}}
		= 41.7^{+6.5}_{-7.7}\mathrm{^\circ}$ and $\mathrm{PA}_{\mathrm{rim}}= 161^{+12}
		_{-10}\mathrm{^\circ}$, though it could again be affected by reduction
		artefacts.

		\subsection{PS~Gem}
		PS~Gem, similar to HR~4049, shows an extremely low metallicity and depleted
		primary at $\mathrm{[Fe/H]\!\sim\!-4.5 \, \mathrm{dex}}$ and
		$\mathrm{[Zn/Ti]\!\sim\!3.0 \, \mathrm{dex}}$. While not a regular RV Tauri variable,
		it does show the RVb phenomenon \citep[][]{Waelkens1991b, Kipper2013, Rao2014}.
		\citet{Kluska2019} fitted a binary model to their PIONIER data, with no
		contribution of any disc signal. In addition to a background flux at $f_{\mathrm{bkg,0}}
		= 4.3^{+0.7}_{-0.8}\, \mathrm{\%}$ , the fitted secondary flux fraction was relatively
		weak at $f_{\mathrm{sec,0}}= 2.4^{+1.6}_{-0.3}\, \mathrm{\%}$, and its
		position relative to the primary was very poorly constrained at
		$\Delta \alpha_{\mathrm{sec}}= 3.3^{+32.5}_{-35.6}\, \mathrm{mas}$ and
		$\Delta \delta_{\mathrm{sec}}= 5.0^{+23.7}_{-21.6}\, \mathrm{mas}$. Indeed,
		we do not recover a significant secondary detection in this study. The lack of
		disc signal in \citet{Kluska2019} was likley caused due to PS~Gem's
		extremely poor $(u, v)$ coverage (a single snapshot at short baselines), as well
		as the generally low flux contribution of the disc (as found in this work; e.g.\ Table
		\ref{table:img_rec_params}).

		\section{Geometric models and synthetic data reconstructions}
		\label{sect:appendix_geom_models_and_synth_reconstructions}

\begin{figure*}[t]
			\centering
			\includegraphics[width=1.0\linewidth]{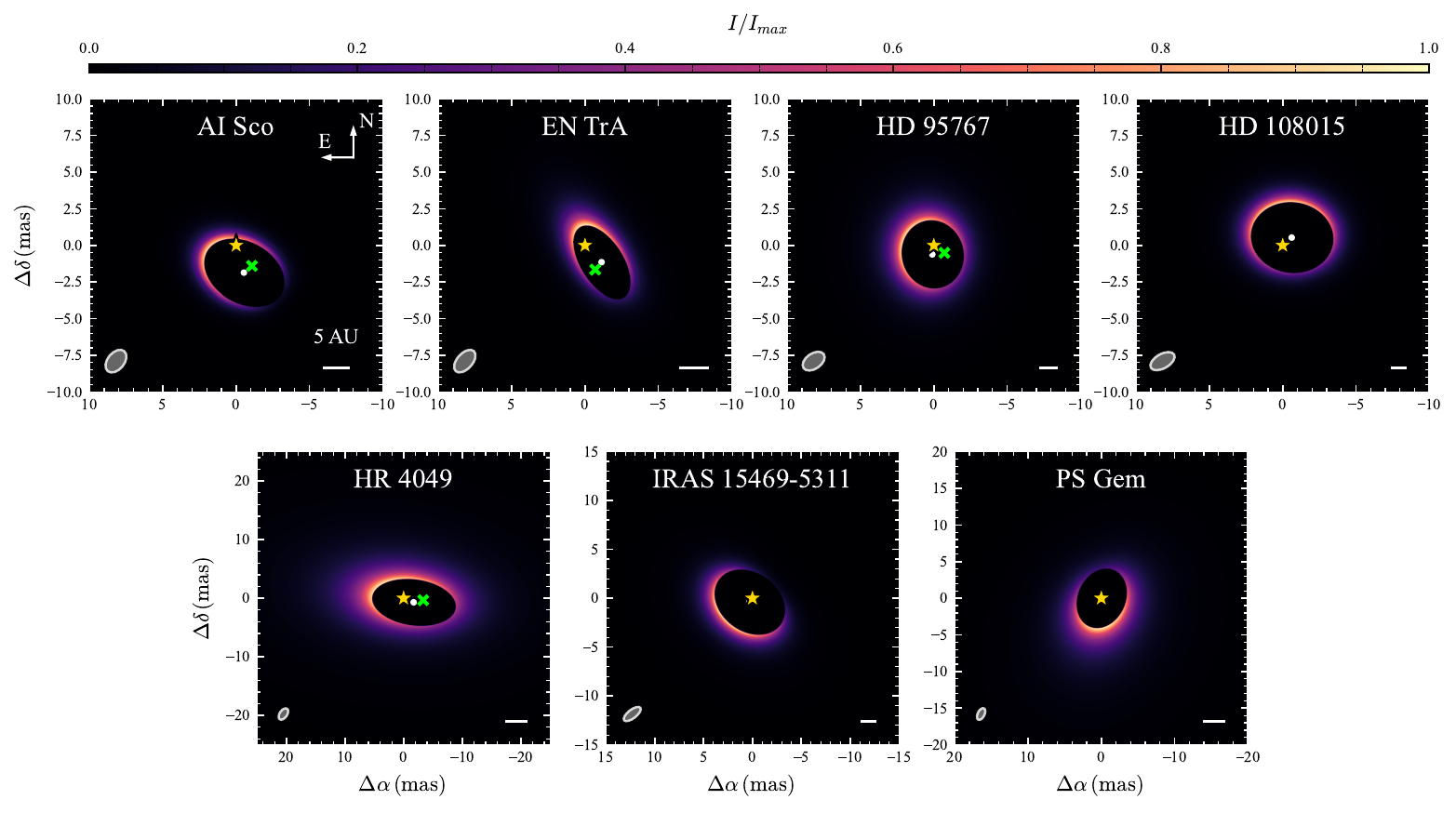}
			\caption{Analytical PMOIRED model disc images (Table \ref{table:geom_params}).
			All other symbols are analogous to those in Figs.\ \ref{fig:organic_imgs_first_four}
			\& \ref{fig:organic_imgs_last_four}.}
			\label{fig:pmoired_images}
		\end{figure*}

		The fitted parameters of the final PMOIRED geometric models are presented in
		table \ref{table:geom_params}. Corresponding images of the disc models are shown
		in Fig.\ \ref{fig:pmoired_images}. These images show the underlying analytical
		intensity distributions used to initially model the inner disc rim in Sect.\ \ref{sect:geometric_modelling}.
		We stress that the goal of these geometric models is not to fully reproduce the
		complex rim emission morphology or act as initial starting images for the image
		reconstructions, but to procure preliminary estimates for the SPARCO
		parameters as initial values for the BADS SPARCO optimisation routine (Sect.\ \ref{sect:sparco_optimisation}).
		As such, we only included up to one order of azimuthal modulation for the rim
		brightness profile, which can cause morphological model biases if the true
		underlying azimuthal brightness profiles have sharper features or are
		generally more complex (e.g.\ AI~Sco's model reaching only a poor goodness
		of fit at $\chi^{2}_{\nu}\sim 5$ and being biased in $\mathrm{PA}$ towards
		the primary; Sect.\ \ref{sect:AI_Sco_detected_features}). Nevertheless, analysing
		the behaviour of reconstructions of these models under analogous circumstances
		to our INSPIRING datasets can help identify potential biases in the imaging procedure
		and rim parameter derivation.

		\begin{sidewaystable}
			\caption{Best fit parameters of the final geometric models.}
			\label{table:geom_params}
			\setlength{\tabcolsep}{2mm}
			\centering
			\resizebox{\textwidth}{!}{
			\renewcommand{\arraystretch}{2.0}
			\begin{tabular}{lcccccccccccccccccc}
				\hline
				\hline
				Target                     & $\chi^{2}_{\nu}$ & $f_{\mathrm{prim},0}$   & $\mathrm{UD}_{\mathrm{prim}}$ & $f_{\mathrm{bg},0}$  & $f_{\mathrm{sec},0}$   & $\Delta \alpha_{\mathrm{sec}}$ & $\Delta \delta_{\mathrm{sec}}$ & $\theta_{\mathrm{rim}}$ & $i_{\mathrm{rim}}$   & $\mathrm{PA}_{\mathrm{rim}}$ & $\Delta \alpha_{\mathrm{rim}}$ & $\Delta \delta_{\mathrm{rim}}$ & $\rho_{\mathrm{rim}}$   & $d_{\mathrm{rim}}$     & $A_{\mathrm{mod},1}$      & $\psi_{\mathrm{mod},1}$ & $r_{\mathrm{hl}}$\tablefootmark{\,a} \\[-1.5ex]
				                           &                  & $(\%)$                  & $(\mathrm{mas})$              & $(\%)$               & $(\%)$                 & $\mathrm{(mas)}$               & $\mathrm{(mas)}$               & $(\mathrm{mas})$        & $(^{\circ})$         & $(^{\circ})$                 & $(\mathrm{mas})$               & $(\mathrm{mas})$               &                         &                        &                           & $(^{\circ})$            & $(\mathrm{mas})$                     \\
				\hline
				AI~Sco                     & $4.98$           & $68.1^{+0.4}_{-0.4}$    & --                            & --                   & $5.9^{+0.2}_{-0.2}$    & $-1.08^{+0.04}_{-0.03}$        & $-1.38^{+0.06}_{-0.06}$        & $5.83^{+0.09}_{-0.07}$  & $45.6^{+2.0}_{-1.6}$ & $60^{+7}_{-5}$               & $-0.53^{+0.07}_{-0.11}$        & $-1.86^{+0.12}_{-0.09}$        & $-6.7^{+0.7}_{-0.9}$    & $0.2^{+0.2}_{-0.2}$    & $0.87^{+0.05}_{-0.05}$    & $-9^{+4}_{-5}$          & $3.41^{+0.09}_{-0.08}$               \\
				EN~TrA                     & $2.40$           & $72.1^{+0.4}_{-0.4}$    & --                            & --                   & $5.0^{+0.2}_{-0.2}$    & $-0.70^{+0.02}_{-0.02}$        & $-1.65^{+0.04}_{-0.04}$        & $5.63^{+0.16}_{-0.14}$  & $58.9^{+1.1}_{-1.2}$ & $32.5^{+1.9}_{-1.7}$         & $-1.13^{+0.07}_{-0.07}$        & $-1.15^{+0.11}_{-0.12}$        & $-3.97^{+0.10}_{-0.10}$ & $4.37^{+0.11}_{-0.11}$ & $0.72^{+0.04}_{-0.04}$    & $-13^{+4}_{-4}$         & $4.02^{+0.09}_{-0.08}$               \\
				HD~95767                   & $1.97$           & $48.5^{+0.7}_{-0.7}$    & --                            & --                   & $21.0^{+0.7}_{-0.7}$   & $-0.719^{+0.013}_{-0.013}$     & $-0.510^{+0.009}_{-0.009}$     & $4.67^{+0.05}_{-0.04}$  & $25.6^{+1.7}_{-1.7}$ & $14^{+4}_{-4}$               & $0.11^{+0.04}_{-0.03}$         & $-0.61^{+0.04}_{-0.03}$        & $-3.83^{+0.03}_{0.03}$  & $6.39^{+0.08}_{-0.08}$ & $0.451^{+0.019}_{-0.020}$ & $43^{+4}_{-4}$          & $3.41^{+0.03}_{-0.03}$               \\
				HD~108015                  & $2.08$           & $61.8^{+0.3}_{-0.3}$    & --                            & $8.6^{+0.5}_{-0.4}$  & --                     & --                             & --                             & $5.63^{+0.03}_{-0.03}$  & $31.2^{+0.7}_{-0.8}$ & $82.5^{+1.2}_{-1.1}$         & $-0.612^{+0.018}_{-0.017}$     & $0.53^{+0.02}_{-0.02}$         & $-5.64^{+0.10}_{-0.12}$ & $7.09^{+0.09}_{-0.08}$ & $0.502^{+0.008}_{-0.007}$ & $-66^{+2}_{-2}$         & $3.419^{+0.014}_{-0.014}$            \\
				HR~4049\tablefootmark{\,b} & $1.71$           & $65.92^{+0.14}_{-0.14}$ & $0.74^{0.02}_{-0.02}$         & --                   & $0.42^{+0.05}_{-0.05}$ & $-1.70^{+0.07}_{-0.08}$        & $-0.69^{+0.10}_{-0.09}$        & $14.33^{+0.16}_{-0.18}$ & $56.8^{+1.0}_{-1.0}$ & $82.32^{+1.0}_{-0.8}$        & $-1.70^{+0.07}_{-0.08}$        & $-0.69^{+0.10}_{-0.09}$        & $-2.73^{+0.04}_{-0.05}$ & $7.90^{+0.07}_{-0.07}$ & $0.42^{+0.03}_{-0.03}$    & $-38^{+3}_{-3}$         & $15.44^{+0.46}_{-0.45}$              \\
				IRAS~15469-5311            & $2.08$           & $59.5^{+0.7}_{-0.6}$    & --                            & $11.1^{+1.2}_{-1.1}$ & --                     & --                             & --                             & $7.80^{+0.05}_{-0.06}$  & $40.2^{+0.8}_{-0.7}$ & $53.5^{+1.0}_{-1.1}$         & $0.31^{+0.03}_{-0.03}$         & $-0.39^{+0.02}_{-0.02}$        & $-4.84^{+0.13}_{-0.14}$ & $8.51^{+0.21}_{-0.19}$ & $0.86^{+0.02}_{-0.02}$    & $74.6^{+1.6}_{-1.5}$    & $5.00^{+0.05}_{-0.05}$               \\
				PS~Gem\tablefootmark{\,b}  & $0.94$           & $89.01^{+0.14}_{-0.14}$ & $0.44^{+0.02}_{-0.02}$        & --                   & --                     & --                             & --                             & $8.4^{+0.6}_{-0.6}$     & $39^{+6}_{-7}$       & $156^{+5}_{-6}$              & --                             & --                             & $-2.79^{+0.22}_{-0.17}$ & $4.3^{+0.2}_{-0.2}$    & $0.65^{+0.06}_{-0.05}$    & $35^{+9}_{-11}$         & $9.1^{+1.3}_{-0.8}$                  \\
				\hline
			\end{tabular}
			\renewcommand{\arraystretch}{1}
			} \tablefoot{\tablefoottext{a}{Different parametrisations can have different biases in the rim diameter. To facilitate comparison with other geometric and RT models (e.g. \citealt{Corporaal2023a}), we also calculated the disc half-light radius, $r_{\mathrm{hl}}$, using elliptical apertures following the inner rim orientation. It is a more robust tracer of the overall rim emission size \citep[cf.][]{Varga2021}. \tablefoottext{b}{For HR~4049 and possibly PS Gem, $r_{\mathrm{hl}}$ is strongly overestimated, likely because the rim model tries partially account for an over-resolved flux component (causing their high $\rho_{\mathrm{rim}}$ values). Their $r_{\mathrm{hl}}$ and $\rho_{\mathrm{rim}}$ values should be treated with caution.}}}
		\end{sidewaystable}

		\begin{table*}
			[t]
			\caption{SPARCO and image-derived rim parameters of the synthetic PMOIRED
			data reconstructions.}
			\label{table:geom_model_reconstruction_rim_params}
			\setlength{\tabcolsep}{1.1mm}
			\centering
			\resizebox{0.6 \textwidth}{!}{
			\renewcommand{\arraystretch}{1.7}
			\begin{tabular}{lcccccccccccccccccc}
				\hline
				\hline
				Target     & $f_{\mathrm{prim,0}}$ & $\mathrm{UD_{prim}}$ & $f_{\mathrm{sec,0}}$ & $d_{\mathrm{rim}}$ & $\theta_{\mathrm{rim}}$ & $i_{\mathrm{rim}}$   & $\mathrm{PA}_{\mathrm{rim}}$ & $\Delta \alpha_{\mathrm{rim}}$ & $\Delta \delta_{\mathrm{rim}}$ & $r_{\mathrm{hl}}$      \\[-0.8ex]
				           & $(\%)$                & $\mathrm{(mas)}$     & $\mathrm{(\%)}$      &                    & $(\mathrm{mas})$        & $(^{\circ})$         & $(^{\circ})$                 & $(\mathrm{mas})$               & $(\mathrm{mas})$               & $(\mathrm{mas})$       \\
				\hline
				AI~Sco     & $67.1$                & --                   & $5.49$               & $0.64$             & $5.8^{+0.5}_{-0.4}$     & $42^{+6}_{-8}$       & $56^{+6}_{-5}$               & $-0.35^{+0.11}_{-0.14}$        & $-1.72^{+0.17}_{-0.20}$        & $3.23^{+0.12}_{-0.12}$ \\
				EN~TrA     & $72.1$                & --                   & $4.99$               & $0.30$             & $6.50^{+0.20}_{-0.16}$  & $52^{+2}_{-2}$       & $30^{+4}_{-5}$               & $-1.04^{+0.04}_{-0.04}$        & $-0.94^{+0.11}_{-0.13}$        & $3.86^{+0.13}_{-0.13}$ \\
				HD~95767   & $47.4$                & --                   & $22.08$              & $1.91$             & $5.66^{+0.06}_{-0.05}$  & $20^{+3}_{-3}$       & $21^{+3}_{-6}$               & $0.02^{+0.03}_{-0.03}$         & $-0.61^{+0.03}_{-0.03}$        & $3.43^{+0.13}_{-0.13}$ \\
				HD~108015  & $56.7$                & --                   & --                   & $1.29$             & $6.31^{+0.06}_{-0.06}$  & $26.3^{+1.6}_{-0.9}$ & $77.3^{+3.2}_{-1.7}$         & $-0.57^{+0.03}_{-0.03}$        & $0.47^{+0.04}_{-0.04}$         & $3.62^{+0.16}_{-0.16}$ \\
				HR~4049    & $65.5$                & $0.74$               & $0.30$               & $2.40$             & $16.6^{+0.7}_{-0.7}$    & $48^{+3}_{-3}$       & $85^{+5}_{-5}$               & $-1.1^{+0.3}_{-0.3}$           & $-0.78^{+0.17}_{-0.17}$        & $16^{+2}_{-2}$         \\
				IRAS       & $52.7$                & --                   & --                   & $2.04$             & $8.76^{+0.09}_{-0.08}$  & $34.8^{+1.1}_{-0.6}$ & $52^{+5}_{-4}$               & $0.38^{+0.06}_{-0.06}$         & $-0.30^{+0.05}_{-0.05}$        & $5.30^{+0.29}_{-0.14}$ \\[-1.0ex]
				15469-5311 &                       &                      &                      &                    &                         &                      &                              &                                &                                &                        \\
				PS~Gem     & $89.1$                & $0.44$               & --                   & $0.67$             & $10.5^{+0.9}_{-0.8}$    & $36^{+8}_{-8}$       & $151^{+11}_{-8}$             & $-0.1^{+0.3}_{-0.3}$           & $0.0^{+0.5}_{-0.4}$            & $9.8^{+0.9}_{-1.1}$    \\
				\hline
			\end{tabular}
			\renewcommand{\arraystretch}{1}
			}
		\end{table*}

		\subsection{Bias assessment through synthetic reconstructions}
		\label{sect:geom_model_synthetic_reconstructions}

		To assess potential biases in our imaging workflow (Sect.\ \ref{sect:img_rec})
		that might affect the interpretation of the final images, we applied our ORGANIC
		workflow on synthetic $V2$ and $\phi3$ datasets derived from the PMOIRED geometric
		models. This includes BADS optimisation of the free SPARCO parameters ($f_{\mathrm{prim,0}}$,
		$\mathrm{ UD_{prim}}$, $f_{\mathrm{sec,0}}$ \& $d_{\mathrm{{rim}}}$;
		starting from their values in Table \ref{table:geom_params}). The synthetic
		datasets use the same $(u,v)$ coverage and noise levels as the observations
		(Figs.\ \ref{fig:uv_coverage}, \ref{fig:observables_v2} \& \ref{fig:observables_t3}).
		The imaging parameters ($\mathrm{SF}$, $\mathrm{PS}$ \& $\mathrm{FOV}$) were
		set to those used in Table \ref{table:img_rec_params}.

		We focus on assessing any potential biases in the image-derived rim
		parameters, as well as the position of morphological features such as azimuthal
		brightness enhancements. The rim parameters themselves, including $\theta_{\mathrm{rim}}$,
		$i_{\mathrm{rim}}$ and $\mathrm{PA}_{\mathrm{rim}}$, are chosen from Table \ref{table:geom_params}
		for the geometric models themselves, and are derived from the
		reconstructions following the rim fitting routine described in Sect.\ \ref{sect:rim_fitting_and_brightness_profile_calculation}.
		Radial brightness profiles were then calculated following Sect.\ \ref{sect:rim_fitting_and_brightness_profile_calculation}.
		In addition, we also calculated the azimuthal brightness profiles by integrating
		the flux along radial rays from the rim centre, separated by
		$0.1^{\mathrm{\circ}}$ in on-sky position angle $\mathrm{PA}$, within elliptical
		apertures following the rim orientation. We show the images reconstructed
		from the synthetic model datasets versus their ground truth counterparts in
		Fig.\ \ref{fig:geom_model_reconstructions}. The corresponding brightness
		profiles are given in Figs.\ \ref{fig:geom_model_reconstructions_rad_profiles}
		\& \ref{fig:geom_model_reconstructions_az_profiles}. The image-derived rim
		parameters are given in Table
		\ref{table:geom_model_reconstruction_rim_params}.

		We find the following potential biases for the rim parameters (we discuss those
		that are of $\geq2\sigma$ significance) and rim morphology of note for each
		of our targets:

		\textit{AI~Sco:} comparing the reconstruction with the model ground truth in
		Fig.\ \ref{fig:geom_model_reconstructions}, we note that the reconstructed disc
		flux close to the position of the primary star is biased in position by about
		$\sim\!1\,\mathrm{mas}$ southwards from where it is supposed to be, and
		seems systematically fainter. This also causes the reconstructed radial brightness
		profile to be slightly shifted and extended inwards, and the peak of the
		azimuthal brightness profile to be slightly shifted (by $\sim\!20^{\circ}$ in
		$\mathrm{PA}$) from its fiducial position (Fig.\ \ref{fig:geom_model_reconstructions_az_profiles}).
		This artefact is likely caused by the close approach of the primary to the
		rim, and the sparsity of north- and north-west oriented long-baseline
		observations (Fig.\ \ref{fig:uv_coverage}). This causes the reconstruction
		routine to be unable to fully distinguish between flux from the primary's
		position and flux from the nearest part of the rim. This artefact can also be
		seen in both our final ORGANIC reconstructions and the additional SQUEEZE reconstructions
		(Figs.\ \ref{fig:organic_imgs_first_four} \& \ref{fig:squeeze_imgs}). Nevertheless,
		this does not affect our main imaging results significantly: the brightness enhancement
		is heavily shifted towards the rim's projected major axis (Sect.\ \ref{sect:AI_Sco_detected_features}).
		All rim parameters are properly retrieved without clear biases (i.e.\ consistency
		within $2 \sigma$ between Tables \ref{table:geom_model_reconstruction_rim_params}
		\& \ref{table:geom_params}).

		\textit{EN~TrA:} comparing Tables \ref{table:geom_model_reconstruction_rim_params}
		\!\&\! \ref{table:geom_params}, we note a bias of
		$0.87^{+0.20}_{-0.16}\,\mathrm{mas}$ in $\theta_{\mathrm{rim}}$ and of
		$-7^{+2}_{-2}\mathrm{^\circ}$ in $i_{\mathrm{rim}}$. These are caused by the
		limited resolution of the interferometer, which smears out the infinitely
		sharp model rim, where the model intensity suddenly jumps from zero before decaying
		with a power law. This pulls the intensity peak slightly backward and blurs
		the sharp ellipticity of the model rim. This induces a systematic positive bias
		for $\theta_{\mathrm{rim}}$, and a negative bias for $i_{\mathrm{rim}}$. These
		biases are a common occurrence for our targets, as seen from the discussion
		of the other targets below. There is also a slight potential bias in $\mathrm{\Delta\alpha_{rim}}$
		of $0.09^{+0.04}_{-0.04}\, \mathrm{mas}$. Nevertheless, these findings do not
		significantly affect our main results: the brightness enhancement is
		positioned along the rim's projected major axis (Sect.\ \ref{sect:EN_TrA_detected_features}).
		Aside from some mild flux leakage into the central cavity, the brightness profiles
		are generally well-recovered (Figs.\ \ref{fig:geom_model_reconstructions_rad_profiles}
		\& \ref{fig:geom_model_reconstructions_az_profiles}).

		\begin{figure}[t]
			\centering
			\includegraphics[width=0.99\linewidth]{
				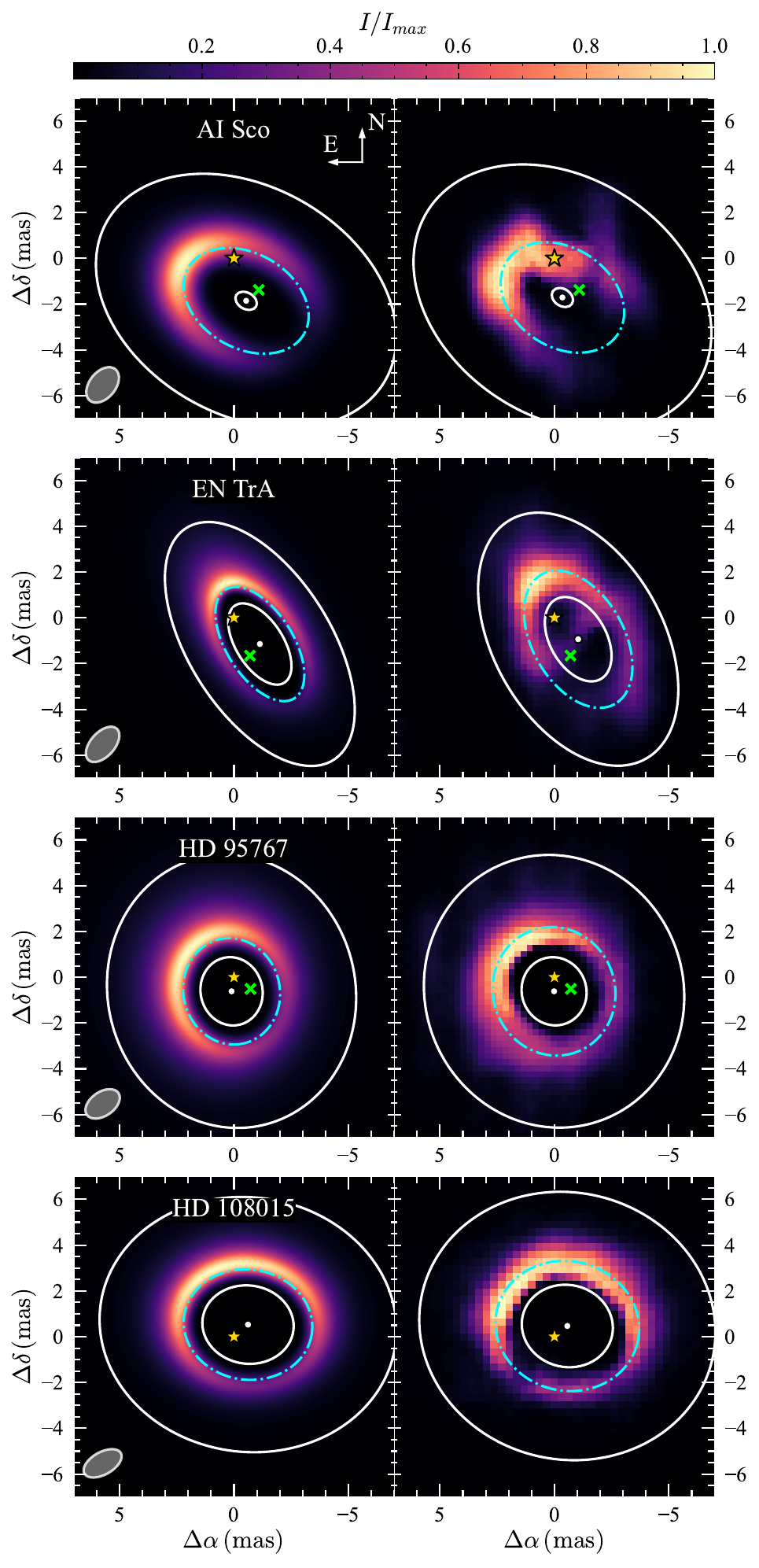
			}

			\caption{Images of Fig.\ \ref{fig:pmoired_images}, Gaussian-blurred to
			match the imaging resolution (left), and their synthetic reconstructions (right).
			White solid lines indicate the elliptical aperture used to calculate the
			azimuthal brightness. Their orientation is taken to be the that of the rim
			orientation from Table \ref{table:geom_params} for the PMOIRED images and from
			Table \ref{table:geom_model_reconstruction_rim_params} (i.e.\ image-derived
			following Sect.\ \ref{sect:rim_fitting_and_brightness_profile_calculation})
			for their reconstructions. These rims are plot in dot-dashed cyan. Other symbols
			analogous to Figs.\ \ref{fig:organic_imgs_first_four} \&
			\ref{fig:organic_imgs_last_four}.}
			\label{fig:geom_model_reconstructions}.
			\vspace{\fill}
		\end{figure}

		\begin{figure}[t]
			\ContinuedFloat
			\centering
			\includegraphics[width=0.99\linewidth]{
				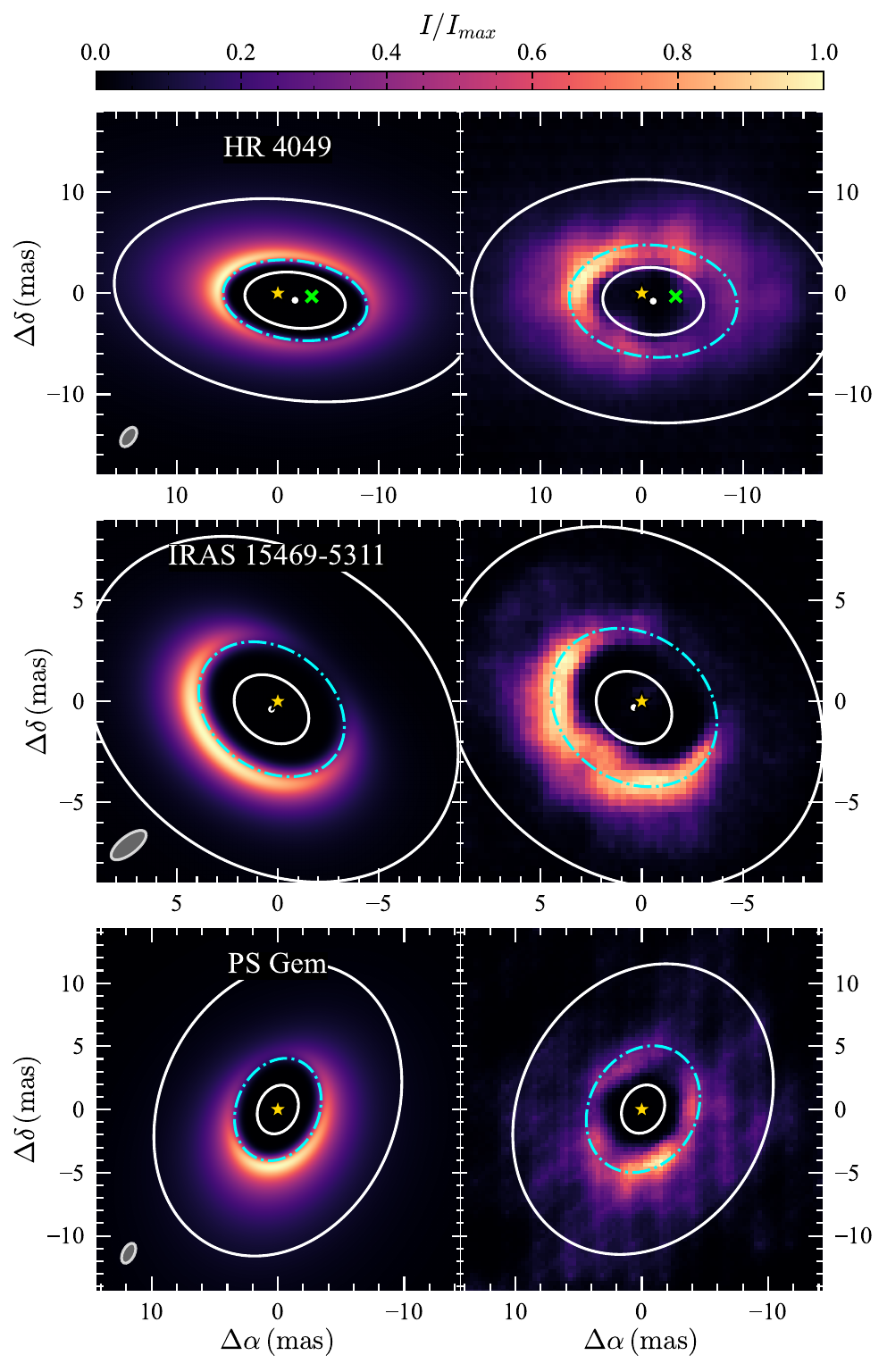
			}

			\caption{continued.}
			\vspace{\fill}
		\end{figure}

		\begin{figure}[t]
			\centering
			\includegraphics[width=1.0\linewidth]{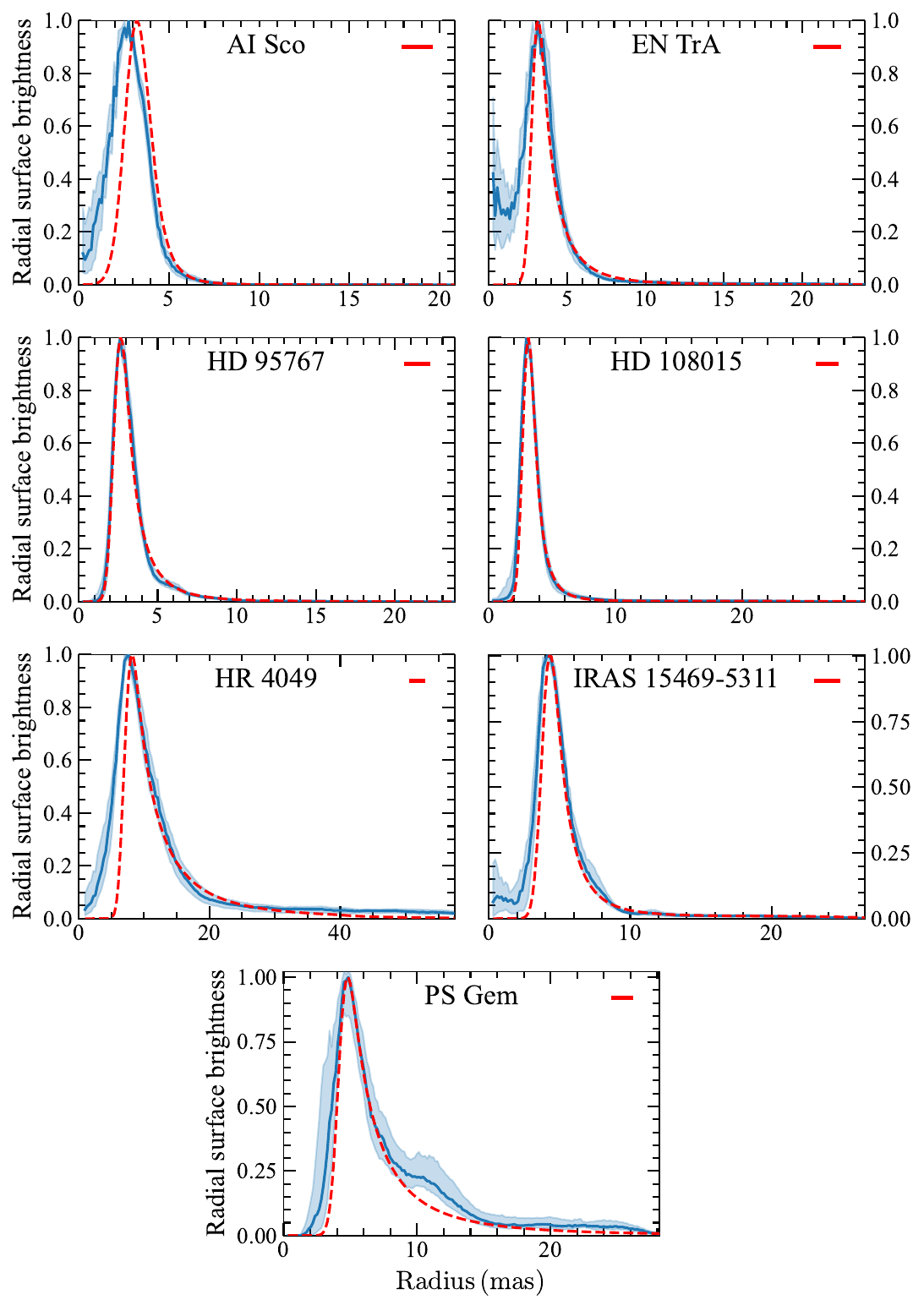}
			\caption{Peak-normalised radial brightness profiles of the convolved model
			images (red dashed) and synthetic reconstructions (blue with $1\sigma$ contours)
			in Fig.\ \ref{fig:geom_model_reconstructions}. Red bars indicate the size
			of the azimuthally averaged interferometric beam.}
			\label{fig:geom_model_reconstructions_rad_profiles}
		\end{figure}

		\begin{figure}[t]
			\centering
			\includegraphics[width=1.0\linewidth]{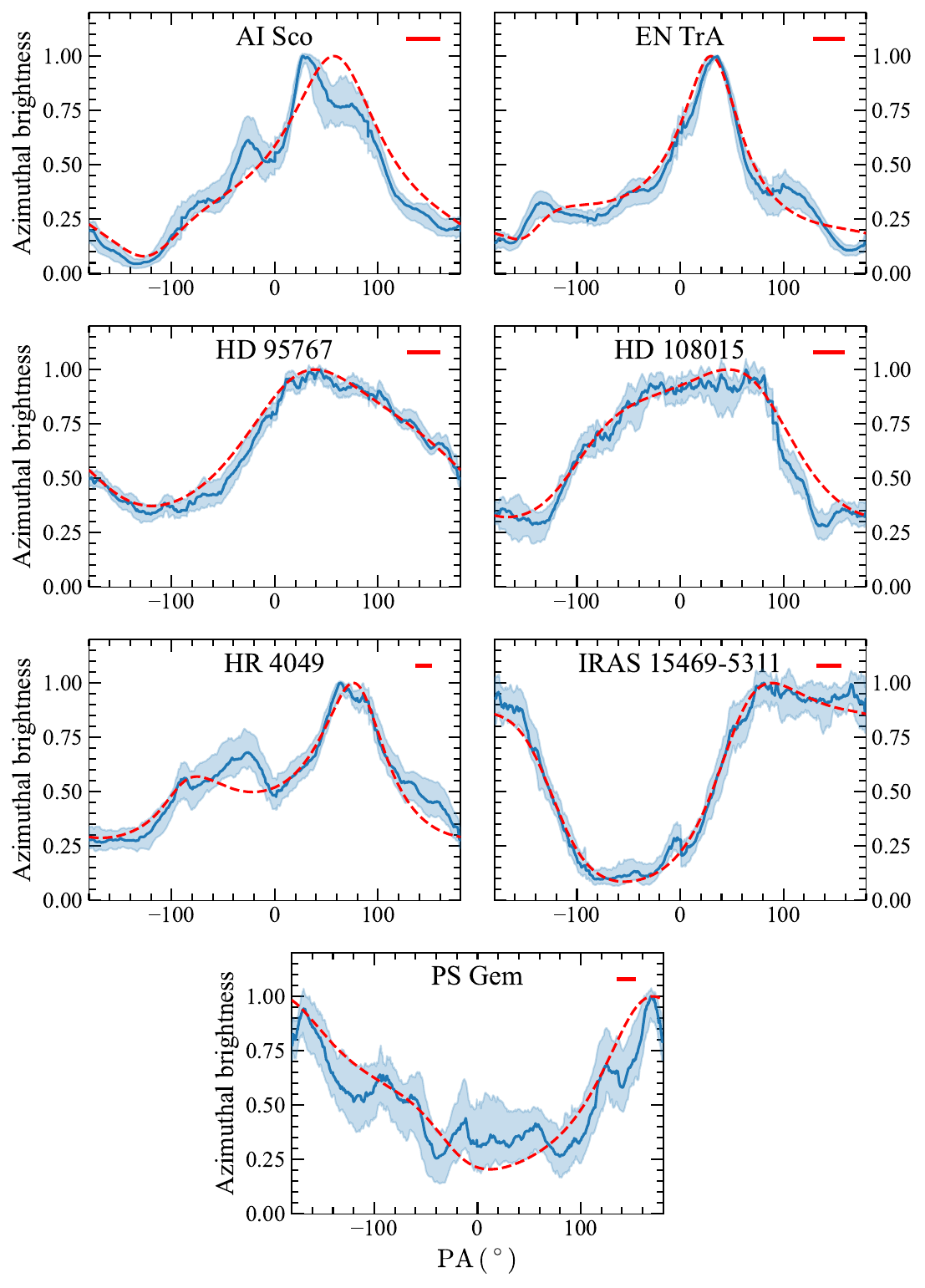}
			\caption{Analogous to Fig.\ \ref{fig:geom_model_reconstructions_rad_profiles}
			but for the peak-normalised azimuthal brightness profiles. The red bars
			now indicate the angle of the azimuthally averaged interferometric beam
			subtended at the reconstruction's rim.}
			\label{fig:geom_model_reconstructions_az_profiles}
		\end{figure}

		\textit{HD~95767:} we note a bias of $0.99^{+0.06}_{-0.05}\,\mathrm{mas}$ in
		$\theta_{\mathrm{rim}}$ and of $-6^{+3}_{-3}\mathrm{^\circ}$ in $i_{\mathrm{rim}}$
		(Tables \ref{table:geom_model_reconstruction_rim_params} \&
		\ref{table:img_rec_params}). There is also a slight potential bias in
		$\mathrm{\Delta\alpha_{rim}}$ of $-0.09^{+0.03}_{-0.03}\, \mathrm{mas}$. These
		do not affect the main results discussed in Sect.\ \ref{sect:HD95767_detected_features}:
		the final image is highly sensitive to $(u,v)$ coverage specifics and is hence
		not included in the discussion on potential substructures. The radial and
		azimuthal profiles are well-recovered (Figs.\ \ref{fig:geom_model_reconstructions_rad_profiles}
		\& \ref{fig:geom_model_reconstructions_az_profiles}).

		\textit{HD~108015:} we note a bias of $0.68^{+0.06}_{-0.06}\,\mathrm{mas}$
		in $\theta_{\mathrm{rim}}$ and of $-4.9^{+1.6}_{-0.9}\mathrm{^\circ}$ in
		$i_{\mathrm{rim}}$ (Tables \ref{table:geom_model_reconstruction_rim_params} \&
		\ref{table:img_rec_params}). There is also a potential bias in $\mathrm{PA_{rim}}$
		of $-5.2^{+3.2}_{-1.6}\mathrm{^\circ}$. These do not affect the main results
		presented in Sect.\ \ref{sect:EN_TrA_detected_features}: the brightness
		enhancement is significantly shifted towards the rim's projected major axis.
		The radial and azimuthal brightness profiles are well-recovered (Figs.\ \ref{fig:geom_model_reconstructions_rad_profiles}
		\& \ref{fig:geom_model_reconstructions_az_profiles}).

		\textit{HR~4049:} we find potential biases of $2.3^{+0.7}_{-0.7}\,\mathrm{mas}$
		in $\theta_{\mathrm{rim}}$ and $-9^{+3}_{-3}\mathrm{^\circ}$ in $i_{\mathrm{rim}}$
		(Tables \ref{table:geom_model_reconstruction_rim_params} \&
		\ref{table:img_rec_params}). These do not significantly affect the main
		results presented in Sect.\ \ref{sect:HR4049_detected_features}: the
		brightness enhancement is significantly shifted towards the rim's projected
		major axis. Aside from a minor amount of flux leakage in the central cavity drawing
		the radial brightness profile slightly inward, the radial and azimuthal
		brightness profiles are recovered (Figs.\ \ref{fig:geom_model_reconstructions_rad_profiles}
		\& \ref{fig:geom_model_reconstructions_az_profiles}).

		\textit{IRAS~15469-5311:} we find potential biases of
		$0.96^{+0.09}_{-0.08}\,\mathrm{mas}$ in $\theta_{\mathrm{rim}}$ and
		$-5.4^{+1.1}_{-0.6}$ in $i_{\mathrm{rim}}$ (Tables \ref{table:geom_model_reconstruction_rim_params}
		\& \ref{table:img_rec_params}). These do not significantly affect the main results
		presented in Sect.\ \ref{sect:IRAS154569-5311_detected_features}: the brightness
		profile is mostly symmetric around rim's projected minor axis, with no other
		robust brightness enhancements detected. The radial and azimuthal brightness
		profiles are recovered (Figs.\ \ref{fig:geom_model_reconstructions_rad_profiles}
		\& \ref{fig:geom_model_reconstructions_az_profiles}).

		\textit{PS~Gem:} we find a potential bias of $2.1^{+0.9}_{-0.8}\,\mathrm{mas}$
		in $\theta_{\mathrm{rim}}$ (Tables
		\ref{table:geom_model_reconstruction_rim_params} \&
		\ref{table:img_rec_params}). The reconstruction shows some pixelation and
		dirty beam artefacts, especially in faint flux regions (Fig.\ \ref{fig:geom_model_reconstructions}),
		causing a bump in the tail of the reconstructed radial brightness profile (Fig.\ \ref{fig:geom_model_reconstructions_rad_profiles}).
		The trend of the radial and azimuthal profiles are otherwise both well-recovered
		(Fig.\ \ref{fig:geom_model_reconstructions}). This does not significantly affect
		the main results presented in Sect.\ \ref{sect:PS_Gem_detected_features}: PS
		Gem's image in Fig.\ \ref{fig:organic_imgs_last_four} is most likely marred
		by the strong flux leakage from the primary, and it is thus excluded from the
		discussion on potential substructures.

		In general, the only bias relative to the geometrical models that seems
		common to all systems is a positive bias of
		$\sim \!1 \!- \!2 \, \mathrm{mas}$ is the rim diameter and a negative bias
		of several degrees in inclination (though always $\lesssim 10^{\circ}$ in magnitude).
		This is a natural result of the infinitely sharp rim assumed in the geometric
		models, which when viewed at the interferometer's angular resolution results
		in a slightly radially outwards shifted peak flux. The rim fitting routine targets
		this peak flux to derive $\theta_{\mathrm{rim}}$. It also slightly blurs the
		apparent ellipticity of the rim, lowering the apparent inclination. It
		should be noted that the biases are only of concern if the true inner rims
		are indeed sharp and wall-like. At least for inner rims set by sublimation physics,
		there is reason to suspect that this is not the case (see the discussion at the
		end of Sect.\ \ref{sect:inner_rim_comparison_to_sublimation}).

		\section{The effect of regularisation weight on ORGANIC reconstructions}
		\label{sect:appendix_regularization_weight}

		\begin{figure*}[t]
			\centering
			\begin{subfigure}
				{\textwidth}
				\includegraphics[width=\linewidth]{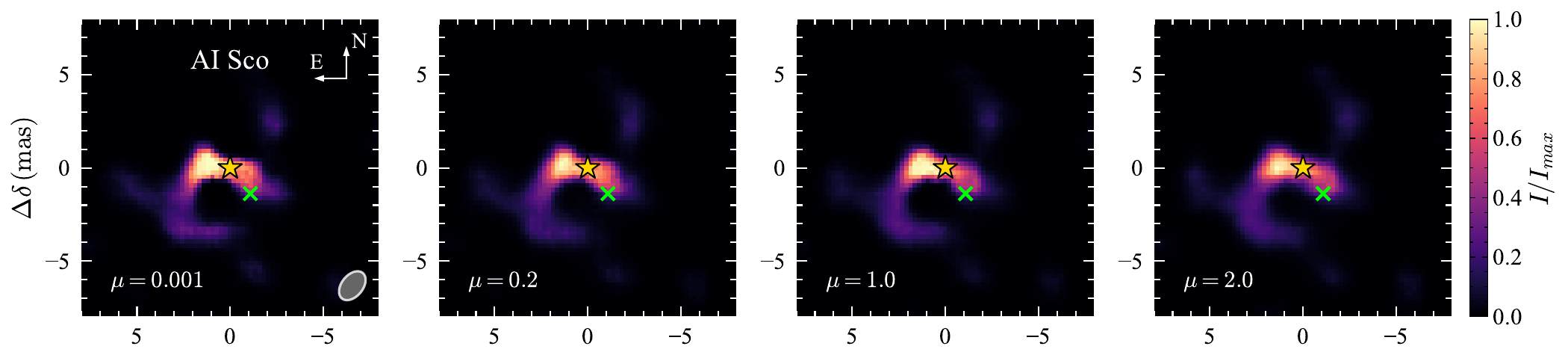}
			\end{subfigure}

			\vspace{-1ex}
			\begin{subfigure}
				{\textwidth}
				\includegraphics[width=\linewidth]{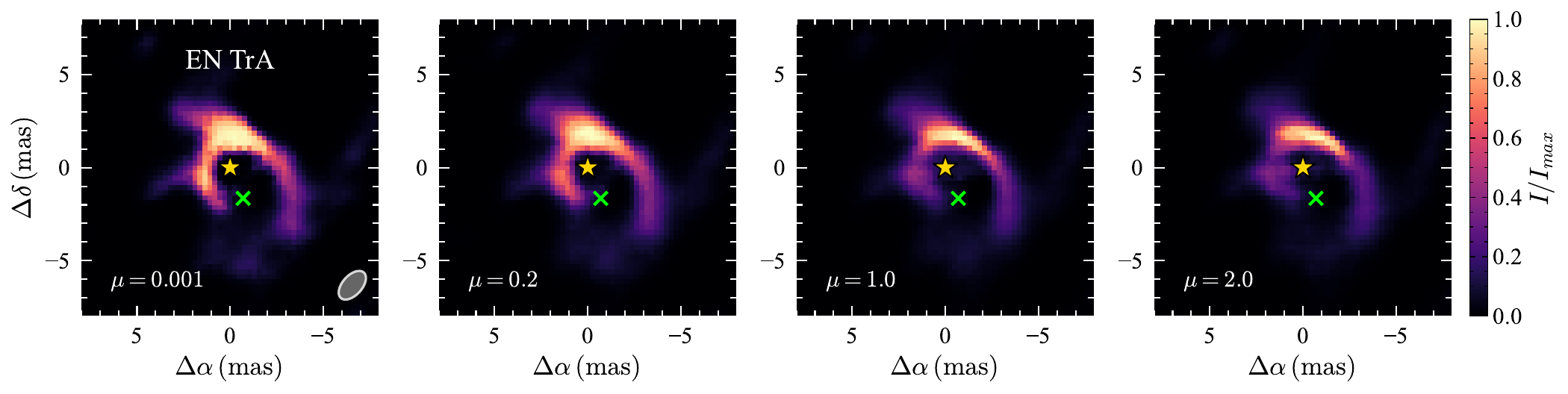}
			\end{subfigure}

			\caption{Illustration of the effect of different values for the
			regularisation weight $\mu$. The interferometric beam size is given in the
			lower-right corner. All other symbols are analogous to Fig.\ \ref{fig:organic_imgs_first_four}.}
			\label{fig:organic_mu_effect}
		\end{figure*}
		In this appendix, we investigate the effect of the regularisation weight $\mu$
		on the ORGANIC images.

		\label{sect:organic_implicit_regularisation} The total reconstruction loss in
		ORGANIC utilises the score output of the GAN discriminator evaluated on the generated
		image, $\hat{s}_{\mathrm{img}}$, as a regularisation term (via
		$\mathcal{L}_{\mathrm{reg}}\!=\!-\ln{\hat s_{\mathrm{img}}}$). Its aim is to
		make the imaged rims look more like the RT model training data. In this section,
		however, we show that most of the regulatory properties of ORGANIC are not caused
		by the numerical weight of $\mathcal{L}_{\mathrm{reg}}$ in the total reconstruction
		loss (Eq.\ \ref{eq:img_rec_tot_loss}), but are instead inherent to the CNN architecture
		itself, as well as the initial training of the generator's convolutional
		kernel weights.

		Fig.\ \ref{fig:organic_mu_effect} shows example ORGANIC images for AI~Sco and
		EN~TrA at various values of the regularisation weight $\mu$ (using the same
		parameters as in Table\ \ref{table:img_rec_params}). Higher $\mu$ generally
		leads to slightly less pixelation, radially sharper rims and more sharply defined
		azimuthal brightness modulations (though over-regularisation tends to occur
		for $\mu \!\gtrsim\! 1$, making the rim too sharp, as seen for EN~TrA).
		Nevertheless, the recovered structures are highly similar, especially on the
		scale of the interferometric beam. This holds even for $\mu\!=\!0.001$,
		which renders the numerical weight of $\mathcal{L}_{\mathrm{reg}}$
		negligible. At very low $\mu$ values, most differences are likely stochastic
		in nature, induced by the random initialisations of the latent noise vector and
		the Monte Carlo dropout layers in ORGANIC \citep[][]{Claes2020}. ORGANIC clearly
		possesses an implicit regularisation floor.

		A similar implicit regularisation was noticed and utilised by the PIRATES\footnote{\url{https://github.com/SAIL-Labs/PIRATES}}
		image reconstruction framework of \citet{Lilley2026}. The authors presented
		a CNN that is applied directly on polarimetric masking interferometry data
		from the VAMPIRES instrument \citep[][]{Lucas2024}. In a first initial
		training step, the convolutional kernel weights of the network are trained to
		map from a set of VAMPIRES observables to a cube of images in Stokes $(I,Q,U)$
		space. The training set for this step consists of sets of Stokes scattered
		light image cubes for various dusty geometries and their corresponding
		VAMPIRES observables, modelled using the MCFOST\footnote{\url{https://github.com/cpinte/mcfost}}
		RT code \citep[][]{Pinte2006, Pinte2009}. This forces the CNN kernels to
		learn the scattering-induced, physically meaningful spatio-polarimetric relationships
		between images of different Stokes components. At reconstruction time, when faced
		with an arbitrary set of VAMPIRES observables, the initially trained CNN
		directly maps to a corresponding $(I,Q,U)$ image cube. In a second step,
		this resulting initial image cube is refined by an iterative fitting routine.
		The initially trained kernel weights of the CNN are trained further in order
		to match the predicted interferometric observables, calculated from the CNN-generated
		image, to those actually observed. The generated image cube at the end of
		this procedure is taken as the final reconstruction. This results in well-regularised
		$(I,Q,U)$ images that provide a good match to the data, but also retain the
		scattering-induced connections between different Stokes components.

		The iterative fitting routine in PIRATES is highly similar to ORGANIC's
		reconstruction algorithm. The ORGANIC reconstruction routine also updates the
		initially trained generator kernel weights in order to make the generated image
		match a set of interferometric data. The main difference between the two is
		that PIRATES acts directly on VAMPIRES observables, while ORGANIC acts on a latent
		noise space (allowing for reconstruction at arbitrary $(u,v)$ coverage, but inducing
		an additional source of uncertainty; Sect.\ \ref{sect:organic_imgrec}). At no
		point does PIRATES include an explicit $\mathcal{L}_{\mathrm{reg}}$ term in its
		reconstruction. Instead, it relies fully on the general inductive bias imparted
		by CNN architectures \citep[such as locality and weight-sharing, e.g.][]{Wang2024, Ulyanov2020}
		-- a property shared by other types architectures, such as neural fields,
		when applied to similar ill-posed inverse reconstruction problems \citep[e.g.][]{Levis2024}
		-- as well as the initial training. The latter pre-emptively forces the weights
		into a subspace that generates images similar to the training data. Together
		with the limited effect of $\mu$ in ORGANIC, this shows that most of ORGANIC's
		regulatory properties share similar origins. They are implicitly imposed by
		the generator's CNN architecture and the initial training phase, not by the numerical
		weight of $\mathcal{L}_{\mathrm{reg}}$ in the total loss.

		The standard GAN architecture of ORGANIC precludes us from determining $\mu$
		through the conventional L-curve method (e.g.\ \citealt{Renard2011}; see Appendix
		\ref{sect:appendix_squeeze_reconstructions_appendix} for a description). This
		is due to the use of the discriminator score output in $\mathcal{L}_{\mathrm{reg}}$.
		The discriminator score estimates the Jensen-Shannon divergence between the
		image during reconstruction and the RT training set \citep[][]{Goodfellow2014}.
		This is a good distance measure during the initial phases of the image
		reconstruction, and does guide towards more regularised final images (as
		seen from Fig.\ \ref{fig:organic_mu_effect}). However, as the reconstruction
		progresses and the similarity between the image and training set decrease, the
		Jensen-Shannon divergence saturates to similar values and becomes uninformative
		as a measure \citep[][]{Arjovsky2017b}. This can be addressed in future ORGANIC
		designs by using a Wasserstein GAN, where the discriminator instead produces
		an actively calculated estimate of the Wasserstein distance \citep[][]{Arjovsky2017a}.
		When used as the regularisation loss $\mathcal{L}_{\mathrm{reg}}$, this Wasserstein
		estimate would remain an informative measure between the reconstructed image
		and the RT training set at all times. Since this would require an
		architectural overhaul of ORGANIC, this is currently beyond the scope of this
		work.

		We performed initial investigations of the image morphologies for all targets
		under varying $\mu$ within a reasonable range ($\mu = 0.1\text{--}2$). The
		main recovered features do not vary significantly under varying $\mu$. As it
		generally provided a good balance between suppression of noisy features and avoiding
		over-regularisation for our PIONIER datasets, we fixed $\mu = 0.2$ for the
		final ORGANIC reconstructions in this article (Figs.\ \ref{fig:organic_imgs_first_four}
		\& \ref{fig:organic_imgs_last_four}).

		\section{SQUEEZE reconstructions}
		\label{sect:appendix_squeeze_reconstructions_appendix}

		\begin{figure}[t]
			\centering
			\includegraphics[width=0.99\linewidth]{
				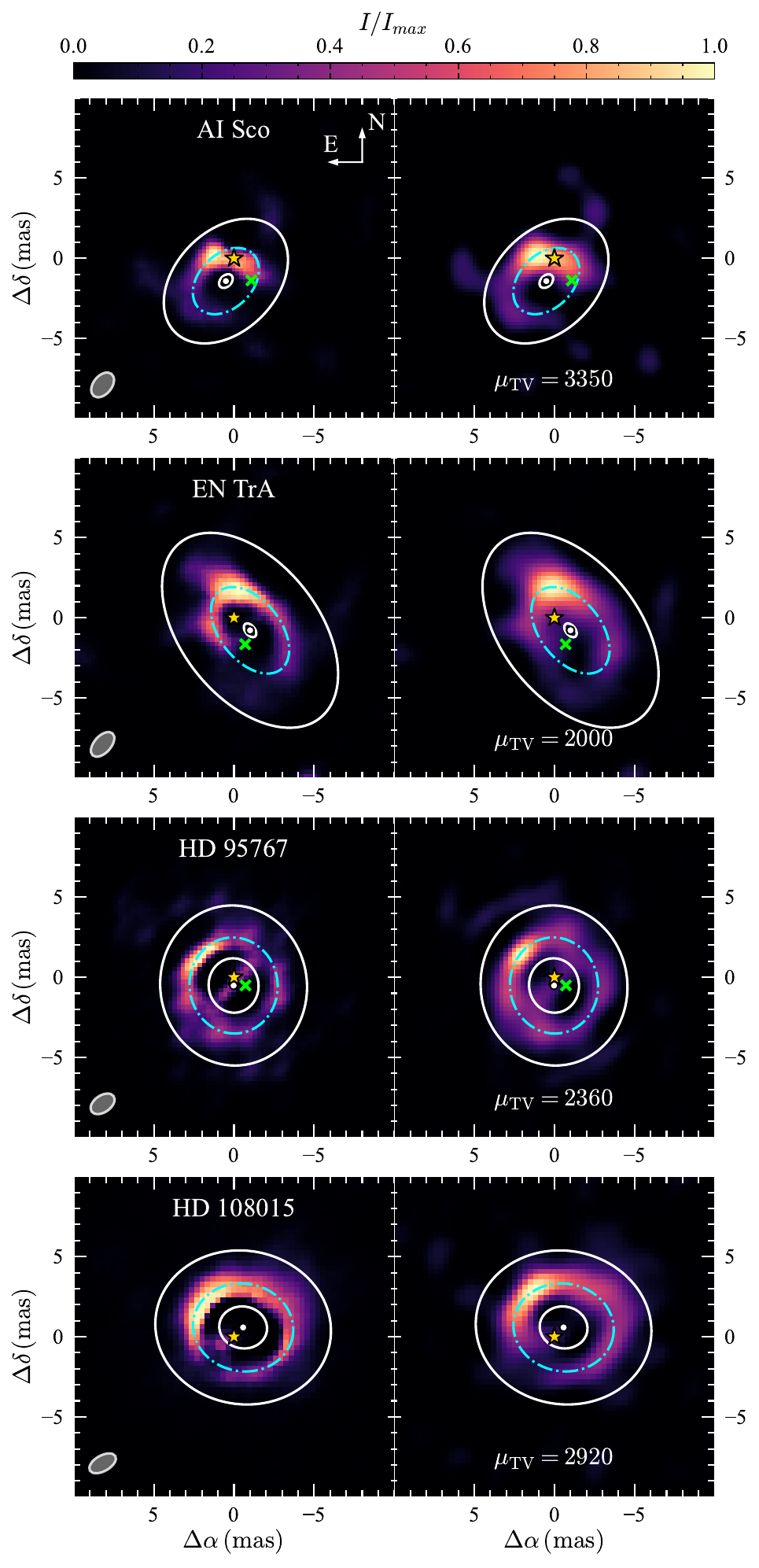
			}

			\caption{Comparison between our final ORGANIC reconstructions (left-hand images;
			Figs.\ \ref{fig:organic_imgs_first_four} \& \ref{fig:organic_imgs_last_four})
			and the alternative SQUEEZE reconstructions (right-hand images). The TV regularisation
			weight $\mu_{\mathrm{TV}}$ is displayed for each SQUEEZE reconstruction.
			White solid lines indicate the elliptical aperture used to calculate the
			azimuthal brightness profiles. In dot-dashed cyan, we display the rim fits
			derived from the ORGANIC images (Table \ref{table:img_rec_params}), whose orientation
			is used to define the azimuthal profile aperture. All other symbols
			analogous to Fig.\ \ref{fig:organic_imgs_first_four}.}
			\label{fig:squeeze_imgs}
		\end{figure}

		\begin{figure}[t]
			\ContinuedFloat
			\centering
			\includegraphics[width=0.99\linewidth]{
				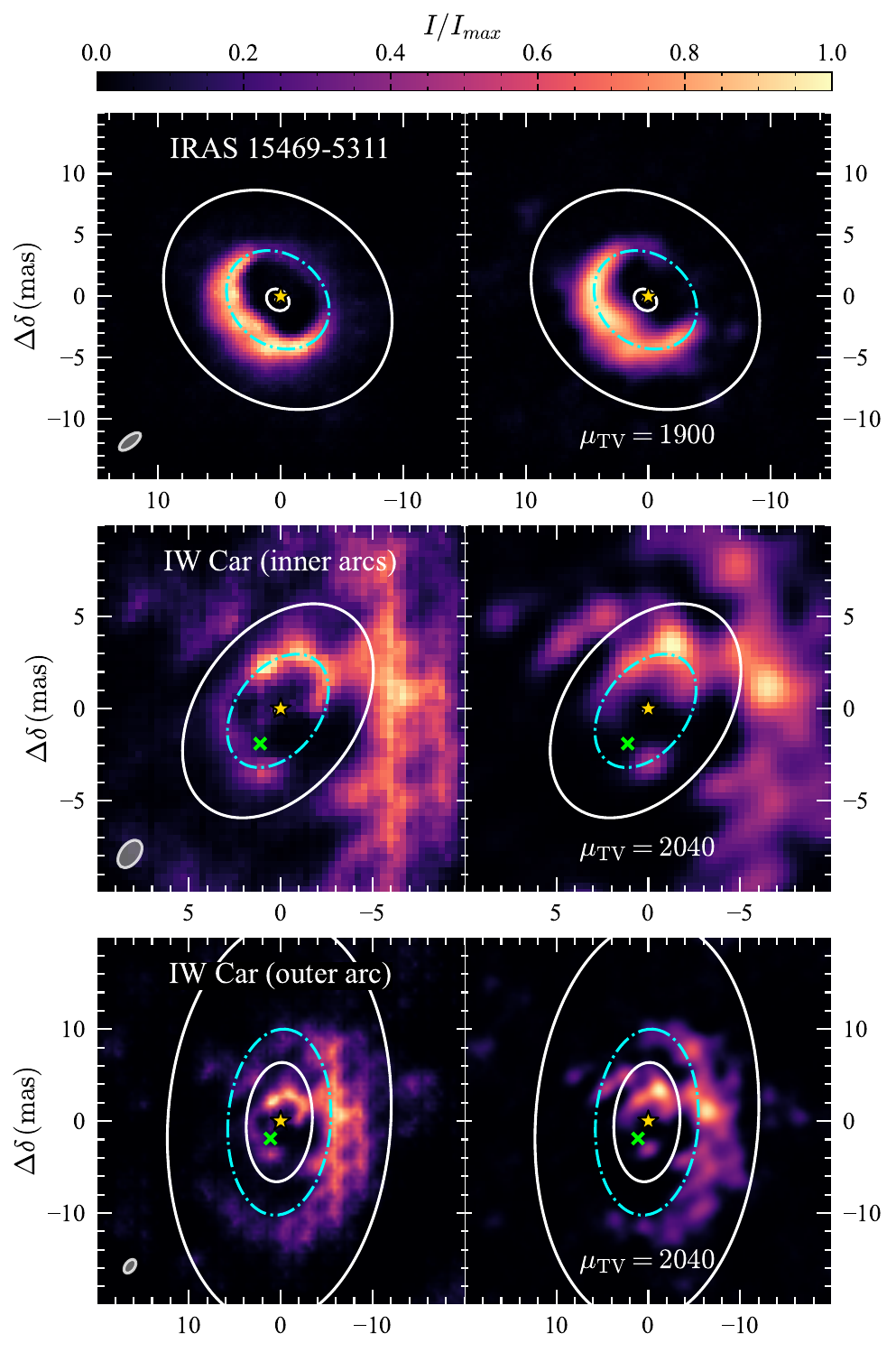
			}
			\caption{ continued. Note that for IW~Car's outer arc, the elliptical aperture
			is centred on $(\mathrm{\Delta\alpha_{rim}}, \mathrm{\Delta\delta_{rim}})$
			of the rim fit to the inner arcs. This is so the inner arcs can be
			properly masked out when calculating the outer arc's azimuthal profile. The
			$\mathrm{PA}$ of the azimuthal profile and the centre of the radial profile
			are thus also defined from this point.}
			\label{fig:squeeze_imgs}
		\end{figure}

		\begin{figure}[t]
			\centering
			\includegraphics[width=1.0\linewidth]{
				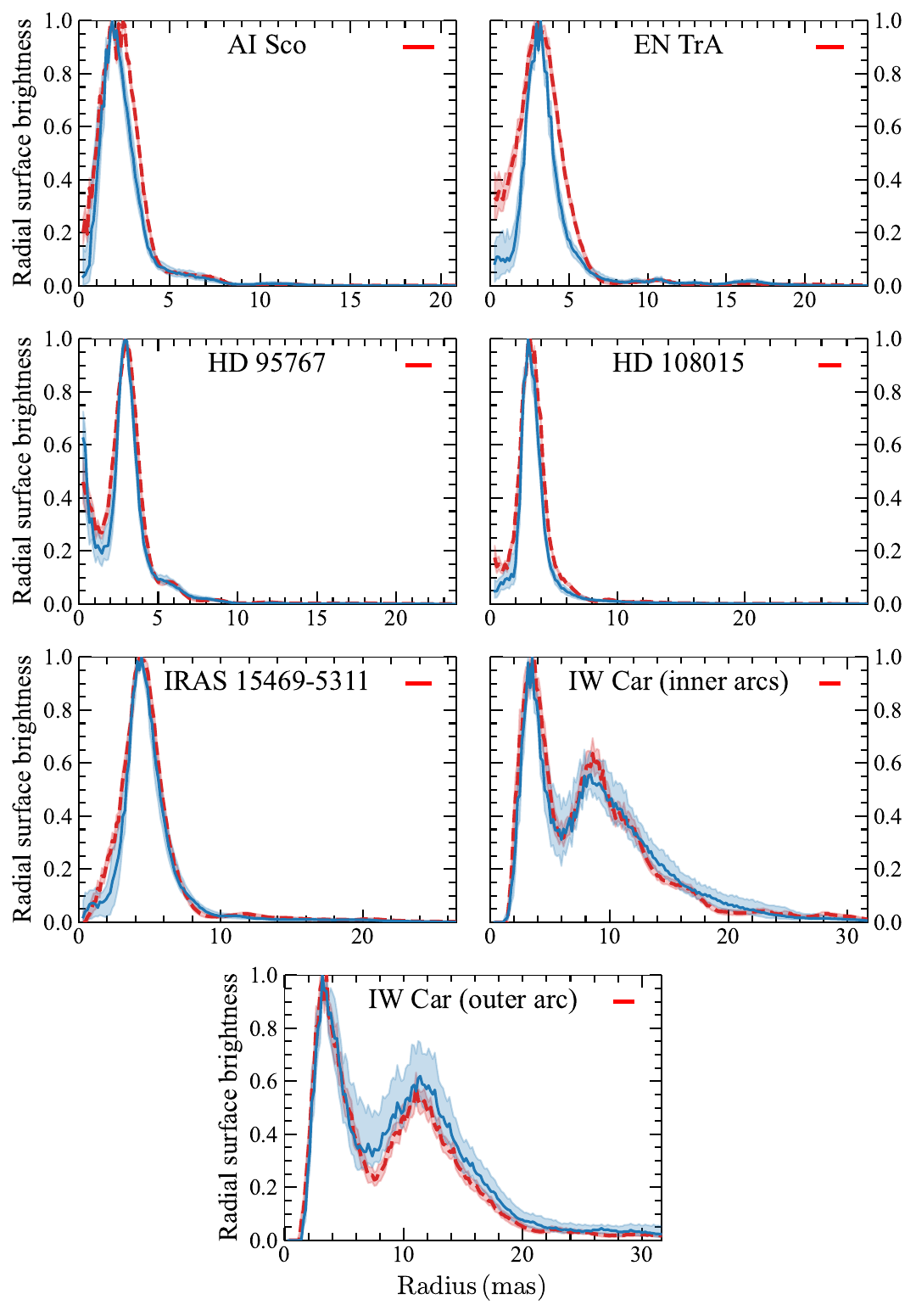
			}
			\caption{Peak-normalised radial brightness profiles of the ORGANIC (solid blue)
			and SQUEEZE (dashed red) image reconstructions displayed in Fig.\ \ref{fig:squeeze_imgs}.
			Both types of images are displayed with their $1\sigma$ significance
			contours. Red bars indicate the size of the azimuthally averaged interferometric
			beam.}
			\label{fig:squeeze_organic_rad_profile_comparison}
		\end{figure}

		\begin{figure}[t]
			\centering
			\includegraphics[width=1.0\linewidth]{
				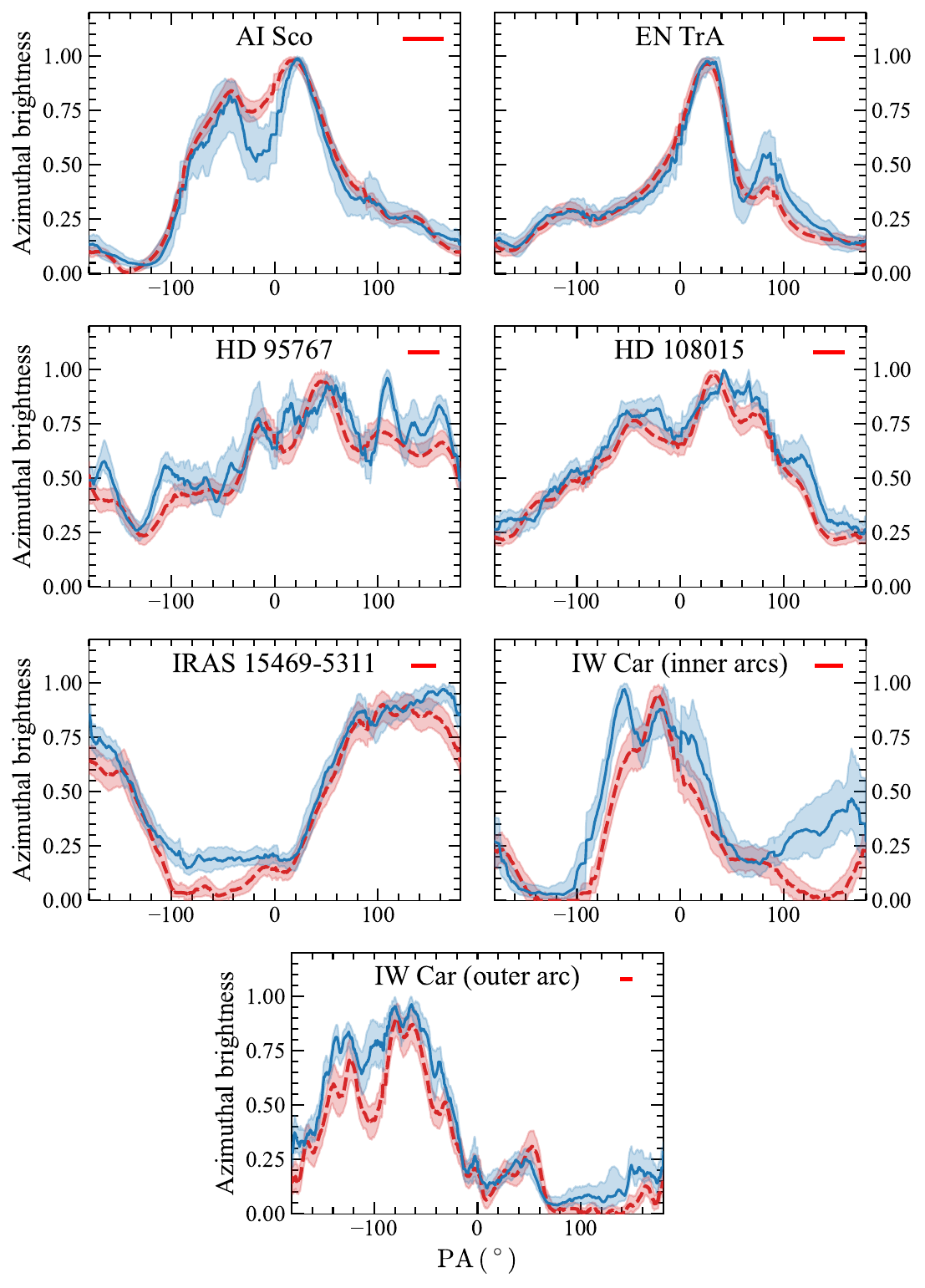
			}
			\caption{Analogous to Fig.\ \ref{fig:squeeze_organic_rad_profile_comparison}
			but for the peak-normalised azimuthal brightness profiles. The red bars
			now indicate the angle of the azimuthally averaged interferometric beam
			subtended at the reconstruction's rim.}
			\label{fig:squeeze_organic_az_profile_comparison}
		\end{figure}

		As further validation of our ORGANIC image reconstruction procedure (Sect.\ \ref{sect:img_rec}),
		we additionally performed image reconstructions on our targets using the
		SQUEEZE\footnote{\url{https://github.com/fabienbaron/squeeze}} package \citep[][]{Baron2010}.
		SQUEEZE uses a more classical minimisation engine and set of regularisations.
		We used SQUEEZE's implementation of SPARCO for a binary star (called \texttt{modelcode\_binary\_bwsmearing}
		internally). We then performed reconstructions using the same imaging and
		SPARCO parameters as in Table \ref{table:img_rec_params}. SQUEEZE uses an annealed
		MCMC routine with Metropolis-Hastings steps in order to minimise the loss,
		and includes various classical regularisation functions. The annealing scheme
		is intended to encourage convergence to the global minimum compared to more prevalent
		gradient-descent-based methods. Starting from different random images and
		using total variation (TV) regularisation -- which tends to produce piece-wise
		flat images \citep[][]{thiebaut2017} -- we ran 5 parallel MCMC chains for 2000
		steps. Annealing generally converged after $\sim \!100$. The mean of the final
		200 images of each chain was then taken as the final image. The
		regularisation weight $\mu$ was set through the L-curve method \citep[e.g.][]{Renard2011}.
		This method consists of performing reconstructions for different values of
		$\mu$, with the resulting plot of $\mathcal{L}_{\mathrm{data}}$ versus $\mathcal{L}
		_{\mathrm{reg}}$ showing an L-like curve. This curve represents a transition
		regime between dominance of one loss term over the other. The optimal $\mu$ value
		is chosen in the knee of this curve. The resulting SQUEEZE images are shown
		in Fig.\ \ref{fig:squeeze_imgs}.

		To provide a quantitative assessment of the similarity between the ORGANIC and
		SQUEEZE reconstructions, we also provide radial (following Sect.\ \ref{sect:rim_fitting_and_brightness_profile_calculation})
		and azimuthal (following Sect.\ \ref{sect:geom_model_synthetic_reconstructions})
		brightness profiles. These are calculated using elliptical apertures following
		the rim orientations given in Table \ref{table:img_rec_params}. The profiles
		are displayed in Figs.\ \ref{fig:squeeze_organic_rad_profile_comparison} \& \ref{fig:squeeze_organic_az_profile_comparison}.
		We find the following for each of our targets:

		\textit{AI~Sco:} Both methods agree on the general shape of the radial and azimuthal
		profile. The SQUEEZE image's TV regularisation does tend to produce a
		slightly broader radial profile and less pronounced azimuthal brightness
		depression of the disc flux close to the primary's position (though this depression
		is likely partially an artefact of the $(u,v)$ coverage; Sect.\ \ref{sect:geom_model_synthetic_reconstructions}).
		This is due to the tendency of TV to prefer broad, piece-wise flux plateaus
		and heavy punishment of flux gradients between pixels within those plateaus
		\citep[e.g.][]{Rudin1992}.

		\textit{EN~TrA:} Both methods agree well on the azimuthal profile. As for AI~Sco,
		the TV regularisation of the SQUEEZE image tends to produce a slightly
		broader radial profile.

		\textit{HD~95767:} Both methods agree well on the radial profile. The
		ORGANIC azimuthal profile, while sharing its main features with the SQUEEZE profile,
		is generally choppier, and unphysical broadening of the rim causes a secondary
		azimuthal brightness spike at $\mathrm{PA \sim 120^\circ}$. This choppy
		nature is likely caused by dirty beam artefacts (see Sect.\ \ref{sect:identifying_beam_artefacts}).
		The TV regularisation more heavily punishes these features, resulting in a smoother
		azimuthal profile.

		\textit{HD~108015:} Both methods agree well on both the radial and azimuthal
		profile. The exact azimuthal position of the brightness enhancement of the rim
		differs by $\sim\!10^{\circ}$ in $\mathrm{PA}$, and the enhancement is
		slightly broader for the ORGANIC image. Nevertheless, this does not
		significantly affect the main results presented in Sect.\ \ref{sect:EN_TrA_detected_features}:
		the brightness enhancement is significantly shifted towards the rim's
		projected major axis.

		\textit{IRAS~15469-5311:} Both methods agree well on the radial profile. The
		azimuthal profiles also match in main features, indicating a rim mostly symmetric
		along its projected minor axis with no other strong brightness enhancements of
		note. The TV regularisation of the SQUEEZE image does discourage narrow flux
		features surrounded by pixels of little flux, resulting in a deeper flux depression
		along the darkest part of the rim.

		\textit{IW~Car:} Both methods agree on the general features of the radial
		and azimuthal profiles for both the inner and outer arcs. However, there are
		mismatches in the specifics of the azimuthal profiles, such as the brightness
		of the inner arcs' southeastern sector. This indicates that the specifics of
		the azimuthal profiles for IW~Car are not entirely well-constrained, and
		should be treated with caution. This nevertheless does not affect the main results
		presented in Sect.\ \ref{sect:IW_Car_detected_features}: both inner arcs and
		a large outer arc are retrieved in the morphology.

		\section{Image robustness against $(u, v)$-coverage artefacts}
		\label{sect:appendix_robustness_against_uv_coverage}

		\begin{figure}[t]
			\centering
			\includegraphics[width=1.0\linewidth]{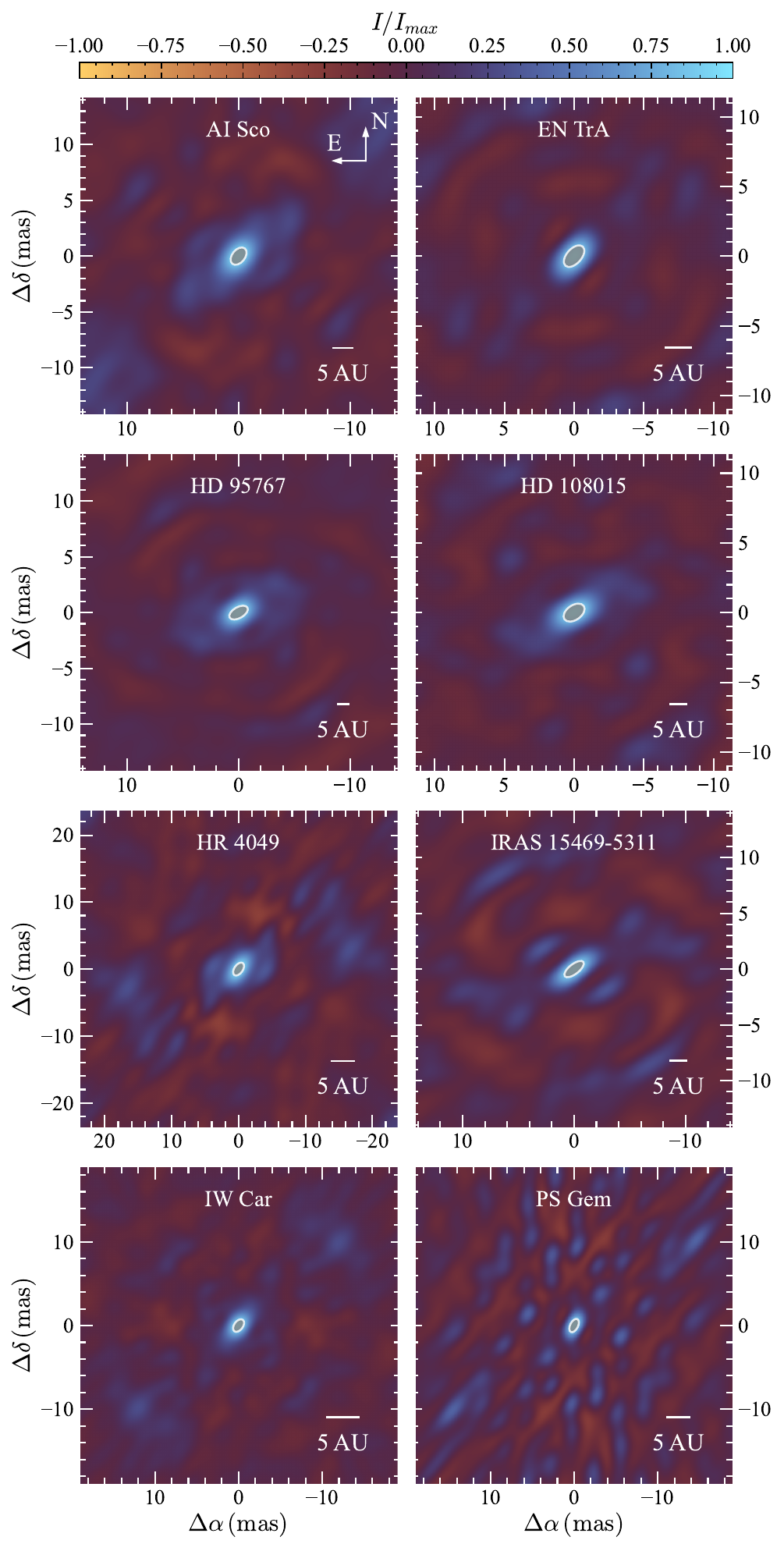}
			\caption{Dirty beams of the observations. Calculated using the sampled $(u,
			v)$ coverages in Fig.\ \ref{fig:uv_coverage}. Central ellipses denote the interferometric
			beam (short side of size
			$\sim \lambda_{\mathrm{min}}/ (2B_{\mathrm{max}})$).}
			\label{fig:dirty_beams}
		\end{figure}

		In this appendix, we assess the robustness of our ORGANIC image
		reconstructions (Figs.\ \ref{fig:organic_imgs_first_four} \&
		\ref{fig:organic_imgs_last_four}) against any dirty beam artefacts potentially
		induced by the sparse $(u, v)$ coverage of our observations. Such artefacts
		are often characteristic of optical interferometry due to the low number of
		telescopes used, and if left unidentified, can significantly bias the
		interpretation of recovered image features.

		\subsection{Dirty beams}
		The sparse $(u, v)$ coverage in optical interferometry can be the cause of
		non-physical artefacts in the reconstructed images that have no basis in the
		actual object's intensity distribution. These artefacts are caused by secondary
		lobes in the interferometric point spread function, also called the dirty
		beam. The dirty beam is calculated using the inverse direct Fourier transform,
		assuming a point source frequency response (i.e.\ a visibility of unity) at the
		sampled $(u,v)$ points and zero visibility elsewhere. In Fig.\ \ref{fig:dirty_beams},
		we show the dirty beams corresponding to our target's $(u,v)$ coverages (Fig.\ \ref{fig:uv_coverage}).
		The centre of the dirty beam consists of an elliptical Gaussian-like peak
		called the primary beam. The centre of this peak is called the interferometric
		beam size -- with a short side is of size
		$\sim \lambda_{\mathrm{min}}/ (2B_{\mathrm{max}})$, with $\lambda_{\mathrm{min}}$
		the shortest wavelength and $B_{\mathrm{max}}$ the longest baseline -- and
		defines the formal resolution element of the observations. Further away from
		the primary beam, one can see multiple other features (even negative flux features
		that can cause subtraction effects) called secondary lobes, which are the cause
		of reconstruction artefacts. In the following, we will apply a test that helps
		identify such artefacts in our final ORGANIC reconstructions.

		\subsection{Identifying potential beam-induced artefacts}
		\label{sect:identifying_beam_artefacts} In Sect.\ \ref{sect:results}, we extensively
		discuss the general features detected in the ORGANIC images. To characterise
		the robustness of these features against dirty beam artefacts, we performed a
		simulation test using a mirrored version of the final ORGANIC images \citep[similar
		to e.g.][]{Planquart2024}. We extracted synthetic $V2$ and $\phi3$ observables
		from horizontally mirrored versions of the final ORGANIC images (Figs.\ \ref{fig:organic_imgs_first_four}
		\& \ref{fig:organic_imgs_last_four}), using the same $(u, v)$ coverage (Fig.\ \ref{fig:uv_coverage}),
		imaging and SPARCO parameters (Table \ref{table:img_rec_params}), and
		observable uncertainties (Figs.\ \ref{fig:observables_v2} \&
		\ref{fig:observables_t3}). Following the procedure outlined in Sect.\ \ref{sect:organic_image_reconstruction},
		we then performed image reconstruction with ORGANIC on these synthetic
		observables. This procedure is equivalent to taking the final ORGANIC image as
		a 'model', and seeing how reliable the features of this model are recovered
		under reconstruction with a mirrored $(u, v)$ coverage. Any general features
		(brightness enhancements, flux arcs, etc.) that retain their integrity in
		the resulting mirror reconstructions -- taking into account that the reconstruction
		induces additional blurring and stretching in the direction of the primary beam
		-- are highly robust against changes in $(u, v)$ coverage and hence dirty
		beam artefacts. Features that disappear are not reliable, as they are highly
		sensitive to the specifics of the $(u, v)$ coverage, and hence likely affected
		by dirty beam artefacts. These mirrored reconstructions are presented in Fig.\ \ref{fig:organic_vs_mirror_reconstructions}.
		We find the following for each of our targets:

		\begin{figure}[t]
			\centering
			\includegraphics[width=1.0\linewidth]{
				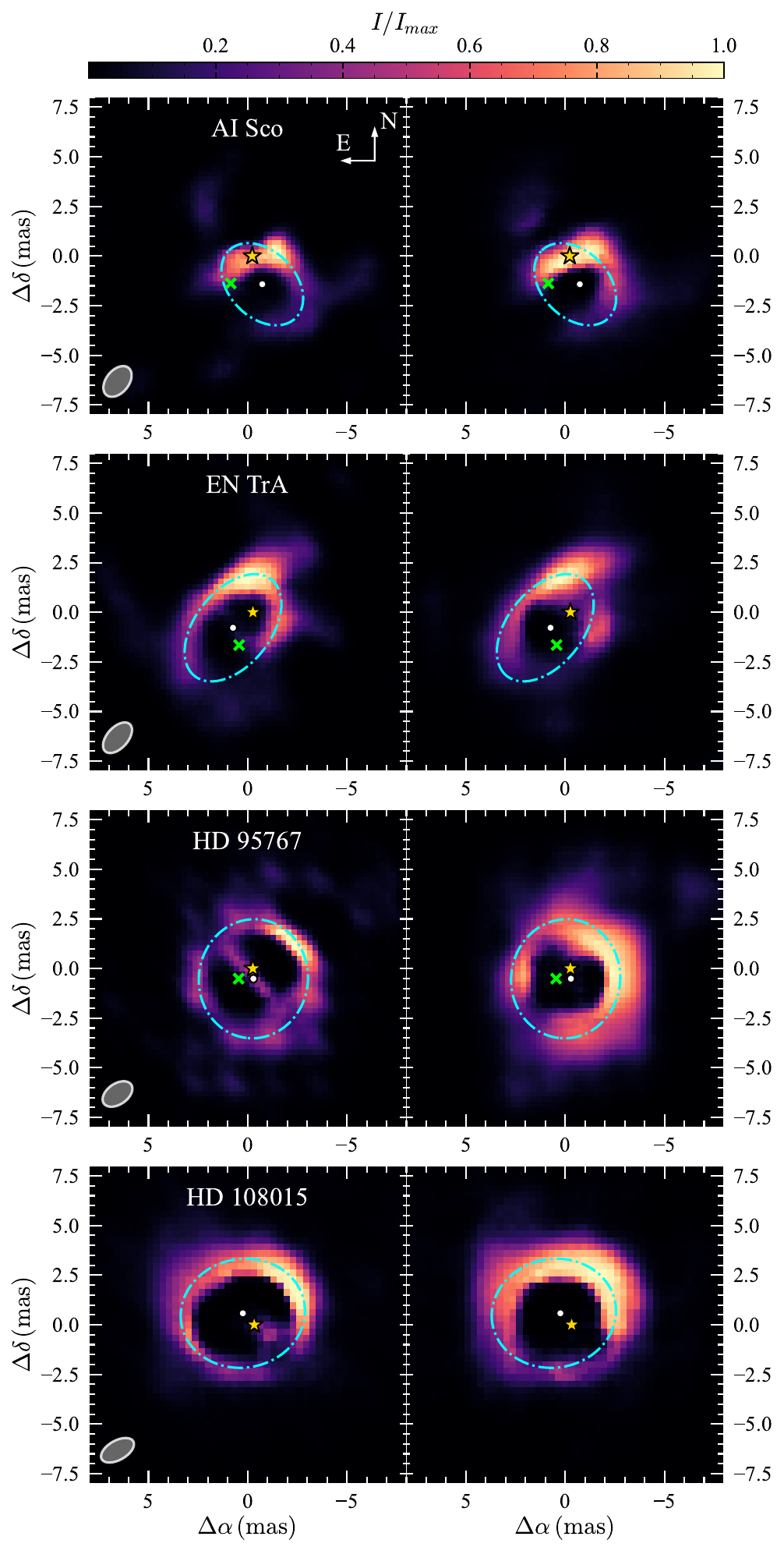
			}

			\caption{Mirrored final ORGANIC images (left) and their simulated image
			reconstructions from synthetic data (right). The displayed rim fits are mirrored
			versions of those derived from the ORGANIC images (Table
			\ref{table:img_rec_params}). All symbols otherwise analogous to Fig.\ \ref{fig:organic_imgs_first_four}.}
			\label{fig:organic_vs_mirror_reconstructions}
		\end{figure}

		\begin{figure}[t]
			\ContinuedFloat
			\centering
			\includegraphics[width=1.0\linewidth]{
				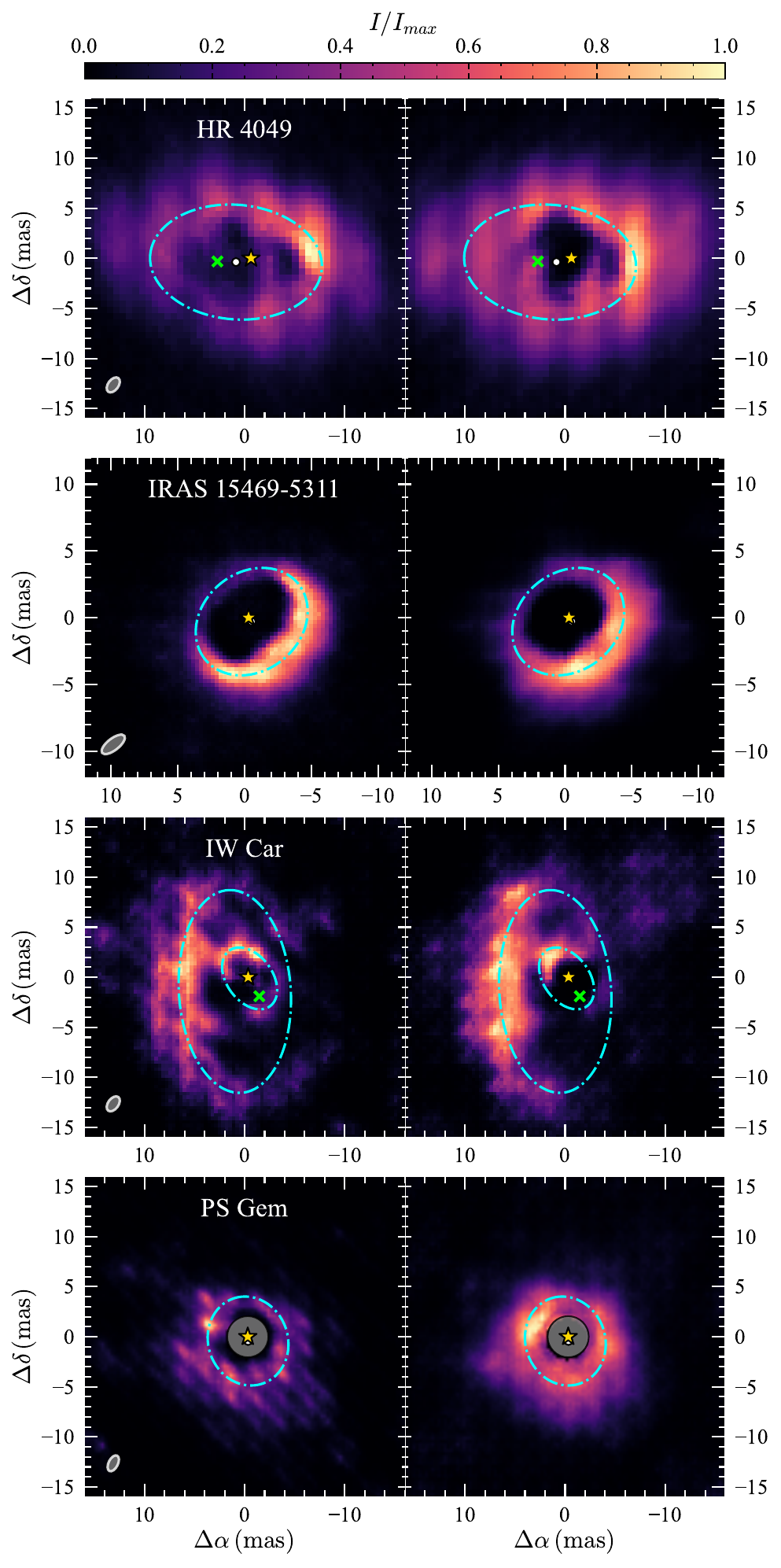
			}

			\caption{continued.}
			\label{fig:organic_vs_mirror_reconstructions}
		\end{figure}

		\textit{AI~Sco.} We recover the expected strong brightness enhancement close
		to the primary star, shifted towards the projected rim's major axis.

		\textit{EN~TrA.} We recover the expected strong brightness enhancement near the
		position of the binary star, shifted towards the projected rim's major axis.

		\textit{HD~95767.} The azimuthal brightness enhancement along the inner rim in
		the original image is heavily smeared out, and its contrast with the rest of
		the inner rim is heavily reduced. This indicates it is highly sensitive to the
		specifics of the $(u,v)$ coverage. Similarly, the central bar of emission crossing
		the inner rim cavity is not recovered, indicating it is an artefact induced
		by the dirty beam. In fact, HD 95765 seems to be the most sensitive to dirty
		beam artefacts out of all targets. A likely culprit is the exceptionally high
		secondary flux fraction -- $f_{\mathrm{sec}}\!\sim\!25\%$, while $f_{\mathrm{sec}}
		\lesssim5\%$ for the others (Table \ref{table:img_rec_params}). This causes
		significant degeneracy between the secondary and the rim's azimuthal
		intensity profile with respect to the $\phi3$ data. Slight over- or
		underestimations in the secondary's SPARCO parameters can then cause
		significant dirty beam artefacts, despite the otherwise excellent $(u,v)$ coverage
		(Fig.\ \ref{fig:uv_coverage}). The reconstructed mirror rim also widens substantially.
		While this is partly to be expected due to the mirror reconstruction
		introducing additional smearing, the effect is stronger than in the other targets,
		indicating that the excessive thinness of the rim at certain positions in
		the original image is unreliable. Due to these complications, reliable
		constraints on the rim morphology of HD~95767 will have to be derived from
		interferometric observations in longer wavelength bands (e.g.\ with VLTI/MATISSE),
		where the central stars' flux contributions are more diminished.

		\textit{HD~108015.} We recover the strong brightness enhancement shifted
		towards the projected rim's major axis. The small amount of flux found inside
		the cavity close to the darkest spot along the rim (where the intensity
		approaches zero) is not recovered. Similarly, the dark spot is filled in.
		These two features are most probably dirty beam artefacts.

		\textit{HR~4049.} The strong brightness enhancement shifted towards the rim's
		major axis is recovered. The deep drop in brightness in the south-western
		part of the original image (south-eastern in the mirrored image) is
		partially filled in, indicating it was slightly affected by dirty beam artefacts.
		Similarly, a non-negligible amount of flux is pulled into the rim's cavity in
		the mirror reconstruction, showing that the small amounts of flux found
		interior of the rim in the original image are unreliable.

		\textit{IRAS~15469-5311.} The generally smooth profile IRAS~15469-5311,
		largely symmetrically distributed around the rim's minor axis, is recovered.

		\textit{IW~Car.} Both the large outer arc as well as the inner arcs closer
		to the central binary are generally recovered, indicating their robustness.
		The lower-significance flux feature immediately south of the secondary star
		is, however, not recovered, marking it as a potential artefact.

		\textit{PS~Gem.} The general 'choppiness' of the rim has disappeared, and instead
		we recover a smoothed, smeared out version of the mirrored image. This is due
		to the fact that that the secondary lobes of PS Gem are relatively anti-symmetric
		under horizontal mirroring (Fig.\ \ref{fig:dirty_beams}). As a result, the
		choppy nature of the original reconstruction caused by dirty beam artefacts
		is mostly cancelled out by similar artefacts of the mirrored dirty beam. Interestingly,
		the bright spot towards the north-west in the original image (north-east in the
		mirrored image) is recovered up to a slight loss in contrast. Nevertheless,
		given that the imaged rim in general is heavily affected by artefacts (Sect.\ \ref{sect:PS_Gem_detected_features}),
		we err on the side of caution, and do not consider it a reliable feature.

		\section{Selected data}
		\label{sect:selected_data_appendix} 
\begin{figure}[h!]
			\centering
			\includegraphics[width=0.93\linewidth]{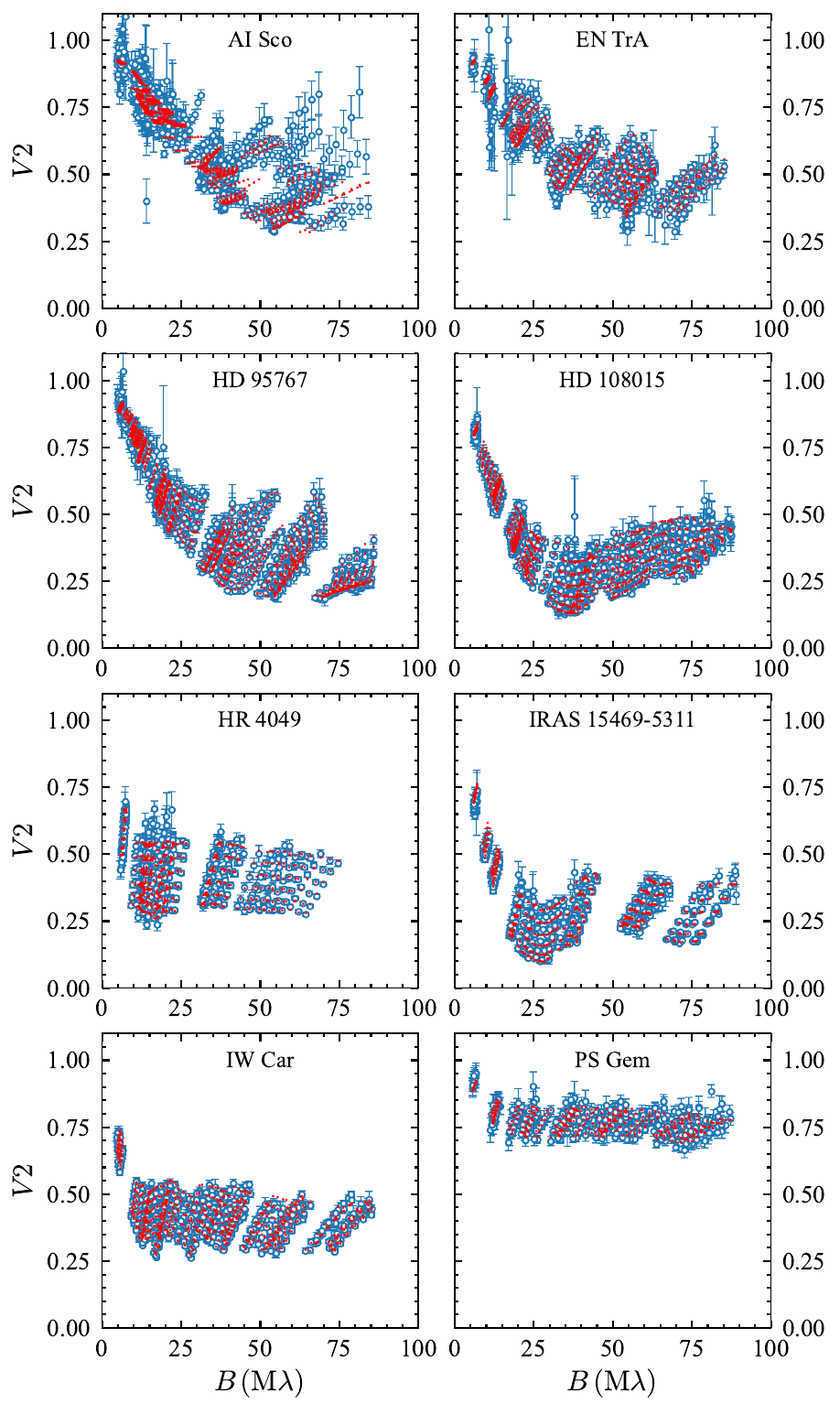}

			\caption{Selected $V2$ data from the INSPIRING PIONIER observations of our
			targets, shown in blue. For comparison, the observables synthesised from the
			final SPARCO+ORGANIC image reconstructions (Figs.\ \ref{fig:organic_imgs_first_four}
			\& \ref{fig:organic_imgs_last_four}) are overlaid as red dots.}
			\label{fig:observables_v2}
		\end{figure}

\begin{figure}[h!]
			\centering
			\includegraphics[width=0.93\linewidth]{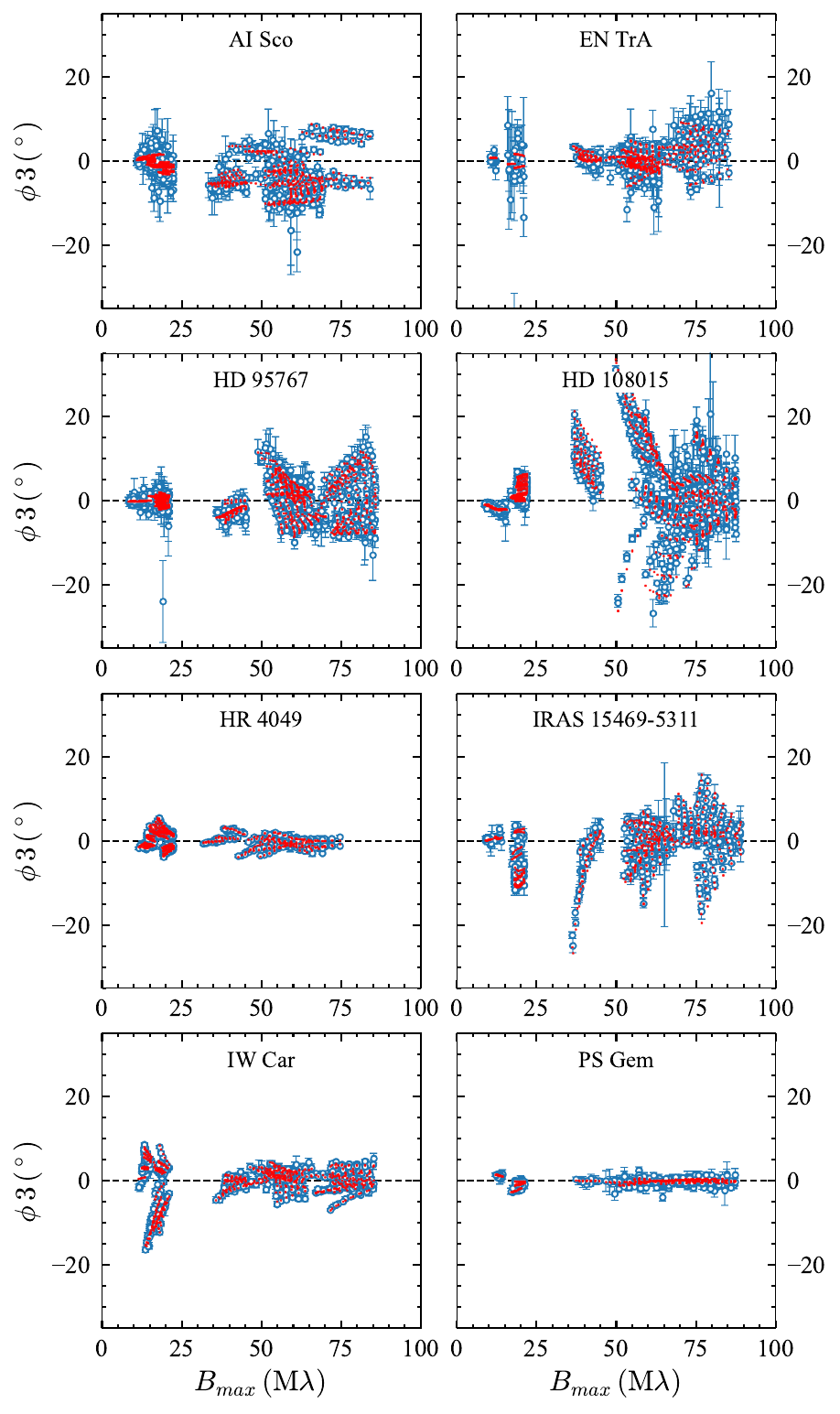}

			\caption{Analogous to Fig.\ \ref{fig:observables_v2} but for the selected $\phi
			3$ data.}
			\label{fig:observables_t3}
		\end{figure}

The $(u,v)$ coverages and observables of
		the selected PIONIER observations are shown in Figs.\ \ref{fig:uv_coverage},
		\ref{fig:observables_v2} \& \ref{fig:observables_t3}. Corresponding
		observation logs are available online on
		\href{https://doi.org/10.5281/zenodo.20155600}{Zenodo}.

	\end{appendix}
\end{document}